\documentclass[sigconf,screen]{acmart} 
\usepackage{todonotes}
\usepackage{soul}
\usepackage{listings}
\usepackage{fancyhdr}
\usepackage[normalem]{ulem}
\usepackage[keeplastbox]{flushend}
\usepackage[english]{babel}
\usepackage[utf8]{inputenc}
\usepackage{algorithm}
\usepackage{float}
\usepackage{wrapfig}
\usepackage[noend]{algpseudocode}
\usepackage{textcomp}
\usepackage{xspace}
\usepackage[binary-units, detect-all]{siunitx}
\usepackage{graphicx}
\usepackage{dblfloatfix} 
\usepackage{booktabs}
\usepackage{longfbox}
\usepackage{enumitem}
\usepackage{pifont}
\usepackage{microtype}
\usepackage[acronym,nonumberlist,nowarn]{glossaries}
    \newacronym{ADI}{ADI}{Alternating Direction Implicit}
\newacronym{AGU}{AGU}{Address Generation Unit}
\newacronym[longplural={Access History Tables}]{AHT}{AHT}{Access History Table}
\newacronym{AHT-R}{AHT-R}{Access History Table - Read}
\newacronym{AHT-W}{AHT-W}{Access History Table - Write-Back}
\newacronym{AIP}{AIP}{Access Interval Predictor}
\newacronym{AL}{AL}{Added Latency for column accesses}
\newacronym{ASIC}{ASIC}{Application Specific Integrated Circuit}
\newacronym{AVX}{AVX}{Advanced Vector Extensions}
\newacronym[longplural={Basic Block Vectors}]{BBV}{BBV}{Basic Block Vector}
\newacronym{BHT}{BHT}{Branch History Table}
\newacronym{HBM}{HBM}{Hybrid Bandwidth Memory}
\newacronym{DDR4}{DDR4}{Double-Data Rate 4}

\newacronym{NN}{NN}{Neural Network}
\newacronym{BTB}{BTB}{Branch Target Buffer}
\newacronym{CAS}{CAS}{Column Address Strobe}
\newacronym{AMAT}{AMAT}{Average Memory Access Time}
\newacronym{CCD}{CCD}{Column to Column Delay}
\newacronym{AI}{AI}{Arithmetic Intensity}
\newacronym[longplural={Columnar Database Systems}]{CDBS}{CDBS}{Columnar Database System}
\newacronym[longplural={Chip Multiprocessors}]{CMP}{CMP}{Chip Multiprocessor}
\newacronym{CMT}{CMT}{Chip Multithreading}
\newacronym[longplural={Coarse-Grain Reconfigurable Arrays}]{CGRA}{CGRA}{Coarse-Grain Reconfigurable Array}
\newacronym{C-RAM}{C-RAM}{Computational-RAM}
\newacronym{CWD}{CWD}{Column Write Delay}
\newacronym{DBI}{DBI}{Dynamic Binary Instrumentation}
\newacronym[longplural={Database Management Systems}]{DBMS}{DBMS}{Database Management System}
\newacronym{DDR}{DDR}{Double Data Rate}
\newacronym{DEWP}{DEWP}{Dead Line and Early Write-Back Predictor}
\newacronym{DRAM}{DRAM}{Dynamic Random Access Memory}
\newacronym{DSBP}{DSBP}{Dead Sub-Block Predictor}
\newacronym{DSM}{DSM}{Decomposed Storage Manage}
\newacronym{FAW}{FAW}{Four row Activation Window}
\newacronym{FP}{FP}{Floating-Point}
\newacronym{EDP}{EDP}{Energy-Delay Product}
\newacronym{FCFS}{FCFS}{First-Come First-Serve}
\newacronym{FIFO}{FIFO}{Fist In, First Out}
\newacronym{FSM}{FSM}{Finite State Machine}
\newacronym{FPGA}{FPGA}{Field-Programmable Gate Array}
\newacronym[longplural={Functional Units}]{FU}{FU}{Functional Unit}
\newacronym{GAg}{GAg}{Global Adaptive branch prediction using one Global PHT}
\newacronym{GAs}{GAs}{Global Adaptive branch prediction using per-Set PHT}
\newacronym{GCC}{GCC}{GNU Compiler Collection}
\newacronym{GEMS}{GEMS}{General Execution-driven Multiprocessor Simulator}
\newacronym[longplural={General Purpose Processors}]{GPP}{GPP}{General Purpose Processor}
\newacronym{HMC}{HMC}{Hybrid Memory Cube}
\newacronym{HIVE}{HIVE}{HMC Instruction Vector Extensions}
\newacronym{IATAC}{IATAC}{Inter-Access Time per Access Count}
\newacronym{ILP}{ILP}{Instruction Level Parallelism}
\newacronym{IPC}{IPC}{Instructions per Cycle}
\newacronym{ISA}{ISA}{Instruction Set Architecture}
\newacronym{IRAM}{IRAM}{Intelligent RAM}
\newacronym{KIPS}{KIPS}{Kilo Instructions per Second}
\newacronym{LDS}{LDS}{Linked Data Structure}
\newacronym{LFSR}{LFSR}{Linear Feedback Shift Register}
\newacronym{LFMR}{LFMR}{Last-to-First Miss-Ratio}
\newacronym{MLP}{MLP}{Memory-Level Parallelism}

\newacronym{LLC}{LLC}{last-level cache}
\newacronym{LRU}{LRU}{Least Recently Used}
\newacronym{LSU}{LSU}{Load Store Unit}
\newacronym{LTP}{LTP}{Last-Touch Predictor}
\newacronym{LvP}{LvP}{Live-time Predictor}
\newacronym{LWP}{LWP}{Last Write Predictor}
\newacronym{MARSS}{MARSS}{Micro Architectural and System Simulator}
\newacronym{McPAT}{McPAT}{Multi-core Power, Area, and Timing}
\newacronym{MIPS}{MIPS}{Microprocessor without Interlocked Pipeline Stages}
\newacronym{MOB}{MOB}{Memory Order Buffer}
\newacronym{MMU}{MMU}{Memory Management Unit}
\newacronym{MPKI}{MPKI}{Misses per Kilo-Instruction}
\newacronym{MSHR}{MSHR}{Miss-Status Handling Registers}
\newacronym{NAS}{NAS}{Numerical Aerodynamic Simulation}
\newacronym{NDP}{NDP}{Near-Data Processing}
\newacronym{NMP}{NMP}{Near Memory Processor}
\newacronym{NoC}{NoC}{Network-on-Chip}
\newacronym{NPB}{NPB}{NAS Parallel Benchmark}
\newacronym{NUCA}{NUCA}{Non-Uniform Cache Architecture}
\newacronym{NUMA}{NUMA}{Non-Uniform Memory Access}
\newacronym{OoO}{OoO}{Out-of-Order}
\newacronym{OpenMP}{OpenMP}{Open Multi-Processing}
\newacronym{OS}{OS}{operating system}
\newacronym{PAg}{PAg}{Per-address Adaptive branch prediction using one Global PHT}
\newacronym{PAs}{PAs}{Per-address Adaptive branch prediction using per-Set PHT}
\newacronym{PC}{PC}{Program Counter}
\newacronym[longplural={Pointer-Chasing Engines}]{PCE}{PCE}{Pointer-Chasing Engine}
\newacronym{PCM}{PCM}{Performance Counter Monitor}
\newacronym{PIM}{PIM}{Processing-in-Memory}
\newacronym[longplural={Pattern History Tables}]{PHT}{PHT}{Pattern History Table}
\newacronym{RAPL}{RAPL}{Running Average Power Limit}
\newacronym{RAS}{RAS}{Row Address Strobe}
\newacronym{RAT}{RAT}{Registers Alias Table}
\newacronym{RC}{RC}{Row Cycle}
\newacronym{RCD}{RCD}{RAS to CAS Delay}
\newacronym[longplural={Row-based Database Systems}]{RDBS}{RDBS}{Row-based Database System}
\newacronym{ROB}{ROB}{Reorder Buffer}
\newacronym{RRD}{RRD}{Row to Row activation Delay}
\newacronym{RP}{RP}{Row Precharge}
\newacronym{RTL}{RTL}{Register Transfer Level}
\newacronym{RTP}{RTP}{Read To Precharge}
\newacronym{RVU}{RVU}{Reconfigurable Vector Unit}

\newacronym{MIG}{MIG}{Majority-Inverter Graph}
\newacronym{AOIG}{AOIG}{AND/OR/Inverter Graph}
\newacronym{TRA}{TRA}{Triple-Row Activation}
\newacronym{AAP}{AAP}{ACTIVATION-ACTIVATION-PRECHARGE}
\newacronym{DCC}{DCC}{Dual-Contact Cell}

\newacronym{AP}{AP}{ACTIVATION-PRECHARGE}
\newacronym{SDP}{SDP}{Skewed Dead-Block Predictor}
\newacronym{SESC}{SESC}{Superescalar Simulator}
\newacronym{SSE}{SSE}{Streaming SIMD Extensions}
\newacronym{SFP}{SFP}{Spatial Footprint Predictor}
\newacronym{SiNUCA}{SiNUCA}{Simulator of Non-Uniform Cache Architectures}
\newacronym{SMT}{SMT}{Simultaneous Multi-Threading}
\newacronym{SIMD}{SIMD}{Single Instruction Multiple Data}
\newacronym{SPEC}{SPEC}{Standard Performance Evaluation Corporation}
\newacronym{SoC}{SoC}{System-on-Chip}
\newacronym{SPP}{SPP}{Spatial Pattern Predictor}
\newacronym{SRAM}{SRAM}{Static Random Access Memory}
\newacronym{SSOR}{SSOR}{Symmetric Successive Over-Relaxation}
\newacronym{SSV}{SSV}{Search Set Vector}
\newacronym{TLB}{TLB}{Translation Look-aside Buffer}
\newacronym{TLP}{TLP}{Thread Level Parallelism}
\newacronym[longplural={Through-Silicon Vias}]{TSV}{TSV}{Through-Silicon Via}
\newacronym{VWQ}{VWQ}{Virtual Write Queue}
\newacronym{WTR}{WTR}{Write Recovery time}
\newacronym{WR}{WR}{Write To Read delay time}

\usepackage{mwe}
\usepackage{mathtools}
    
    \DeclarePairedDelimiter\floor{\lfloor}{\rfloor}
\usepackage{multirow}
\usepackage[compact]{titlesec} 
\usepackage{subcaption}
\usepackage{cleveref}
\usepackage{tikz}
\usepackage{changepage}
\usepackage{textcomp}
\usepackage{hhline}
\usepackage[compatibility=false, skip=2pt]{caption}
\usepackage{hyperref}
\usepackage{cleveref}

\crefformat{section}{\S#2#1#3}
\crefformat{subsection}{\S#2#1#3}
\crefformat{subsubsection}{\S#2#1#3}

\usetikzlibrary{decorations.pathreplacing,calc}
\newcommand{\tikzmark}[1]{\tikz[overlay,remember picture] \node (#1) {};}

\newcommand*{\AddNoteRed}[4]{%
    \begin{tikzpicture}[overlay, remember picture]
        \draw [decoration={brace,amplitude=0.5em},decorate,ultra thick,red]
            ($(#3)!(#1.north)!($(#3)-(0,1)$)$) --  
            ($(#3)!(#2.south)!($(#3)-(0,1)$)$)
                node [align=center, text width=2.5cm, pos=0.5, anchor=west] {#4};
    \end{tikzpicture}
}%
\newcommand*{\AddNoteBlue}[4]{%
    \begin{tikzpicture}[overlay, remember picture]
        \draw [decoration={brace,amplitude=0.5em},decorate,ultra thick,blue]
            ($(#3)!(#1.north)!($(#3)-(0,1)$)$) --  
            ($(#3)!(#2.south)!($(#3)-(0,1)$)$)
                node [align=center, text width=2.5cm, pos=0.5, anchor=west] {#4};
    \end{tikzpicture}
}%
\newcommand*{\AddNoteGreen}[4]{%
    \begin{tikzpicture}[overlay, remember picture]
        \draw [decoration={brace,amplitude=0.5em},decorate,ultra thick,black]
            ($(#3)!(#1.north)!($(#3)-(0,1)$)$) --  
            ($(#3)!(#2.south)!($(#3)-(0,1)$)$)
                node [align=center, text width=2.5cm, pos=0.5, anchor=west] {#4};
    \end{tikzpicture}
}%

\AtBeginDocument{%
  \providecommand\BibTeX{{%
    \normalfont B\kern-0.5em{\scshape i\kern-0.25em b}\kern-0.8em\TeX}}}
    
\setcopyright{acmlicensed}
\acmPrice{15.00}
\acmDOI{10.1145/3445814.3446749}
\acmYear{2021}
\copyrightyear{2021}
\acmSubmissionID{asplos21main-p926-p}
\acmISBN{978-1-4503-8317-2/21/04}
\acmConference[ASPLOS '21]{Proceedings of the 26th ACM International Conference on Architectural Support for Programming Languages and Operating Systems}{April 19--23, 2021}{Virtual, USA}
\acmBooktitle{Proceedings of the 26th ACM International Conference on Architectural Support for Programming Languages and Operating Systems (ASPLOS '21), April 19--23, 2021, Virtual, USA}

\definecolor{dgreen}{rgb}{0.00, 0.75, 0.00}
\definecolor{dollarbill}{rgb}{0.52, 0.73, 0.4}
\definecolor{internationalorange}{rgb}{1.0, 0.31, 0.0}
\definecolor{islamicgreen}{rgb}{0.0, 0.56, 0.0}
\definecolor{amethyst}{rgb}{0.6, 0.4, 0.8}
\definecolor{mygreen}{rgb}{0,0.6,0}
\definecolor{mygray}{rgb}{0.5,0.5,0.5}
\definecolor{mymauve}{rgb}{0.58,0,0.82}
\definecolor{backcolour}{rgb}{0.95,0.95,0.92}
\definecolor{bluehl}{rgb}{0.8,0.874,1}
\definecolor{pinkhl}{rgb}{0.992156863,0.847058824,1}
\definecolor{greenhl}{rgb}{0.835,0.996,0.839}
\definecolor{yellowhl}{rgb}{0.996,0.957,0.8}
\newcommand\aap{\texttt{AAP}/\texttt{AP}\xspace}
\newcommand\aaps{\texttt{AAP}s/\texttt{AP}s\xspace}
\newcommand\uop{\textmu{}Op}
\newcommand\uprog{\textmu{}Program}
\newcommand\ureg{\textmu{}Register}
\newcommand\upc{\textmu{}PC}
\newcommand\uprogc{\textmu{}Program counter}

\newcommand{\revdel}[1]{}
\newcommand{\revdelrefa}[1]{}  
\newcommand{\revdelrefr}[1][0]{}  

\newcommand{\revonuriii}[1]{{#1}}
\newcommand{\omiii}[1]{\textcolor{black}{#1}} 
\newcommand{\omiv}[1]{\textcolor{black}{#1}} 
\newcommand{\omv}[1]{\textcolor{black}{#1}} 
\newcommand{\omvi}[1]{\textcolor{black}{#1}} 
\newcommand{\omvuii}[1]{\textcolor{black}{#1}} 
\newcommand{\omviii}[1]{\textcolor{black}{#1}}
\newcommand{\omix}[1]{\textcolor{black}{#1}}
\newcommand{\omdef}[1]{\textcolor{black}{#1}}
\newcommand{\omdefi}[1]{\textcolor{black}{#1}}
\newcommand{\omdefii}[1]{\textcolor{black}{#1}}
\newcommand{\omdefiii}[1]{\textcolor{black}{#1}}

\newcommand{\geraldorevii}[1]{{#1}}

\newcommand{\sgii}[1]{{#1}}
\newcommand{\nasrev}[1]{{#1}}
\newcommand{\om}[1]{{#1}}
\newcommand{\omi}[1]{{#1}} 
\newcommand{\omii}[1]{{#1}} 
 
\newcommand{\cutdel}[1]{}  
\newcommand{\sr}[1]{{#1}}

\newcommand{\cmr}[1]{\textcolor{black}{#1}} %

\newif\ifrevision
\revisionfalse

\ifrevision
     \newcommand\geraldorevi[1][0]{}
     
     \newcommand{\sgii}[1]{#1}
\else
    \newcommand{\geraldorevi}[1]{{#1}}
    \newcommand{\jgll}[1]{}
    
    \newcommand{\textfromsl}[1]{{\color{black}#1}}
    \newcommand{\gfrev}[1]{}
    
    \newcommand{\revonur}[1]{{#1}} 
    \newcommand{\revonurii}[1]{{#1}} 
    \newcommand{\revsgii}[1]{{#1}}
    \newcommand{\revonuri}[1]{{#1}} 
\fi

\ifrevision
     \newcommand\nasirevi[1][0]{}
\else
    \newcommand{\nasirevi}[1]{{#1}}
\fi

\newif\ifsubmission
\submissiontrue

\ifsubmission
    \newcommand\jgl[1]{}
    \newcommand\juang[1][0]{}
    \newcommand\juan[1][0]{}
    \newcommand\juangr[1][0]{}
    \newcommand\geraldo[1][0]{}
    \newcommand\revGeraldo[1][0]{}
    \newcommand\jr[1]{}
    \newcommand\gf[1]{}
    \newcommand\jds[1]{}
    \newcommand\joao[1][0]{}
    \newcommand\nas[1]{}
    \newcommand\nasi[1][0]{}
    \newcommand\nass[1][0]{}
    \newcommand\nastaran[1][0]{}
    \newcommand\nasii[1][0]{}
    \newcommand\onur[1][0]{}
    \newcommand\minp[1][0]{}
    \newcommand\mpr[1][0]{}
    \newcommand\mpi[1][0]{}
    \newcommand{\sg}[1][0]{}
    \newcommand\new[1][0]{}
    \newcommand\juangg[1][0]{}
    \newcommand\edit[1][0]{}
    \newcommand\rev[1][0]{}
\else
    \newcommand{\jgl}[1]{[{\color{dgreen}JGL: #1}]}
    \newcommand{\juang}[1]{{\color{black}#1}}
    \newcommand{\juan}[1]{{\color{black}#1}}
    \newcommand{\juangr}[1]{\textcolor{black}{#1}}
    \newcommand{\juangg}[1]{{\color{dgreen}#1}}
    \newcommand{\geraldo}[1]{{\color{blue}#1}}
    \newcommand{\revGeraldo}[1]{\textcolor{blue}{#1}}
    \newcommand{\jr}[1]{[{\color{blue}Geraldo: #1}]}
    \newcommand{\gf}[1]{[{\color{blue}GF: #1}]}
    \newcommand{\jds}[1]{[{\color{black}JDS: #1}]}
    \newcommand{\joao}[1]{{\color{black}#1}}
    \newcommand{\nas}[1]{[{\color{teal}Nas: #1}]}
    \newcommand{\nasi}[1]{{\color{black}#1}}
    \newcommand{\nass}[1]{#1}
    \newcommand{\nastaran}[1]{\textcolor{black}{#1}}
    \newcommand{\nasii}[1]{{\color{VioletRed}#1}}
    
    \newcommand{\onur}[1]{\textcolor{Orange}{#1}}
    \newcommand{\minp}[1]{\textcolor{black}{#1}}
    \newcommand{\mpr}[1]{\textcolor{dgreen}{#1}}
    \newcommand{\mpi}[1]{{\textcolor{dgreen}#1}}
    \newcommand{\sg}[1]{\textcolor{MidnightBlue}{#1}}
    \newcommand{\new}[1]{\textcolor{black}{#1}}
    \newcommand{\edit}[1]{\textcolor{black}{#1}}
    \newcommand{\rev}[1]{\textcolor{teal}{#1}}
\fi
\newcommand{\revAMicro}[1]{\textcolor{black}{#1}}
\newcommand{\revBMicro}[1]{\textcolor{black}{#1}}

\newcommand{\revCMicro}[1]{\textcolor{black}{#1}}
\newcommand{\revDMicro}[1]{\textcolor{black}{#1}}

\newcommand{\mech}{{SIMDRAM}\xspace} 
\newcommand\tempcommand[1]{\renewcommand{\arraystretch}{#1}}

\newcommand{\circled}[1]{\tikz[baseline=(char.base)]{\node[shape=circle,draw,inner sep=0pt,fill=black, text=white] (char) {#1};}}

\newcommand{\ignore}[1]{}

\algnewcommand\algorithmicforeach{\textbf{for each}}
\algdef{S}[FOR]{ForEach}[1]{\algorithmicforeach\ #1\ \algorithmicdo}
\algrenewcommand\algorithmicindent{0.5em}%

\lstdefinestyle{myC}{
  backgroundcolor=\color{backcolour},  
  basicstyle=\ttfamily\footnotesize,      
  breakatwhitespace=false,  
  breaklines=true,       
  captionpos=b,                   
  commentstyle=\color{mygreen},    
  deletekeywords={...},           
  escapechar=\%,
  xleftmargin=0pt,
  xrightmargin=0pt,
  aboveskip=\medskipamount,
  belowskip=\medskipamount,
  extendedchars=true,            
  keepspaces=true,               
  keywordstyle=\color{blue},      
  language=C++,              
  morekeywords={bbop_trsp_init, bbop_add, bbop_sub, bbop_greater, bbop_if_else, bbop_trsp_cpy, malloc, *,...},          
  numbers=left,                   
  numbersep=1pt,                   
  numberstyle=\tiny\color{mygray}, 
  rulecolor=\color{black},     
  showspaces=false,              
  showstringspaces=false,        
  showtabs=false,                 
  stepnumber=1,                    
  stringstyle=\color{mymauve},     
  tabsize=2,	                   
  title=\lstname                   
}

\newcommand{\HilightBlue}{\makebox[0pt][l]{\color{bluehl}\rule[-4pt]{\linewidth}{10pt}}}
\newcommand{\HilightPink}{\makebox[0pt][l]{\color{pinkhl}\rule[-4pt]{\linewidth}{10pt}}}
\newcommand{\HilightYellow}{\makebox[0pt][l]{\color{yellowhl}\rule[-4pt]{\linewidth}{10pt}}}
\newcommand{\HilightGreen}{\makebox[0pt][l]{\color{greenhl}\rule[-4pt]{\linewidth}{10pt}}}

\AtBeginDocument{\DeclareCaptionSubType{lstlisting}}
\crefname{sublstlisting}{listing}{listings}
\Crefname{sublstlisting}{Listing}{Listings}

 \titlespacing\section{0pt}{5pt plus 2pt minus 2pt}{0pt plus 2pt minus 2pt}
 \titlespacing\subsection{0pt}{5pt plus 2pt minus 2pt}{0pt plus 2pt minus 2pt}
 \titlespacing\subsubsection{0pt}{5pt plus 2pt minus 2pt}{0pt plus 2pt minus 2pt}

\begin{document}

\newif\ifcameraready
    \camerareadytrue
\newif\ifasploscr
    \asploscrfalse

\title[SIMDRAM: An End-to-End Framework for Bit-Serial SIMD Computing in DRAM]{SIMDRAM: An End-to-End Framework for \\ Bit-Serial SIMD Computing in DRAM}

\newcommand{\tsc}[1]{\textsuperscript{#1}} 
\newcommand{\affilETH}{\tsc{1}}
\newcommand{\affilSFU}{\tsc{2}}
\newcommand{\affilUIUC}{\tsc{3}}
\settopmatter{authorsperrow=1} 

\author{\vspace{-10pt}%
 {%
     *Nastaran Hajinazar\affilETH$^,$\affilSFU \qquad 
     *Geraldo F. Oliveira\affilETH \qquad 
     Sven Gregorio\affilETH \qquad 
     João Dinis Ferreira\affilETH
 }
}
\author{
 {
     Nika Mansouri Ghiasi\affilETH\qquad 
     Minesh Patel\affilETH \qquad 
     Mohammed Alser\affilETH \qquad 
     Saugata Ghose\affilUIUC
 }
}
\author{
 {
     Juan Gómez-Luna\affilETH\qquad 
     Onur Mutlu\affilETH 
 }
}
\affiliation{
\institution{
      \vspace{8pt}
      \affilETH ETH Zürich \qquad
      \affilSFU Simon Fraser University \qquad 
      \affilUIUC University of Illinois at Urbana–Champaign
  }
  \country{}
 }

\thanks{*Nastaran Hajinazar and Geraldo F. Oliveira are co-primary authors.}

\renewcommand{\authors}{Nastaran Hajinazar, Geraldo F. Oliveira, Sven Gregorio, João Ferreira, Nika Mansouri Ghiasi, Minesh Patel, Mohammed Alser, Saugata Ghose, Juan Gómez Luna, and Onur Mutlu}

\renewcommand{\shortauthors}{N. Hajinazar and G. F. Oliveira et al.}

\begin{abstract}
    Processing-using-DRAM \juangr{has been proposed for a limited set of basic operations (i.e., logic operations, addition).} 
However, in order to enable full adoption of processing-using-DRAM, it is necessary to provide support for more complex operations. In this paper, we propose \mech, \nasrev{\sgii{a flexible general-purpose processing-using-DRAM framework that (1)~enables the efficient implementation of complex operations, and (2)~provides a flexible mechanism to support the implementation of arbitrary user-defined operations.}} 
\nasrev{\sgii{The SIMDRAM framework comprises three key steps. The first step builds an efficient MAJ/NOT representation of a given desired operation. The second step allocates DRAM rows that are reserved for computation to the operation's input and output operands, and generates the required sequence of DRAM commands to perform the MAJ/NOT implementation of the desired operation in DRAM. The third step uses the SIMDRAM \emph{control unit} located inside the memory controller to manage the computation of the operation from start to end, by executing the DRAM commands generated in the second step of the framework.}}
\omii{We design the hardware and
ISA support for \mech framework to (1) address key system integration challenges, and (2) allow programmers to employ new SIMDRAM operations
without hardware changes.}


We evaluate SIMDRAM for reliability, area overhead, throughput, and energy efficiency using \nasii{a wide range of operations and \geraldo{seven} real-world applications \nasii{to demonstrate \omiv{SIMDRAM's} generality.}} 
\omiv{Our evaluations using a single DRAM bank show that (1)~over 16 operations,
SIMDRAM provides 2.0$\times$ the throughput and 2.6$\times$ the energy efficiency of Ambit, a \onur{state-of-the-art} \geraldorevii{processing-using-DRAM} mechanism; } 
\omiv{(2)~over seven real-world applications, SIMDRAM provides 2.5$\times$ the performance of Ambit.} 
\omiv{Using 16 DRAM banks, SIMDRAM provides (1)~88$\times$ and 5.8$\times$ the throughput, and 257$\times$ and \geraldo{31$\times$} the energy efficiency, of a CPU and a high-end GPU, respectively, over 16 operations; (2)~21$\times$ and 2.1$\times$ the performance of the CPU and GPU, over seven real-world applications.} \omiv{SIMDRAM incurs an area overhead of \sgii{only 0.2\% in a high-end CPU.}}
\end{abstract}


\renewcommand{\shortauthors}{N. Hajinazar and G. F. Oliveira et al.}
\renewcommand{\shorttitle}{SIMDRAM: An End-to-End Framework for Bit-Serial SIMD Computing in DRAM}

\maketitle

\section{Introduction}
\label{intro}
The increasing prevalence and growing size of data in modern applications
has led to high \omiv{energy and latency} costs for computation in traditional computer architectures.
Moving large \om{amounts} of data between memory (e.g., DRAM) and the CPU across 
bandwidth-limited memory channels can consume more than
60\% of the total energy in modern systems~\cite{mutlu2019, boroumand2018google}.
To mitigate \omiv{such} costs, \om{the} \emph{processing-in-memory} (PIM) \om{paradigm moves} computation closer to where the 
data resides, reducing (and in some cases eliminating) the
need to move data between memory and the processor.

There are two main approaches to PIM~\cite{ghoseibm2019, mutlu2020modern}:
(1)~processing-near-memory, where PIM logic is added to the same die as memory or to the logic layer of 3D-stacked memory~\cite{deoliveira2021, lee2016simultaneous, ahn2015scalable, nai2017graphpim, boroumand2018google, lazypim, top-pim, gao2016hrl, kim2018grim, drumond2017mondrian, santos2017operand, NIM, PEI, gao2017tetris, Kim2016, gu2016leveraging, HBM, HMC2, boroumand2019conda, hsieh2016transparent, cali2020genasm,Sparse_MM_LiM, NDC_Micro_2014, farmahini2015nda,loh2013processing,pattnaik2016scheduling,akin2016data, hsieh2016accelerating,babarinsa2015jafar,lee2015bssync, devaux2019true,ghose2018enabling}; and (2)~processing-using-memory, which makes use of the operational principles of the memory cells themselves to perform computation by enabling interactions between cells~\cite{Chi2016, Shafiee2016, seshadri2017ambit, seshadri2019dram, li2017drisa, seshadri2013rowclone, seshadri2016processing, deng2018dracc, xin2020elp2im, song2018graphr, song2017pipelayer,gao2019computedram, eckert2018neural, aga2017compute,dualitycache}. Since processing-using-memory operates directly in the memory \om{arrays}\revdel{cells}, it benefits from the large internal bandwidth and parallelism available inside the memory arrays, which processing-near-memory solutions \omiv{cannot take advantage of}.

A common approach for processing-using-memory architectures is to make use of bulk bitwise computation. Many widely-used data-intensive applications (e.g., databases, neural networks, graph analytics) heavily rely on a broad set of simple (e.g., AND, OR, XOR) and complex (e.g., equality check, multiplication, addition) bitwise operations. Ambit~\cite{seshadri2017ambit, seshadri2015fast}, an in-DRAM processing-using-memory accelerator, was the first work to propose exploiting DRAM's analog operation\om{al principles} to perform bulk bitwise AND, OR, and NOT logic operations. Inspired by Ambit, many prior works have explored DRAM (as well as NVM) designs that are capable of performing in-memory bitwise operations~\omiv{\cite{angizi2019graphide, angizi2018imce, ali2019memory, pinatubo2016, gao2019computedram, xin2020elp2im, angizi2020pim, he2020sparse, angizi2019redram, deng2018dracc,imani2019floatpim,angizi2018dima}}.
However, a major shortcoming prevents these proposals from becoming widely applicable: 
they support only basic operations (e.g., Boolean operations, addition) and fall short on flexibly \om{and easily} supporting new and more 
complex operations. Some prior works propose processing-using-DRAM designs that support more complex operations~\cite{li2017drisa, deng2018dracc}. However, such designs (1)~require significant changes to the DRAM subarray, and (2)~support only a limited and specific set of operations, lacking the flexibility to support new operations and cater to the wide variety of applications that can potentially benefit from in-memory computation. 

\textbf{Our goal} in this paper is to design a framework that aids the adoption of processing-using-DRAM by efficiently implementing complex operations and providing the flexibility to support new desired operations. To this end, we propose SIMDRAM, an end-to-end processing-using-DRAM framework that provides the programming interface, the ISA, and the hardware support for (1)~efficiently computing \emph{complex} operations, and (2)~providing the ability to implement \emph{arbitrary} operations as required, all in an in-DRAM massively-parallel SIMD substrate.
At its core, we build the SIMDRAM framework around a DRAM substrate that enables two previously-proposed techniques:
(1)~vertical data layout in DRAM, and 
(2)~majority-based logic for computation.

\textbf{Vertical Data Layout.} Supporting bit-shift operations is essential for implementing complex computations, such as addition or multiplication. Prior works show that employing a vertical layout~\omiv{\cite{batcher1982bit,shooman1960parallel, gao2019computedram,ali2019memory,eckert2018neural, dualitycache, cmhill, kahle1989connection,hillis1993cm, tucker1988architecture}} for the data in DRAM,  such that all bits of an operand are placed in a single DRAM column (i.e., in a single bitline), eliminates the need for adding extra logic in DRAM to implement shifting~\cite{deng2018dracc, li2017drisa}. Accordingly, \mech supports efficient bit-shift operations by storing operands in a vertical fashion in DRAM. This provides \mech with two key benefits. First, a bit-shift operation can be performed by simply copying a DRAM row into another row (using RowClone~\cite{seshadri2013rowclone}, LISA~\cite{chang2016low}, NoM~\cite{nom20} or FIGARO~\cite{wang2020figaro}). For example, SIMDRAM can perform a left-shift-by-one operation by copying the data in DRAM row $j$ to DRAM row \textit{j+1}. (Note that while SIMDRAM supports bit shifting, we can optimize many applications to avoid the need for explicit shift operations, by simply changing the row indices of the SIMDRAM commands that read the shifted data). Second, \mech enables massive parallelism, wherein each DRAM column operates as a SIMD lane by placing the source and destination operands of an operation on top of each other in the same DRAM column.

\textbf{Majority-Based Computation.} Prior works use majority operations to implement basic logical operations~\cite{seshadri2017ambit,gao2019computedram,seshadri2015fast,li2017drisa} (e.g., AND, OR) or addition~\cite{angizi2019graphide,ali2019memory,gaillardon2016programmable,li2017drisa,deng2018dracc,gao2019computedram}. These basic operations are then used as basic building blocks to implement the target in-DRAM computation. \mech extends the use of the majority operation by directly using \omiv{only} the logically complete set of majority (MAJ) and NOT operations to implement in-DRAM computation. Doing so enables \mech to achieve higher performance, throughput, and reduced energy consumption compared to using basic logical operations as building blocks for in-DRAM computation. We find that a computation typically requires fewer DRAM commands using MAJ and NOT than using basic logical operations AND, OR, and NOT.

\nasirevi{To aid the adoption of processing-using-DRAM by flexibly supporting new and more complex operations, \mech addresses \revonurii{two key} challenges: (1)~how to synthesize new arbitrary in-DRAM operations, and (2)~how to exploit an optimized implementation and control flow for such \revonurii{newly-added operations} \revonurii{while} taking into account key limitations of in-DRAM processing (e.g., DRAM operations \revonurii{that destroy input data}, limited number of DRAM rows that are capable of \revonurii{processing-using-DRAM}, and the need to avoid costly in-DRAM copies).}
As a result, \nasii{\mech is the first end-to-end framework for \revonurii{processing-using-DRAM}. \mech provides (1)~\onur{a\omiv{n} \omiv{effective algorithm}} to generate an efficient MAJ/NOT-based implementation of \revonurii{a given desired operation}; (2)~\omiv{\onur{\omiv{an} algorithm to} appropriately allocate DRAM rows \omiii{to} the operands of the operation and an algorithm to map the computation \nasirevi{to an efficient sequence of DRAM commands to execute \emph{any} MAJ-based computation;}} 
and (3)~the \revonurii{programming} interface, ISA support and hardware components required to }\revonurii{(i)~}\nasirevi{compute any new user-\omiv{defined} in-DRAM operation without hardware modifications\revonurii{, and (ii)~}program the memory controller for issuing DRAM commands to the corresponding DRAM rows and correctly performing the computation.}
\textfromsl{\omiv{Such} end-to-end support enables \mech as a holistic approach that facilitates the adoption of processing-using-DRAM \nasirevi{through (1) enabling the flexibility to support new in-DRAM operations by providing the user with a simplified interface to add desired operations, and (2) eliminating the need for adding extra logic to DRAM}.}

\textfromsl{The \mech framework efficiently supports a wide range of operations of different types. In this work, we demonstrate the functionality of the \mech framework using an example set of \omiv{16} operations including  (1)~\emph{N}-input logic operations (e.g., AND/OR/XOR of more than 2 input bits); (2)~relational operations (e.g., equality/inequality check, greater than, maximum, minimum); (3)~arithmetic operations (e.g., addition, subtraction, multiplication, division); (4)~predication (e.g., if-then-else); and (5)~other complex operations such as bitcount and ReLU~\revdelrefa{\hl{[40]}}\revdelrefr{\cite{goodfellow2016deep}}. The SIMDRAM framework is not limited to these \omiv{16} operations, and can enable processing-using-DRAM for other existing and future operations. 
\geraldorevi{\mech is well-suited to application classes that (i) \omiv{are SIMD-friendly, (ii)} have a regular access pattern, and (\omiv{iii}) are memory bound. Such applications are common in domains such as database analytics, \omiv{high-performance computing}, \omiv{image processing,} and machine learning.}}

We compare the benefits of \mech to different state-of-the-art computing platforms (CPU, GPU, and the Ambit~\cite{seshadri2017ambit} in-DRAM computing mechanism). We comprehensively evaluate SIMDRAM's reliability, area overhead, throughput, and energy efficiency. We leverage the \mech framework to accelerate seven application kernels from machine learning, databases, and image processing (VGG-13~\cite{simonyan2014very}, VGG-16~\cite{simonyan2014very}, LeNET~\cite{lecun2015lenet}, kNN~\cite{lee1991handwritten}, TPC-H~\cite{tpch}, BitWeaving~\cite{li2013bitweaving}, \omii{b}rightness~\cite{gonzales2002digital}). 
\omiv{Using a single DRAM bank, SIMDRAM provides
(1)~2.0$\times$ the throughput and 2.6$\times$ the energy efficiency of Ambit~\cite{seshadri2017ambit}, averaged across the 16 implemented operations; and
(2)~2.5$\times$ the performance of Ambit, averaged across the seven application kernels.}
\omiv{Compared to a CPU and a high-end GPU, SIMDRAM using 16 DRAM banks provides (1)~257$\times$ and 31$\times$ the energy efficiency, and \omi{88}$\times$ and \omi{5.8}$\times$ the throughput of the CPU and GPU, respectively, averaged across the 16 operations; and (2)}~\omiv{21$\times$ and 2.1$\times$ the performance of the CPU and GPU, respectively, averaged across the \omv{seven} application kernels.}  \omiv{SIMDRAM incurs no additional area overhead on top of Ambit\omiii{, and a total area overhead of \sgii{only 0.2\% in a high-end CPU.}}} We also evaluate the reliability of \mech under different degrees of manufacturing process variation, and observe that it guarantees correct operation as the DRAM process technology node scales down to smaller sizes. 

We make the following key contributions:
\begin{itemize}[noitemsep,topsep=0pt,parsep=0pt,partopsep=0pt,labelindent=0pt,itemindent=0pt,leftmargin=*]
\item To our knowledge, this is the first work to propose a framework to enable efficient computation of a flexible \om{set} and wide range of operations in a massively parallel SIMD substrate built via processing-using-DRAM. 
\item \mech provides a three-step framework to develop efficient and reliable MAJ/NOT-based implementations of a wide range of operations.  We design this framework, and add hardware\omiv{, programming,} and ISA support, to (1)~address key system integration challenges and (2)~allow programmers to \omiv{define and} employ new \mech operations without hardware changes.
\item We provide a detailed reference implementation of \mech, including required changes to applications, ISA, and hardware.
\item We evaluate the reliability of \mech under different degrees of process variation and observe that it guarantees correct operation as the DRAM technology scales to smaller node sizes.
\end{itemize}

\section{Background}
\label{background}
%
%
%
\om{W}e \om{first} briefly explain the architecture of \geraldorevii{a typical} DRAM \geraldorevii{chip}. \nasrev{Next, we describe prior processing-using-DRAM works that \mech builds on top of (RowClone~\cite{seshadri2013rowclone} and Ambit~\om{\cite{seshadri2017ambit, seshadri2015fast,seshadri2019dram}})}\revdel{We then} \om{and} \nasii{explain the implications of majority-based computation.}
%

\subsection{DRAM {Basics}}
\label{sec:dram-basics}

\begin{figure}[ht]
    \centering
    \includegraphics[width=0.95\linewidth]{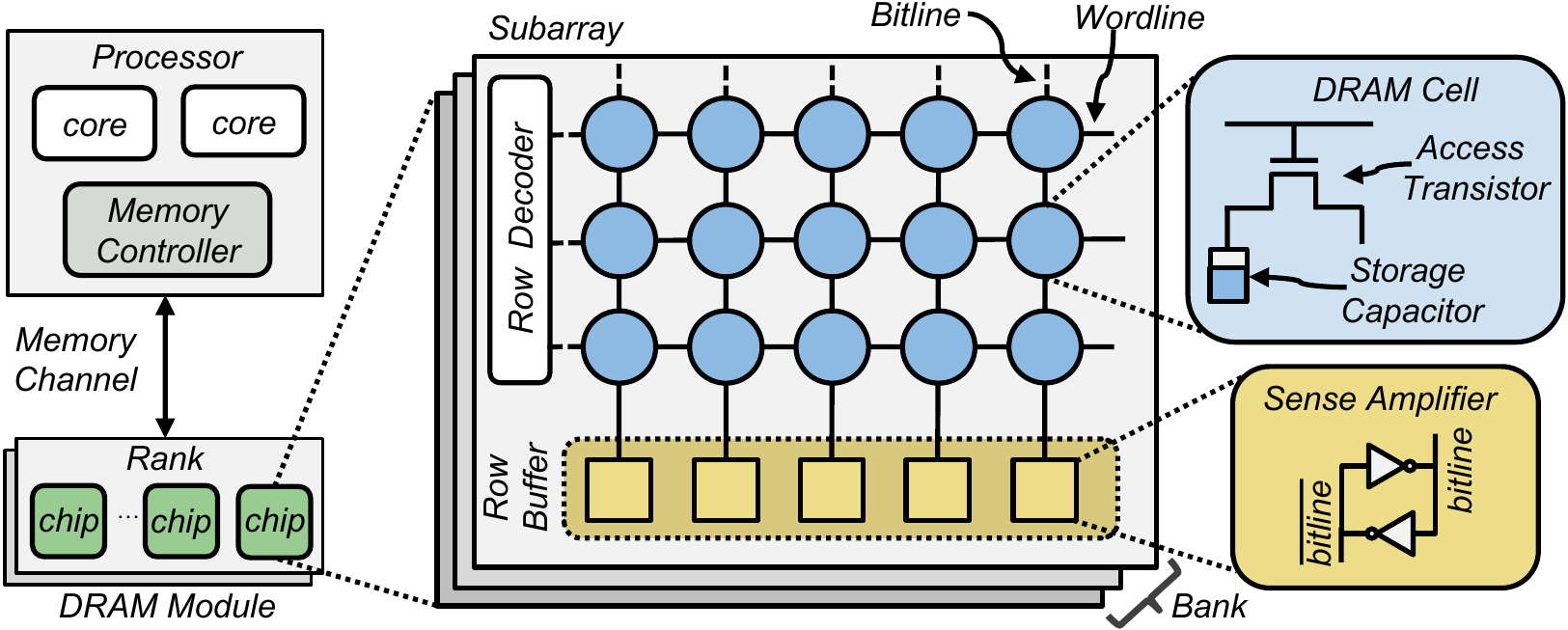}
    \caption{\geraldorevii{High-level overview of DRAM organization.}}
    \label{fig_subarray_dram}
\end{figure}

A DRAM \revdel{module}\om{system} comprises a hierarchy of \revdel{2D }components, \geraldorevii{as \cmr{Fig.}~\ref{fig_subarray_dram} shows,} starting with \emph{channels} at the highest level. A channel is\revdel{ further} subdivided into \emph{ranks}, and a rank is subdivided into multiple \emph{banks} \om{(e.g., 8-16).}\revdel{, which are e}\om{~Each bank is} composed of \om{multiple (e.g., 64-128)} 2D arrays of cells known as \emph{subarrays}. Cells within a subarray are organized into \geraldorevii{multiple} \emph{rows} \geraldorevii{(e.g., 512-1024)} and \geraldorevii{multiple} \emph{columns}\revdel{,} \geraldorevii{(e.g., 2-8~kB)}~\cite{lee2015adaptive, lee2017design, kim2018solar}. A cell consists of an \emph{access transistor} and a \emph{storage capacitor} that encodes a single bit of data using its voltage level. The source \nasrev{node\om{s} of the} access transistors of all the cells in the same column connect the cells' storage capacitor\om{s} to the same \emph{bitline}. Similarly, the gate \nasrev{node\om{s} of} the access transistor\om{s} of all the cells in the same row connect the\revdel{ir} \om{cells'} \revdel{storage capacitor\om{s}}\geraldorevii{access transistors} to the same \emph{wordline}. 

\omi{When a wordline is asserted, all cells along the wordline are connected to their corresponding bitlines, which perturbs the voltage of each bitline depending on the value stored in each cell's capacitor. A two-terminal \emph{sense amplifier} connected to each bitline senses the voltage difference between the bitline (connected to one terminal) and a reference voltage (typically  $\frac{1}{2}V_{DD}$; connected to the other terminal) and amplifies it to a CMOS-readable value. In doing so, the sense amplifier terminal connected to the reference voltage is amplified to the \emph{opposite} (i.e., \emph{negated}) value, which is shown as the $\overline{\mbox{bitline}}$ terminal in \cmr{Fig.}~\ref{fig_subarray_dram}.}
The set of sense amplifiers in each subarray forms a logical \emph{row buffer}, which \omi{maintains the sensed data for as long as the row is \emph{open} (i.e., the wordline continues to be asserted).}
\nasii{\sgii{A read or write} operation \nasii{in DRAM} includes \emph{three} steps:
\begin{enumerate}[noitemsep,topsep=0pt,parsep=0pt,partopsep=0pt,labelindent=0pt,itemindent=0pt,leftmargin=*]
    \item \nasrev{\texttt{ACTIVATE}}. The \emph{wordline} of the target row is asserted, which connects all cells along the row to their respective bitlines. Each \omi{bitline shares charge with its corresponding cell capacitor, and the resulting bitline voltage shift is sensed and amplified by the bitline's sense amplifier. Once the sense amplifiers \omiii{finish amplification}, 
    the row buffer contains the values originally stored within the cells along the asserted wordline.}
    \item \texttt{\sgii{\om{RD/WR}}}. The memory controller then \revdel{performs}\geraldorevii{issues} read or write commands to columns within the activated row \om{\omi{(i.e., the data within the row buffer)}}.
    \item \nasrev{\texttt{PRECHARGE.}} \revdel{\nasrev{In this step, s}\om{S}ense amplifiers are disabled, and each bitline is restored to its quiescent state (e.g., typically $\frac{1}{2}V_{DD}$).} The capacitor is disconnected from the bitline by disabling the wordline, \omi{and the bitline voltage is restored to its quiescent state (e.g., typically $\frac{1}{2}V_{DD}$)}.  
\end{enumerate}
}

\subsection{Processing-using-DRAM}
\label{sec:in-dram}

\subsubsection{In-DRAM Row Copy.} 
\label{sec_rowclone}

~RowClone~\cite{seshadri2013rowclone} is a\revdel{n in-DRAM} mechanism that exploits the vast internal DRAM bandwidth to \nasrev{efficiently} copy rows \revdel{within a subarray}\om{inside DRAM} without CPU intervention. RowClone enables copying \juang{a source} row~$A$ to \juang{a destination} row~$B$ \om{in the same subarray} by issuing two consecutive \texttt{ACTIVATE}
commands to these two rows, \juang{followed by a \texttt{PRECHARGE} command}. 
\juang{This \om{command} sequence is called \texttt{AAP}~\om{\cite{seshadri2017ambit}}.} \juang{The first \texttt{ACTIVATE} command copies the contents of the source row $A$ into the row buffer. \geraldorevii{The second \texttt{ACTIVATE} command connects the cells in the destination row~$B$ to the bitlines.} \revdel{When the second \texttt{ACTIVATE} command is issued,} 
\omi{Because the sense amplifiers \omiii{have already sensed \omiv{and amplified} the source data } 
by the time row~$B$ is activated, the data (i.e., voltage level) in each cell of row~$B$ is overwritten by the data stored in the row buffer (i.e., row~$A$'s data).}}
\omi{Recent work~\cite{gao2019computedram} experimentally demonstrates the feasibility of executing in-DRAM row copy operations in unmodified off-the-shelf DRAM chips.}


\subsubsection{In-DRAM Bitwise Operations.} 
\label{sec_ambit_logic} 

Ambit~\om{\cite{seshadri2017ambit, seshadri2015fast, seshadri2019dram}} shows that simultaneously activating \emph{three} DRAM \juang{rows} \juang{(\omi{via a DRAM operation} called \emph{Triple Row Activation, TRA})} can be used to perform \om{bitwise} Boolean \omi{AND, OR, and NOT operations} on the values contained within the cells \omi{of the three rows}. \juang{\omi{When activating three rows}, three cells connected to each bitline share charge simultaneously and contribute to the perturbation of the bitline. 
\omi{Upon sensing the perturbation, the sense amplifier amplifies the bitline voltage to}
\geraldorevii{$V_{DD}$ or 0 if at least two of the capacitors of the three DRAM cells are charged or discharged, respectively.}} \juang{As \omi{such},} a TRA results in a Boolean \emph{majority operation} ($MAJ$) \juang{among the three DRAM cells on each bitline}. A majority operation MAJ \omi{outputs a 1 (0)} only if more than half of \omi{its} inputs are \omi{1 (0)}. \juang{In terms of AND ($\cdot$) and OR (+) operations, a 3-input majority operation can be expressed as \texttt{MAJ(A, B, C) = A $\cdot$ B + A $\cdot$ C + B $\cdot$ C.}}

\omi{Ambit implements MAJ by introducing a custom row decoder (discussed in \cmr{\cref{subarray}}) that can perform a TRA by simultaneously addressing three wordlines. To use this decoder, Ambit defines a new command sequence called \texttt{AP}, which issues (1)~a TRA to compute the MAJ of three rows, followed by (2)~a \texttt{PRECHARGE} to close all three rows.\footnote{\omi{Although the `\texttt{A}' in \texttt{AP} refers to a TRA operation instead of a conventional \texttt{ACTIVATE} command, we use this terminology to remain consistent with the Ambit paper~\cite{seshadri2017ambit}\omiii{, since an \texttt{ACTIVATE} command can be internally translated to a TRA operation by the DRAM chip~\cite{seshadri2017ambit}}.}}}
Ambit uses \omi{\texttt{AP} command sequences} to implement Boolean \om{AND} and \om{OR} operations by simply setting one of the inputs (e.g., $C$) to \omi{1} or \omi{0}. The AND operation is computed by setting $C$ to 0 (i.e., \texttt{MAJ(A, B, 0) = A AND B})\revdel{, while t}\om{. T}he OR operation is computed by setting $C$ to 1 (i.e., \texttt{MAJ(A, B, 1) = A OR B}).

\omi{To achieve functional completeness alongside AND and OR operations, Ambit implements NOT operations by exploiting the differential design of DRAM sense amplifiers. As \cmr{\cref{sec:dram-basics}} explains, the sense amplifier already generates the complement of the sensed value as part of the activation process ($\overline{\mbox{bitline}}$ in \cmr{Fig.}~\ref{fig_subarray_dram}). Therefore, Ambit simply forwards $\overline{\mbox{bitline}}$ to a special DRAM row in the subarray that consists of DRAM cells with \emph{two} access transistors, called \emph{dual-contact cells} (DCCs). Each access transistor is connected to one side of the sense amplifier and is controlled by a separate wordline (\emph{d-wordline} or \emph{n-wordline}). By activating \omiii{either} the d-wordline or the n-wordline, the row of DCCs can \omiii{provide} the true or negated value stored in the \omiii{row's} cells, respectively.}



\subsubsection{Majority-Based Computation.}
\label{sec_maj-based}


\juangr{~Activating multiple rows simultaneously reduces the reliability of the value read by the sense amplifiers \mpi{due to manufacturing process variation, which} introduces non-uniformities in circuit-level electrical characteristics (e.g., \omi{variation in} cell capacitance \omi{level\omiii{s}})~\om{\cite{seshadri2017ambit}}. This effect worsens with (1) an increased  number of simultaneously activated rows, and (2) more advanced technology nodes with smaller sizes. Accordingly, \geraldorevii{although}\revdel{while} \geraldorevii{processing-using-DRAM} can potentially support majority operations with more than \nasrev{three} inputs 
(as proposed by prior works~\cite{ali2019memory, angizi2019graphide,pinatubo2016}) \mpi{our realization of 
\geraldorevii{processing-using-DRAM} uses} the minimum number of inputs required for a majority operation ($N$=3) to maintain \nasrev{the reliability of the} computation. 
\ifasploscr
\cmr{In the \omvuii{definitive} version of the paper~\cite{extendedsimdram},}
\else
In \cmr{\cref{sec_reliability}}, 
\fi
we \om{demonstrate via SPICE simulations}\revdel{show} that using 3-input MAJ \nasrev{operations} \revdel{improves}\om{provides} \revdel{the}\om{higher} reliability \revdel{of \gfrev{in-DRAM operations?}\mech }compared to designs with more than \nasrev{three} inputs per MAJ \nasrev{operation}.} Using 3-input MAJ, \revdel{\mech}\geraldorevii{a processing-using-DRAM substrate} does not require modifications to the subarray organization (\cmr{Fig.}~\ref{fig_subarray}) beyond the ones proposed by Ambit \omi{(\cmr{\cref{subarray}})}. 
\mpi{\om{R}ecent work~\cite{gao2019computedram} experimentally demonstrates the feasibility of executing MAJ operations by activating three rows \nasrev{in \om{unmodified off-the-shelf}\revdel{commodity DRAM} \omi{DRAM} chips}.}

\section{SIMDRAM Overview}
\label{mechanism}


\sgii{SIMDRAM is a processing-using-DRAM framework whose goal is to (1)~enable the efficient implementation of complex operations and (2)~provide a flexible mechanism to support the implementation of arbitrary user-defined operations.}
\nasii{\revdel{In this section, w}\om{W}e present the subarray organization in SIMDRAM, describe an overview of the SIMDRAM framework, and explain how \sgii{to integrate SIMDRAM} into \sgii{a} system.}

\subsection{Subarray Organization}
\label{subarray}

\begin{figure}[ht]
    \centering
    \includegraphics[width=0.8\linewidth]{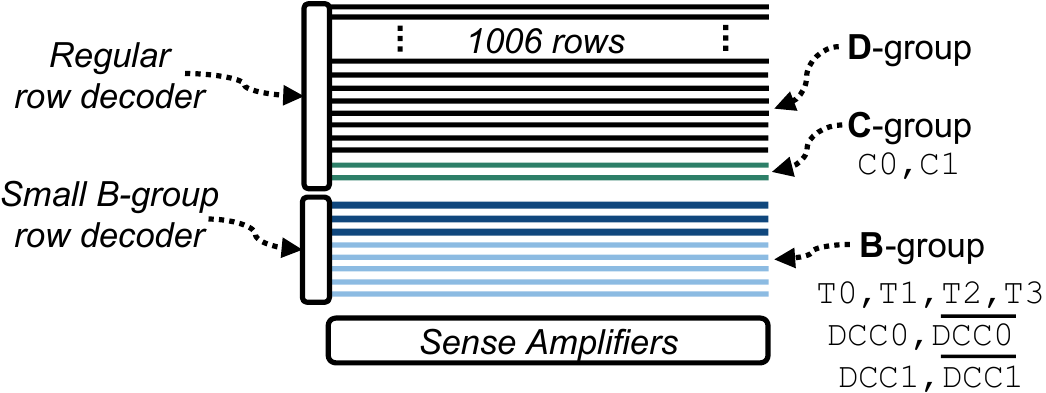}
    \caption{\geraldorevi{\mech subarray organization~\cite{seshadri2017ambit}. \revdel{Adapted from~\cite{seshadri2017ambit}.}}}
    \label{fig_subarray}
\end{figure}

\sgii{In order to perform processing-using-DRAM, SIMDRAM makes use of a subarray organization that incorporates additional functionality to perform logic primitives (i.e., MAJ and NOT). This subarray organization is \emph{identical} to Ambit's~\cite{seshadri2017ambit} and is similar to DRISA's~\cite{li2017drisa}.}
\textfromsl{\cmr{Fig.}~\revdelrefr{\ref{fig_subarray}}\revdelrefa{\hl{1}} illustrates the internal organization of a subarray in \mech, which \juang{resembles a conventional DRAM subarray}. \juang{\mech requires \om{only} minimal modifications to the DRAM subarray (namely, a small row decoder that can activate three rows simultaneously) to enable computation}. \om{Like Ambit~\cite{seshadri2017ambit},} \juang{\mech divides DRAM rows} into \emph{three groups}: \sgii{the} \textbf{D}ata group (D-group), \sgii{the} \textbf{C}ontrol group (C-group) and \sgii{the} \textbf{B}itwise group (B-group).}

The D-group contains regular rows that store \revdel{users'}\om{program or system} data. The \sgii{C-group} \juang{consists of two constant} \sgii{rows, called C0 and C1,} that \juang{contain} \om{all-}0 and \om{all-}1 \juang{values}, respectively. \geraldo{These rows are used (1)~\sgii{as} initial input values for a given \sgii{SIMDRAM} \geraldorevii{operation}
(e.g., the initial carry-in bit in a full \omi{addition}), or (2)~to perform operations that naturally require\revdel{s} AND/OR operations (e.g., AND/OR reductions)}. The D-group and the C-group are connected to the regular row decoder, which selects a single row \juang{at a time}.

The \sgii{B-group} contains six \sgii{regular rows, called T0, T1, T2, and T3;} and two rows of dual-contact cells \sgii{(see \cmr{\cref{sec_ambit_logic}}), 
whose d-wordlines are called DCC0 and DCC1, and whose n-wordlines are called $\overline{\mbox{DCC0}}$ and $\overline{\mbox{DCC1}}$, respectively}.
\sgii{The B-group} rows\om{, called \emph{compute rows}, are} designated to perform bitwise operations. They are all connected to a special row decoder that can simultaneously activate three rows using a single address 
(i.e., perform a TRA) 

\geraldorevi{\revonur{Using} a typical subarray size of 1024 rows~\revdelrefr{\cite{chang2014improving, kim2012case, kim2018solar,Tiered-Latency_LEE,kim2019d}}\revdelrefa{\hl{[19,50,53,57,64]}}, \mech splits the row addressing into 1006 \sgii{D-group rows, 2 C-group rows, and 16 B-group rows}.}

\subsection{Framework Overview}
\label{overview}

\begin{figure}[t!]
\begin{subfigure}{\linewidth}
    \centering
    \includegraphics[width=\textwidth]{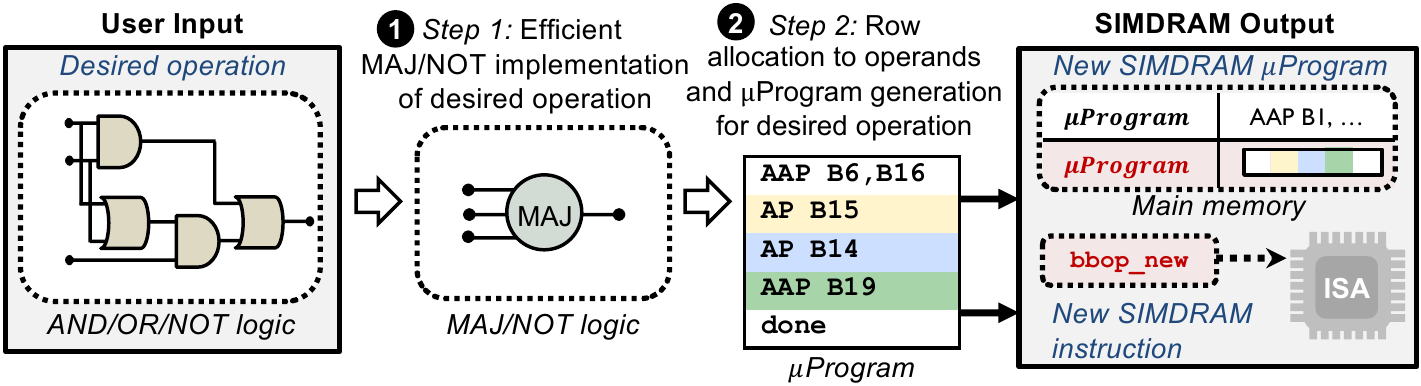}%
    \vspace{-4pt}
    \caption{\geraldorevi{SIMDRAM Framework: Steps 1 and 2}}
    \label{fig_framework_1_2}
\end{subfigure}
\par\bigskip 
\begin{subfigure}{\linewidth}
  \centering
    \includegraphics[width=\textwidth]{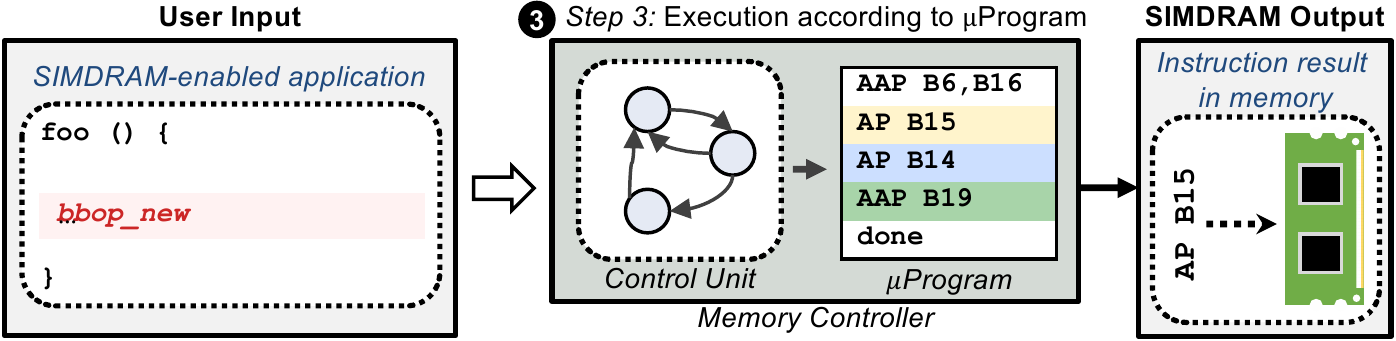}%
    \vspace{-4pt}
    \caption{\geraldorevi{SIMDRAM Framework: Step 3}}
    \label{fig_framework_3}
\end{subfigure}
\caption{\geraldorevi{Overview of the \mech framework.}}
\label{fig:figoverview}
\end{figure}

\sgii{SIMDRAM \omi{is} an end-to-end framework that provides the user with the ability to implement an \emph{arbitrary} operation in DRAM using the \aap command sequences.
The framework comprises}
three key steps, \geraldorevi{which \sgii{are illustrated in \cmr{Fig.}~\ref{fig:figoverview}.}
\geraldo{The first two steps \sgii{of the framework} \sg{give} \omi{the} user the 
\juangg{ability} to efficiently implement 
any desired \revonuri{operation} in DRAM, while the third step controls the execution flow of the in-DRAM computation transparently from the user. 
\revonuri{We} briefly describe these steps \sgii{below, and discuss each step in detail in \cmr{\cref{sec:operations}}}.}}

\om{T}he first step \geraldorevi{(\ding{182} in \cmr{Fig.}~\ref{fig_framework_1_2}; \sgii{\cmr{\cref{sec:framework:step1}}})} \sgii{builds an efficient MAJ/NOT representation of \revonuri{a given} desired operation 
\omi{from its AND/OR/NOT-based implementation}.  Specifically, this step} \om{takes as input a desired operation and} use\om{s} logic optimization to minimize the number of \sgii{logic primitives}
(and, therefore, the \om{computation} latency) required to perform \revdel{a specific}\om{the} operation. \sg{Accordingly, for a desired \om{operation input into the \mech framework by the user}, \jgll{Weird writing.} the first step 
derive\geraldorevii{s} its \emph{optimized} MAJ/NOT-based implementation.  }

The second step (\geraldorevi{\ding{183} in \cmr{Fig.}~\ref{fig_framework_1_2}; \sgii{\cmr{\cref{sec:framework:step2}}}}) 
\omi{allocates DRAM rows \omiii{to}} 
\sgii{the operation's inputs and outputs and generates the required sequence of DRAM commands to execute the desired operation. Specifically, this step}
translates the 
MAJ/NOT-based implementation \nasrev{of the operation} 
\sg{into} \geraldorevii{\aaps}. 
This step 
\sgii{involves}
(1)~\omi{allocating the designated compute rows in DRAM \omiii{to} the operands}, and (2)
\juangg{~determining} the \geraldorevii{optimized} sequence of \geraldorevii{\aaps} that are required to perform the \geraldorevii{operation}. \juangg{While doing so, SIMDRAM minimizes} the number of \geraldorevii{\aaps}\revdel{AAPs/TRAs} required for a specific operation. \geraldorevii{\cmr{This step's output} is a \uprog{}, i.e., the optimized sequence of \aaps that \omi{is stored in main memory and} will be used to execute the  operation \sgii{at} runtime. }

\sg{The third step \geraldorevi{(\ding{184} in \cmr{Fig.}~\ref{fig_framework_3}; \sgii{\cmr{\cref{sec:framework:step3}}})} 
\sgii{executes the \uprog{} to perform the operation.
\omiii{Specifically, when a user program encounters a \emph{bbop} instruction (\cmr{\cref{sec:bbop}}) associated with a SIMDRAM operation, the \emph{bbop} instruction triggers the execution of the SIMDRAM operation by performing its \uprog{} in the memory controller.} }
SIMDRAM uses a \emph{control unit} in the memory controller that transparently \sgii{issues} the sequence of \geraldorevii{\aaps}\revdel{AAPs/TRAs} 
\sgii{to DRAM, as dictated by the \uprog{}}.}
\sgii{Once the \uprog{} is complete, the result of the operation is held in DRAM.}

\subsection{Integrating SIMDRAM in a System}
\label{sec:integ-overview}

\sg{As we discuss \mpi{in \cmr{\cref{intro}}}, SIMDRAM operates on data using a vertical layout. \geraldorevii{\cmr{Fig.}~\ref{fig:datalayout} illustrates how data is organized within a DRAM subarray when employing a horizontal data layout (\cmr{Fig.}~\ref{fig:datalayout}a) and a vertical data layout (\cmr{Fig.}~\ref{fig:datalayout}b). We assume that each data element is \sgii{four bits} wide, and \sgii{that} there are four data elements (each one represented by a different color). In a conventional horizontal data layout, data elements are stored in different DRAM rows, \sgii{with the contents of each data element ordered} from the most significant bit to the least significant bit (or vice versa) \sgii{in a single row}. In contrast, in a vertical data layout, \sgii{the} DRAM row holds \sgii{only} the $i$-th bit of \sgii{\emph{multiple}} data elements \sgii{(where the number of elements is determined by the bit width of the row)}. Therefore, when activating a single DRAM row in a vertical data layout organization, a \emph{single} bit of \sgii{data from each} data \sgii{element is} read at once, which enables \cmr{in\omvuii{-}DRAM} bit-serial parallel computation~\omiv{\cite{batcher1982bit,shooman1960parallel,seshadri2019dram,li2017drisa,ali2019memory,gu2016leveraging}}.} 

\begin{figure}[ht]
    \centering
    \includegraphics[width=0.8\linewidth]{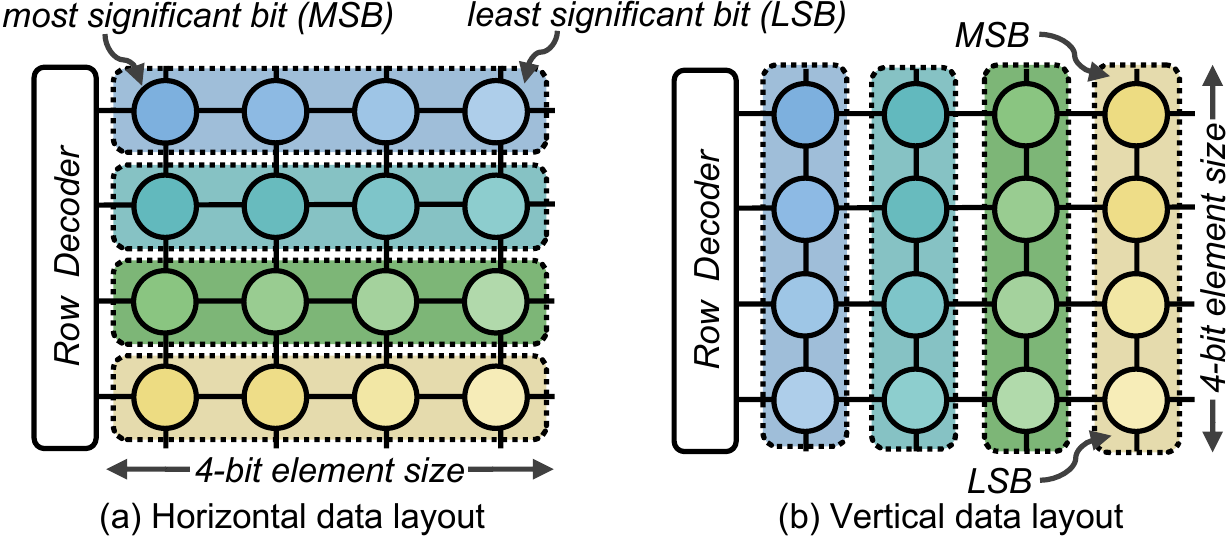}
    \caption{\geraldorevi{Data layout\omi{: horizontal vs.\ vertical}}.}
    \label{fig:datalayout}
\end{figure}

To maintain compatibility with traditional system software, we store regular data in the conventional horizontal layout and provide hardware support (explained in \cmr{\cref{sec:transposing}}) to transpose horizontally\omi{-}laid\omi{-}out data into \omi{the} vertical layout for in-DRAM computation. To simplify program integration, we provide ISA extensions that
expose SIMDRAM operations to the programmer (\cmr{\cref{sec:bbop}}).}

\section{SIMDRAM Framework}
\label{sec:operations}
\label{sec:framework}
\cmr{\omi{\omvuii{\omiii{We describe}
the three steps of the \mech framework introduced in \cmr{\cref{overview}}, using the full addition operation as a running example.}}}

\subsection{Step 1: Efficient MAJ/NOT Implementation}
\label{sec:aoi-mi}
\label{sec:framework:step1}

\nasrev{SIMDRAM implements in-DRAM computation using the logically\omi{-}complete set of MAJ and NOT logic primitives, which requires fewer \aap \omiii{command sequences} to perform a given operation when compared to using AND/OR/NOT. As a result, the goal of the first step in the \mech framework is to build an optimized MAJ/NOT implementation of a given operation that executes the operation using as few \aap \omiii{command sequences} as possible, thus \omi{minimizing the operation's latency}. To this end, Step 1 transforms an AND/OR/NOT representation of a given operation to an optimized MAJ/NOT representation using \sgii{a transformation process} formalized by prior work~\cite{epflmaj}.}

\nasrev{\sgii{The transformation process uses a graph-based representation of the logic \sgii{primitives}, called \sgii{an} \textit{AND--OR--Inverter Graph\revdel{s}} (AOIG) \sgii{for AND/OR/NOT logic, and a} \textit{Majority--Inverter Graph\revdel{s}} (MIG) \sgii{for MAJ/NOT logic}. An AOIG is a logic representation structure in the form of a directed acyclic graph \omi{where} each node represents an AND or OR \sgii{logic primitive}. Each edge in an AOIG represents an input/output dependency between nodes. The incoming edges to a node represent input operands of the node and the outgoing edge of a node represent\omi{s} the output of the node. The edges in \sgii{an} AOIG can be either regular or complemented (\sgii{which represents \omi{an inverted} input operand; \omi{\omiii{denoted by}}} \omiii{a bubble on the} edge).
The direction of the edges follows the natural direction of computation from inputs to outputs. 
Similarly, a MIG is a directed acyclic graph in which each node represents a three-input MAJ \sgii{logic primitive}, and each regular/complemented edge represents one input or output to the MAJ \sgii{primitive} that the node represents.
The transformation process consists of two parts that operate on an input AOIG.}}

\sgii{The first part of the transformation process naively substitutes AND/OR primitives with MAJ primitives.
Each two-input AND or OR primitive is simply replaced with a three-input MAJ primitive, where one of the inputs is tied to logic 0 or logic 1, respectively.
This naive substitution yields \omi{a} MIG that \emph{correctly} replicates the functionality of the input AOIG, but the MIG is \omi{likely} \emph{inefficient}.}

\begin{figure}[ht]
    \centering
    \includegraphics[width=1\linewidth]{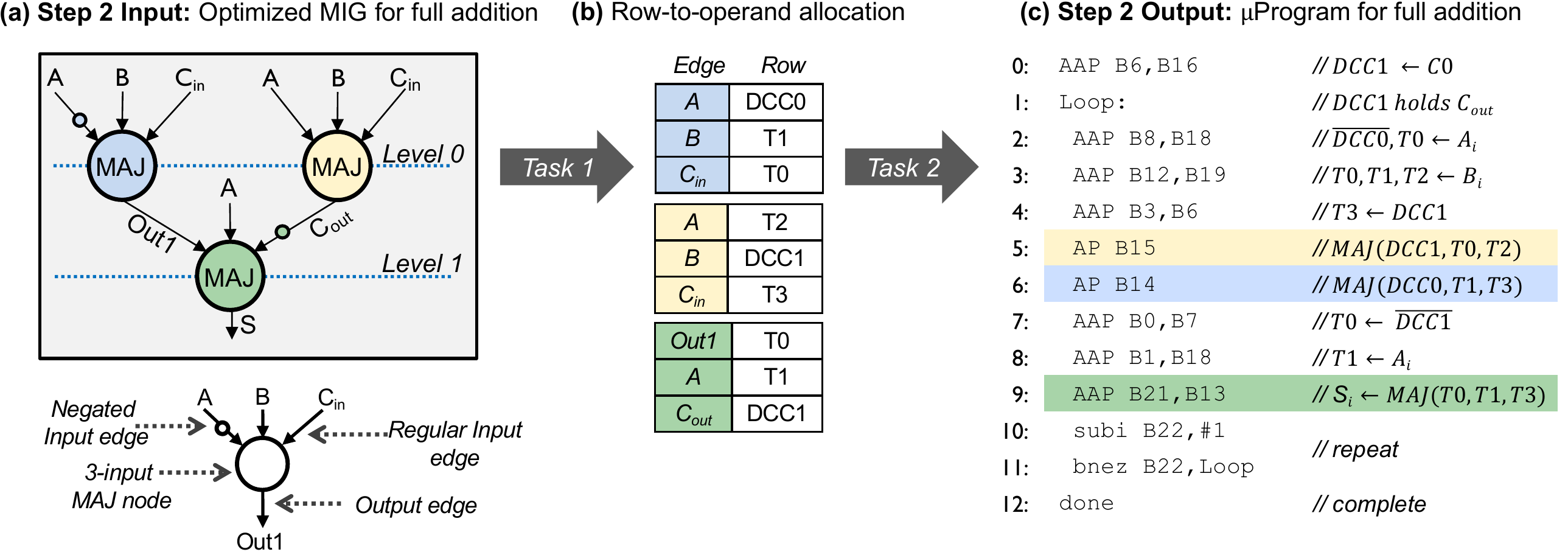}
    \caption{\sgii{(a)~\omiv{Optimized \cmr{MIG}}\cmr{;} (b)~\omi{row-to-operand allocation}\cmr{;} (c)~\uprog{}} for full \omi{addition.}}
    \label{fig_output_mapping}
\end{figure}

\sgii{The second part of the transformation process takes the inefficient MIG and uses a greedy algorithm 
\omi{(\omdef{see Appendix~\ref{apdx:aoi-to-mig})}} to apply a series of transformations that identifies how to consolidate multiple MAJ primitives into a smaller number of MAJ primitives with identical functionality. This yields a smaller MIG, which in turn requires fewer logic primitives to perform the same operation that the unoptimized MIG (and, thus, the input AOIG) perform\omi{s}. \cmr{Fig.}~\ref{fig_output_mapping}a shows the optimized MIG produced by the transformation process for a full addition operation.}

\subsection{Step 2: \sgii{\uprog{} Generation}}
\label{sec:mi-aap}
\label{sec:framework:step2}


\nasrev{\sgii{Each \mech operation is stored as a \emph{\uprog{}}, which consists of a series of microarchitectural operations (\uop{}s) that \mech uses to execute the \mech operation in DRAM.
The goal of the second step is to take the optimized MIG generated in Step~1 and generate a \uprog{} that executes the 
\omi{\mech operation that the MIG represents}.
To this end, \omi{as shown in \cmr{Fig.}~\ref{fig_output_mapping}}, the second step of the framework performs two key tasks on the optimized MIG:}
(1)~\emph{\omi{allocating DRAM rows \omiii{to} the operands}}, \sgii{which assigns each input operand (i.e., an incoming edge) of each MAJ node in the MIG to a DRAM row \omi{(\cmr{Fig.}~\ref{fig_output_mapping}b)};} and (2)~\emph{generating the \uprog{}}, \sgii{which creates the \omiii{series} of \uop{}s that perform the MAJ and NOT logic primitives (i.e., nodes) in the MIG, while maintaining} the correct flow of the computation \omi{(\cmr{Fig.}~\ref{fig_output_mapping}c)}. In this section, we first describe the \uop{}s used in \mech (\cmr{\cref{sec:framework:step2:uops}}). \sgii{Second}, we explain the process of \omi{allocating DRAM rows \omiii{to}} the input operands of the MAJ nodes in the MIG to DRAM rows (\cmr{\cref{sec:framework:step2:mapping}}).  \sgii{Third, we explain the process of generating} the \uprog{} (\cmr{\cref{sec:framework:step2:generating}}).}

\subsubsection{\textbf{\mech \uop{}s.}}
\label{sec:framework:step2:uops}

\begin{figure}[!ht]
    \centering
    \includegraphics[width=0.9\linewidth]{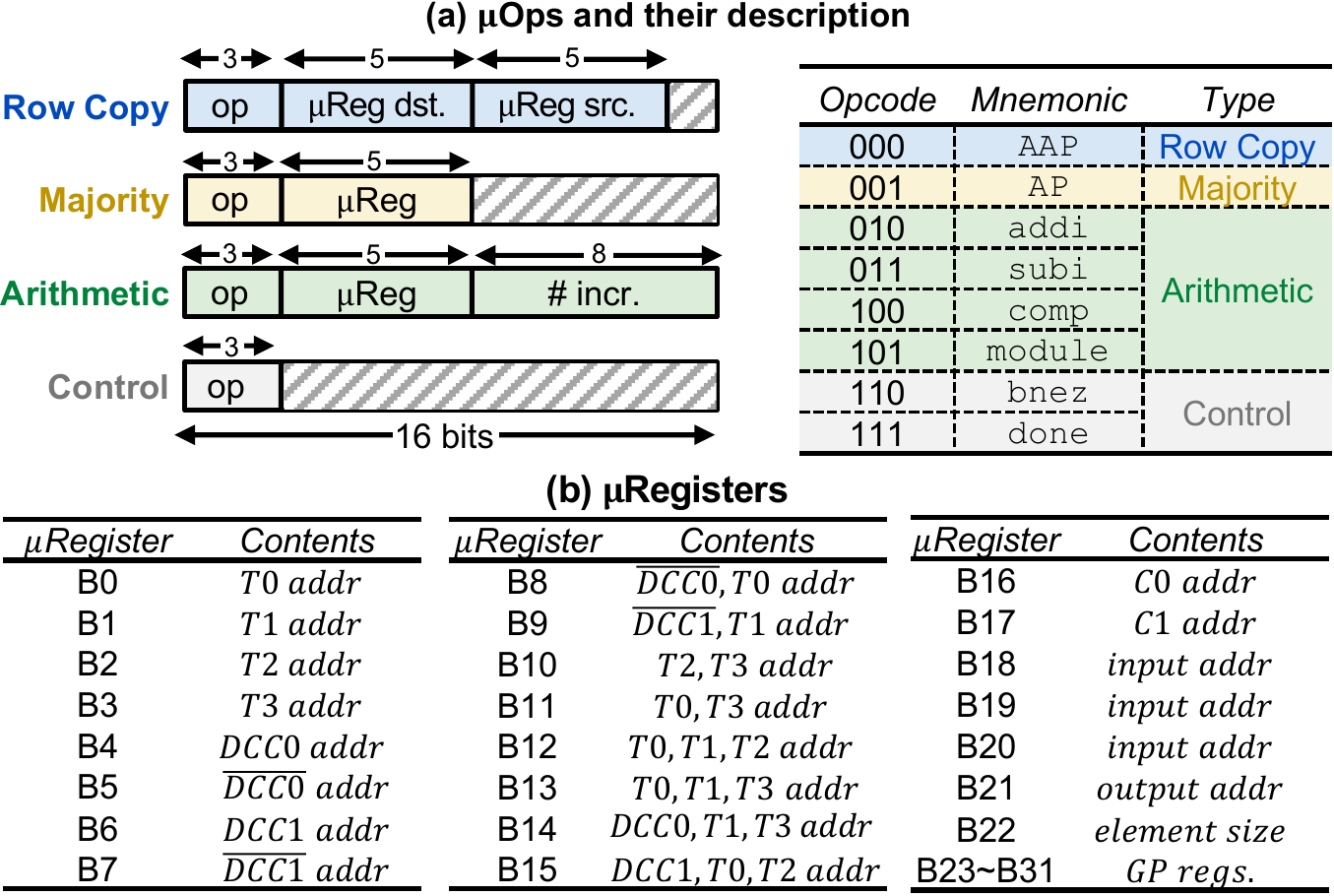}%
    \caption{\sgii{\uop{}s and \ureg{}s in \mech.} 
    \jgll{Keep this in mind when improving the figure: A MAJ can be executed with an AP or with an AAP -- AP if the destination row is among the source rows. AAP if not.}}
    \label{fig_opcodes}
\end{figure}

\nasrev{~\cmr{Fig.}~\ref{fig_opcodes}a shows the set of \uop{}s \sgii{that we implement in \mech. Each \uop{} is either
(1)~a \emph{command sequence} that is issued by \mech to a subarray to perform a portion of the in-DRAM computation, or
(2)~a \emph{control} operation that is used by the \mech control unit (see \cmr{\cref{sec:framework:step3}}) to manage the execution of the \mech operation.}
We further break down the command sequence \uop{}s into one of three types:
(1)~\textit{row copy}, \sgii{a \uop{} that} performs in-DRAM copy from a source memory address to a destination memory address using an \texttt{AAP} \omi{command sequence};
(2)~\textit{majority}, \sgii{a \uop{} that performs a majority logic primitive on three DRAM rows using an \texttt{AP} \omi{command sequence} (i.e., it performs a TRA); and} 
(3)~\textit{arithmetic}, \sgii{four \uop{}s that} \omi{perform simple arithmetic operations \omi{on \mech control unit registers} required to control the execution of the operation (\texttt{addi}, \texttt{subi}, \texttt{comp}, \texttt{module})}.}
\omi{We provide two control operation \uop{}s to support loops and termination in the \mech control flow (\texttt{bnez}, \texttt{done}).}

\nasrev{\sgii{During \uprog{} generation, the \mech framework converts the MIG into a series of \uop{}s.  \cmr{Note that MIG represents a 1-bit\omi{-wide} computation of an operation.}} \sgii{\cmr{Thus, to} implement a multi-bit\omi{-wide} \mech operation, the framework} needs to repeat the \omiii{series} of the \uop{}s that implement the MIG $n$ times, 
where $n$ is the number of bits in the operands of the \sgii{\mech} operation. To this end, SIMDRAM uses the arithmetic and control \uop{}s to 
repeat the 1-bit\omi{-wide} computation $n$ times\omi{, transparently to the programmer}}.

\nasrev{To support the execution of \uop{}s, SIMDRAM utilizes a set of \sgii{\emph{\ureg{}s}} (\cmr{Fig.}~\ref{fig_opcodes}b) located in the \omi{\mech} control unit \omi{(\cmr{\cref{sec:mc}})}. \sgii{The framework uses \ureg{}s}
(1)~to store the memory addresses of DRAM rows in the B-group and C-group (\cmr{Fig.}~\ref{subarray}) of the subarray (\sgii{\ureg{}s B0--B17}),
(2)~to store the memory addresses of input and output rows for the computation (\sgii{\ureg{}s B18--B22}), and
(3)~as general-purpose registers during the execution of arithmetic and control operations (\sgii{\ureg{}s B23--B31}).
\jgll{
We also need to justify why 32 registers and why 32-bit.}}

\subsubsection{Task 1: \omi{Allocating DRAM Rows \omiii{to} the Operands}.}
\label{sec:framework:step2:mapping}

\nasrev{~The goal of this task is to \omi{allocate DRAM rows \omiii{to}} the  
input operands (i.e., incoming edges) of each MAJ node in the operation's MIG, \sgii{such that we} minimize the total number of \uop{}s \cmr{needed} to compute the operation. To this end, we present a new \omi{allocation} algorithm inspired by the linear scan register allocation algorithm~\cite{poletto1999}. However, unlike register allocation algorithms, our \omi{allocation algorithm} considers two extra constraints that are specific to processing-using-DRAM:
(1)~performing MAJ in DRAM has destructive behavior, i.e., a TRA overwrites the original values of the three input rows with the MAJ output; and 
(2)~the number of compute rows (i.e., B-group in \cmr{Fig.}~\ref{fig_subarray}) that are designated to perform bitwise operations is limited 
(\sgii{\omi{there are only} six compute rows in \omi{each} subarray, as discussed in \cmr{\cref{subarray}}}).}

\nasrev{\cutdel{Algorithm~\ref{alg_mapping} describes the \emph{mapping algorithm} used in \mech to map the input operands of each MAJ node in the operation's MIG to DRAM rows. The algorithm takes the optimized MIG of the operation as input, and generates an optimized mapping 
of the input operands of each MAJ node in the MIG to DRAM rows as the output. The \mech \emph{mapping algorithm} assumes that (1)~C0 and C1 rows in the C-group in the subarray (\cmr{Fig.}~\ref{fig_subarray}) contain all-0 and all-1 values, respectively; (2)~the input operands of the SIMDRAM operation are already stored in separate rows of the D-group in the subarray using vertical layout \sgii{(\cmr{\cref{sec:integ-overview}})}, before the computation of the operation starts; and (3)~the final output of the MIG is also stored in a D-group row after the computation of the entire MIG ends.}} 

\omi{The \mech \emph{\omdefi{row-to-operand} allocation algorithm} receives the operation's MIG as input. The algorithm assumes that the input operands of the operation are already stored in separate rows of the D-group in the subarray using vertical layout \sgii{(\cmr{\cref{sec:integ-overview}})}, before the computation of the operation starts. The algorithm then does a topological traversal starting with the leftmost MAJ node \omiii{at} the highest level of the MIG (e.g., level 0 in \cmr{Fig.}~\ref{fig_output_mapping}a), 
\omi{allocating compute rows \omiii{to}} the input operands of each MAJ node in the current level of the MIG, before moving to the next lower level of the graph. The algorithm finishes once \omi{DRAM rows are allocated \omiii{to}} all the input operands of all the MAJ nodes in the MIG. \cmr{Fig.}~\ref{fig_output_mapping}b shows these \omi{allocations} as the output of Task 1 for the full addition example. \sgii{The resulting} \omi{row-to-operand allocation} is then used in the second task in step two (\cmr{\cref{sec:framework:step2:generating}}) to generate the \omiii{series} of \uop{}s to compute the operation that the MIG represents. \omdef{We describe our \omdefi{row-to-operand} \omi{allocation} algorithm in Appendix~\ref{apdx:row-to-op}}. 
}




\subsubsection{Task 2: Generating \omi{a \uprog{}}.}
\label{sec:framework:step2:generating}
~\geraldorevii{\sgii{The goal of this task is} to \sgii{use the MIG and the DRAM row \omi{allocations} from Task~1 to generate the \uop{}s of the \uprog{} for our \mech operation.}
\sgii{To this end, Task~2}
(1)~translates the MIG into \sgii{a \omiii{series} of} row copy and majority \omiii{\uop{}s} \omi{(i.e., \aaps)}, 
(2)~optimizes \sgii{the \omiii{series of \uop{}s to reduce the number of \aaps}}, and 
(3)~\sgii{generalizes} the \sgii{one-bit bit-serial operation described by the MIG into an} $n$-bit operation by utilizing SIMDRAM's arithmetic and control \uop{}s.}

\omi{\textbf{\emph{Translating the MIG into \omiii{a Series of Row Copy and Majority \uop{}s.}}}} The \omi{allocation} produced \omi{during Task~1}
dictates how \omi{DRAM rows are allocated \omiii{to}} each edge in the MIG \sgii{during the \uprog{}}. With this information, \sgii{the framework} can generate the appropriate \omiii{series} of row copies and majority \omiii{\uop{}s} to reflect the MIG's computation in DRAM. To do so, we \sgii{traverse} the input MIG in topological order. For \emph{each} node, we \emph{first} assign row copy \omiii{\uop{}s} (using the \texttt{AAP} \omiii{command sequence}) to the node's edges. Then, we assign a majority \omiii{\uop{}} (using the \texttt{AP} \omiii{command sequence}) to execute the current MAJ node, following the DRAM row \omi{allocation} assigned to each edge of the node.  The \sgii{framework repeats this procedure} for all the nodes in the MIG. To illustrate, we assume that 
\omi{the SIMDRAM \omi{allocation} algorithm}
\omi{allocates DRAM rows DCC0, T1, and T0 \omiii{to} edges A, B, and C$_{in}$, respectively, \omi{of the blue node} in the full \omi{addition} MIG (\omi{\cmr{Fig.}~\ref{fig_output_mapping}a}).} 
Then, when \omviii{visiting} this node, we generate the following \omiii{series} of \uop{}s:

\begin{center}
\tempcommand{.8}
\begin{tabular}{ll}
\texttt{AAP DCC0, A};      & // DCC0 $\leftarrow$ A      \\
\texttt{AAP T1, B};        & // T1 $\leftarrow$ B        \\
\texttt{AAP T0, C$_{in}$}; & // T0 $\leftarrow$ C$_{in}$ \\
\texttt{AP $\overline{\mbox{DCC0}}$, T1, T0}   & // MAJ(NOT(A), B, C$_{in}$)
\end{tabular}
\end{center}

\textbf{\emph{\omiii{Optimizing the Series of \uop{}s.}}} 
\geraldorevii{After traversing all of the nodes in the MIG and generating the appropriate \omiii{series} of \uop{}s, we optimize the \omiii{series} of \uop{}s by \sgii{coalescing \aap \omiii{command sequences}, which we can do in one of two cases}. 

\noindent\sgii{\textbf{Case 1:}} we can coalesce a \omiii{series} of row copy \sgii{\omiii{\uop{}s} if all of the \omiii{\uop{}s} have the same \ureg{} source as an input}. For example, consider a \omiii{series} of two \omiii{\texttt{AAP}s} that copy data array $A$ into rows T2 and T3. We can coalesce this \omiii{series} of \omiii{\texttt{AAP}s} into a single \texttt{AAP} issued to the wordline address stored in \ureg{}~B10 (see \cmr{Fig.}~\ref{fig_opcodes}a). This wordline address leverages the special row decoder in the B-group \omi{(which is part of the Ambit subarray structure~\cite{seshadri2017ambit})} to activate \sgii{multiple DRAM rows in the group} \omvuii{at once} with a single activation command.
\sgii{For our example, activating \ureg{}~B10 allows the \texttt{AAP} command \omiii{sequence} to copy array $A$ into both rows T2 and T3 at once.}

\noindent\sgii{\textbf{Case 2:}} we can coalesce an \texttt{AP} command \omiii{sequence} (i.e., a majority \omiii{\uop{}}) followed by an \texttt{AAP} \sgii{command sequence (i.e., \omiii{a} row copy \omiii{\uop{}}) when the destination of the \texttt{AAP} is one of the rows used by the \texttt{AP}.
For example, consider an \texttt{AP} that performs a MAJ logic primitive on DRAM rows T0, T1, and T2 (storing the result in all three rows), followed by an \texttt{AAP} that copies \ureg{}~B12 (which refers to rows T0, T1, and T2) to row~T3.  \omiii{The \texttt{AP} followed by the \texttt{AAP}} puts the majority value in all four rows (T0, T1, T2, T3).
The two command \omiii{sequences} can be coalesced into a single \texttt{AAP} (\texttt{AAP} T3, B12), as the first ACTIVATE would automatically perform the majority on rows T0, T1, and T2 by activating all three rows simultaneously. The second ACTIVATE then copies the value from those rows into T3.}}

\omi{\textbf{\emph{Generalizing the Bit-Serial Operation into an $n$-bit Operation\omiii{.}}}} 
\sgii{Once all potential \uop{} coalescing is complete, the framework now has an optimized 1-bit version of the computation.}
We generalize \sgii{this 1-bit \uop{} \omiii{series}}
into a loop body that repeats $n$ times to implement an $n$-bit operation. We leverage the arithmetic and control \uop{}s available in SIMDRAM to orchestrate the $n$-bit computation. \omi{Data produced by the computation of one bit that \omviii{needs} to be used for computation of the next bit (e.g., the carry bit in full addition) is kept in a B-group row across the two computations, allowing for bit-to-bit data transfer without the need for dedicated shifting circuitry.}

\geraldorevii{The final \omiii{series} of  \uop{}s produced after this step is then packed into a  \uprog{} and stored in DRAM for future use.\footnote{\label{footnote:upgram}\sgii{In our example implementation of \mech, a \uprog{} has a maximum size of 128~bytes, as this is enough to store the largest \uprog{} generated in our evaluations (the division operation, which requires 56 \uop{}s, each two bytes wide, resulting in a total \uprog{} size of 112~bytes.)}} \cmr{Fig.}~\ref{fig_output_mapping}c shows the final \uprog{} produced at the end of Step 2 for the full \omi{addition} operation. The figure shows the optimized \omiii{series} of \uop{}s that generates the 1-bit implementation of the full \omi{addition} (lines 2--9), and the arithmetic and control \uop{}s included to enable the $n$-bit implementation of the operation (lines 10--11).}


\omi{\textbf{\emph{Benefits of the \uprog{} Abstraction\omiii{.}}}}
The \uprog{} abstraction \sgii{that we use to store SIMDRAM operations} provides three main advantages to the framework. First, it allows SIMDRAM to minimize the total number of new \omi{CPU} \omiii{instructions} required to implement SIMDRAM operations, \revonur{thereby} reducing SIMDRAM's impact on the ISA. While a different implementation could use more \revonur{new} \omi{CPU} \omiii{instructions} to express finer-grained operations (e.g., an \texttt{AAP}), we believe that using a minimal set of \omiii{new \omiv{CPU} instructions} simplifies adoption and software design. Second, the \uprog{} \revonur{abstraction} enables a \revonur{smaller} application binary size since the \color{black}only information that needs to be placed in the application's binary is the address of the \uprog{} \nasirevi{in main memory}. Third, \revonur{the \uprog{}} provides an abstraction to relieve the end user \revonur{from} low-level programming 
with MAJ/NOT operations \nasirevi{that} is equivalent to programming with Boolean logic. \omi{We discuss how a user program invokes \mech \uprog{}s in \cmr{\cref{sec:bbop}}.}

\subsection{\nasii{Step 3: \sgii{Operation} Execution}}
\label{sec:mc}
\label{sec:framework:step3}

\sgii{Once the framework stores the generated \uprog{} for a \mech operation in DRAM, \omi{the} SIMDRAM \omi{hardware} can now receive program requests to execute the operation.  To this end, we discuss the \mech \emph{control unit}, which handles the execution of the \uprog{} at runtime.}
\nasii{The control unit is designed} as an extension of the memory controller, \sgii{and is transparent to the programmer}. 
\sgii{A program issues a request to perform a \mech operation using a \emph{bbop} instruction \omi{(introduced by Ambit~\cite{seshadri2017ambit})}, which is one of \omi{the} \omiii{CPU} ISA extensions to allow programs to interact with the \mech framework (see \cmr{\cref{sec:bbop}}). Each \mech operation corresponds to a different \emph{bbop} instruction. Upon receiving the request,}
the control unit loads the \uprog{} corresponding to the \sgii{requested} \emph{bbop} \sgii{from memory,} and \sgii{performs the \uop{}s in the \uprog{}}. 
\omi{Since all input data elements of a \mech operation may not fit in one DRAM row, the control unit repeats the \uprog{} $i$ times, where $i$ is the total number of data elements divided by the number of elements in a single DRAM row.}

\begin{figure}[ht]
    \centering
    \includegraphics[width=\linewidth]{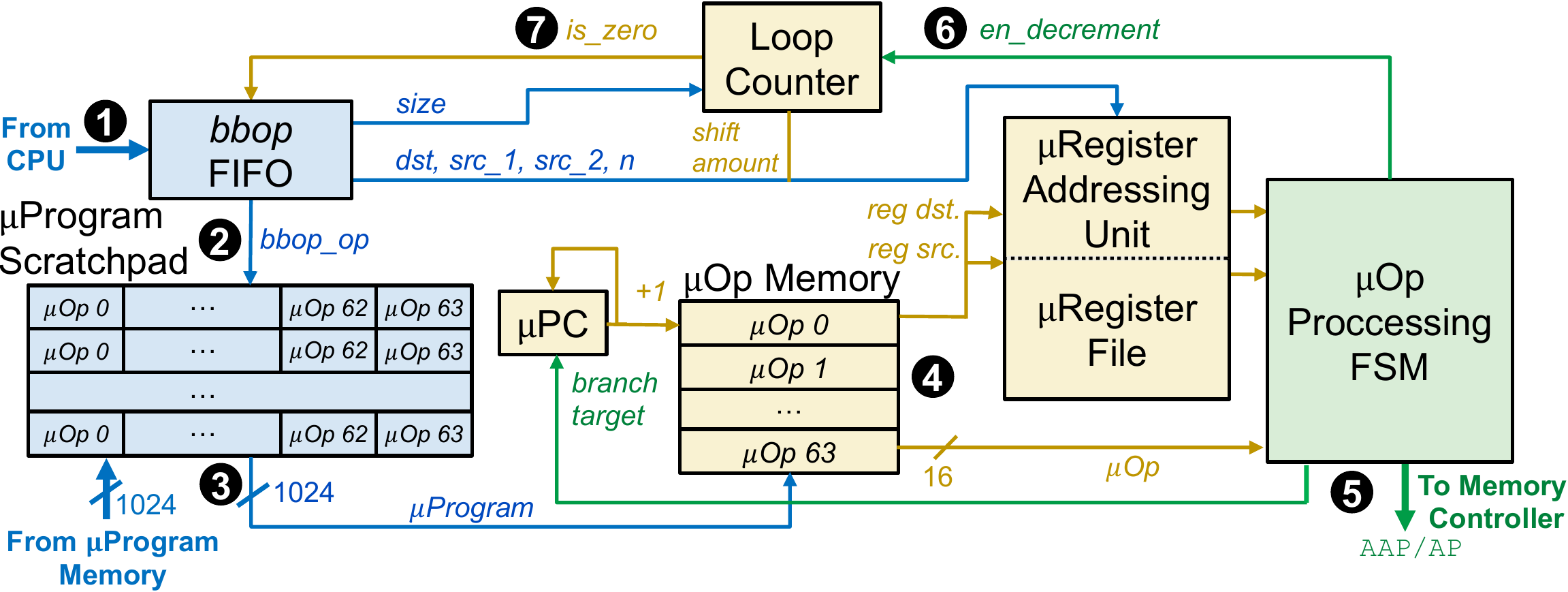}%
    \caption{\geraldo{\mech control unit.}}
    \label{fig_control}
\end{figure}

\cmr{Fig.}~\ref{fig_control} shows a block diagram of the \mech control unit\geraldorevii{, which} consists of \nasrev{nine} main components:
(1)~a \emph{bbop FIFO} that receives the \emph{bbop}s from the program,
(2)~a \emph{\uprog{} Memory} allocated in DRAM (not shown in the figure), 
(3)~\geraldo{a \emph{\uprog{} Scratchpad} \sgii{that holds commonly-used \uprog{}s}},
(4)~a \emph{\uop{} Memory} that holds the \uop{}s of the currently running \geraldorevii{\uprog{}},
\sgii{(5)~a \emph{\ureg{} Addressing Unit} that generates the physical row addresses being used by the \ureg{}s that map to DRAM rows (based on the \ureg{}-to-row assignments for B0--B17 in \cmr{Fig.}~\ref{fig_opcodes}),
(6)~a \emph{\ureg{} File} that holds the non-row-mapped \ureg{}s (B18--B31 in \cmr{Fig.}~\ref{fig_opcodes}),
(7)~a \emph{Loop Counter} that tracks the number of remaining \omi{data elements that the \uprog{} needs to be performed on,}}
(8)~\omi{a \uop{} Processing FSM}  that controls the execution flow and issues \geraldorevii{\aap} \sgii{command sequences},
\sgii{and (9)~a \omi{\uprogc{}~(\upc{})}}. 
\geraldo{\mech reserves a region of DRAM for the \uprog{} Memory to store \uprog{}s corresponding to \emph{all} \mech operations. \geraldorevii{At runtime,} the control unit 
\cmr{stores the most commonly used \uprog{}s in the \uprog{} Scratchpad, to reduce} 
the overhead of fetching \uprog{}s from DRAM.
}

\sgii{At runtime, when a CPU running a user program reaches a \emph{bbop} instruction, it forwards the \emph{bbop}}
to the \mech control unit (\circled{1} \geraldorevii{in \cmr{Fig.}~\ref{fig_control}}). 
The \sgii{control unit enqueues the \emph{bbop}} in the \emph{bbop} FIFO.
\sgii{The control unit goes through a four-stage procedure to execute the queued \emph{bbop}s one at a time.}

\sgii{In the first stage,} 
the control unit fetches and decodes the \emph{bbop} at the \sgii{head} of the FIFO (\circled{2}).
Decoding a \emph{bbop} involves
(1)~\sgii{setting the index of the \uprog{} Scratchpad to the \emph{bbop} opcode};
(2)~writing the number of \sgii{loop iterations required to perform the operation on all elements (i.e., the number of data elements divided by the number of elements in a single DRAM row) into the Loop Counter;} and
(3)~writing the base DRAM addresses of the source and destination arrays involved in the computation, and \sgii{the \omiii{size} of each data element, to the \ureg{} Addressing Unit}.

\sgii{In the second stage,}
the control unit \sgii{copies} the 
\uprog{} \sgii{currently indexed in the \uprog{} Scratchpad} to the \uop{} Memory (\circled{3}).
\geraldorevii{At this point}, the control unit is ready to start executing the \uprog{}, one \uop{} \sgii{at a} time. 

\sgii{In the third stage,}
\sgii{the current \uop{} is fetched} from the \uop{} Memory, \sgii{which is} indexed by the \omi{\upc{}}. The \omi{\uop{} Processing FSM} decodes the \uop{},
\sgii{and determines which \ureg{}s are needed (\ding{185}). For \ureg{}s B0--B17, the \omviii{\ureg{} Addressing Unit} generates the DRAM addresses that correspond to the requested registers (see \cmr{Fig.}~\ref{fig_opcodes}) and sends the addresses to the \omi{\uop{} Processing FSM}. For \ureg{}s B18--B31, the \omviii{\ureg{} File} provides the register values to the \omi{\uop{} Processing FSM}.}

\sgii{In the fourth stage,}
\sgii{the \omi{\uop{} Processing FSM} executes the \uop{}. If the \uop{} is a command sequence, the corresponding commands are sent to the memory controller's request queue (\circled{5}) and the \omi{\upc{}} is incremented.  If the \uop{} is a \texttt{done} control operation, this indicates that all of the command sequence \uop{}s have been performed for the current iteration.
The \omi{\uop{} Processing FSM} then decrements the Loop Counter (\circled{6}).
If the \omi{decremented} Loop Counter is greater than zero, the \omi{\uop{} Processing FSM}
shifts the base source and destination addresses stored in the \ureg{} Addressing Unit to move onto the next set of data elements,\footnote{The \omi{source and destination} base addresses are \omi{incremented} by $n$ rows, where $n$ is the data element \omiii{size}. \omi{This is because each DRAM row contains one bit of a set of elements, so \mech uses $n$ consecutive rows to hold all $n$~bits of the set of elements.}} and resets the \omi{\upc{}} to the first \uop{} in the \uop{} Memory.}
\sgii{If the \omi{decremented} Loop Counter equals zero, this indicates that the control unit has completed executing the current \emph{bbop}.  The} control unit \sgii{then} fetches the next \emph{bbop} from the \emph{bbop} FIFO (\circled{7}), 
\sgii{and repeats all four stages for the next \emph{bbop}}.

\subsection{Supported Operations}
\label{sec:supported:ops}
\revGeraldo{We use our 
\nasii{framework} to} efficiently support a wide 
\juang{range} of operations of different types. \omi{In this paper, we evaluate (in \cmr{\cref{sec_evaluation}}) a set of 16 \mech operations of five different types for \sgii{$n$-bit data elements}:} (1)~\sgii{$N$}-input logic operations (OR-/AND-/XOR-reduction \sgii{across $N$ inputs}); (2)~relational operations (equality/inequality \omi{check}, greater\omi{-/less-than check, greater-than-or-equal-to check}, \omi{and maximum/minimum element in a set}); (3)~arithmetic operations (addition, subtraction, multiplication, division\omi{, and absolute value}); (4)~predication (if-then-else); and (5)~other complex operations (bitcount, and Re\geraldo{LU}). \omi{We support four different \omiv{element sizes} \omiii{that} correspond to \omiii{data type sizes} in popular programming languages (8-bit, 16-bit, 32-bit, 64-bit). }

\section{System Integration of SIMDRAM}
\label{implementation}
\nasrev{
\omi{We} discuss \sgii{several} challenges of integrating \mech in a real system, and how we address them: 
(1)~data \sgii{layout} and how \mech manages storing the data required for in-DRAM computation in a vertical layout (\cref{sec:transposing}); 
(2)~ISA extensions \omiii{for} and programm\omiii{ing interface of SIMDRAM} (\cref{sec:bbop}); 
(3)~how \mech handles 
page faults, address translation, \omi{coherence,} and interrupts (\cref{sec:pagefaults});
(4)~how \mech manages computation \omiii{on} large amounts of data (\cref{sec:limited});
\omi{(5)~security implications of \mech
(\cref{sec:security})};
and (6)~current limitations of the \mech framework (\cref{sec:limitations}).}

\subsection{Data Layout}
\label{sec:transposing}

\nasrev{We envision \mech as \emph{supplementing} (not \emph{replacing}) the traditional processing elements. As a result, a program in a \mech-enabled system can have a combination of CPU instructions and \mech instructions, with possible data sharing between the two. However, while \mech operates on vertically-laid-out data \omi{(\cmr{\cref{sec:integ-overview}})}, the other system components (including the CPU) expect the data to be laid out in the traditional horizontal format, making it challenging to share data between \mech and CPU instructions. To address this challenge, memory management in \mech needs to (1)~support both horizontal and vertical data layouts in DRAM \emph{simultaneously}; and (2)~\omi{transform} vertically-laid-out data used by \mech to 
\omi{a} horizontal layout for CPU use, and vice versa. 
\sgii{We cannot rely on software (e.g., compiler or application support) to handle the data layout transformation, as this would go through the on-chip memory controller, and would introduce significant data movement, \omi{and thus latency}, between the DRAM and CPU during the transformation.
To avoid data movement during transformation,}
\mech uses a specialized hardware unit placed between the last-level cache (LLC) and the memory controller, called the \emph{\omi{data} transposition unit}, to transform 
data from horizontal data layout to vertical data layout, and vice versa.}
\omi{The transposition unit ensures that for every \mech object, its corresponding data is in a horizontal layout whenever the data is in the cache, and in a vertical layout whenever the data is in DRAM.}

\begin{figure}[ht]
    \centering
    \includegraphics[width=0.85\linewidth]{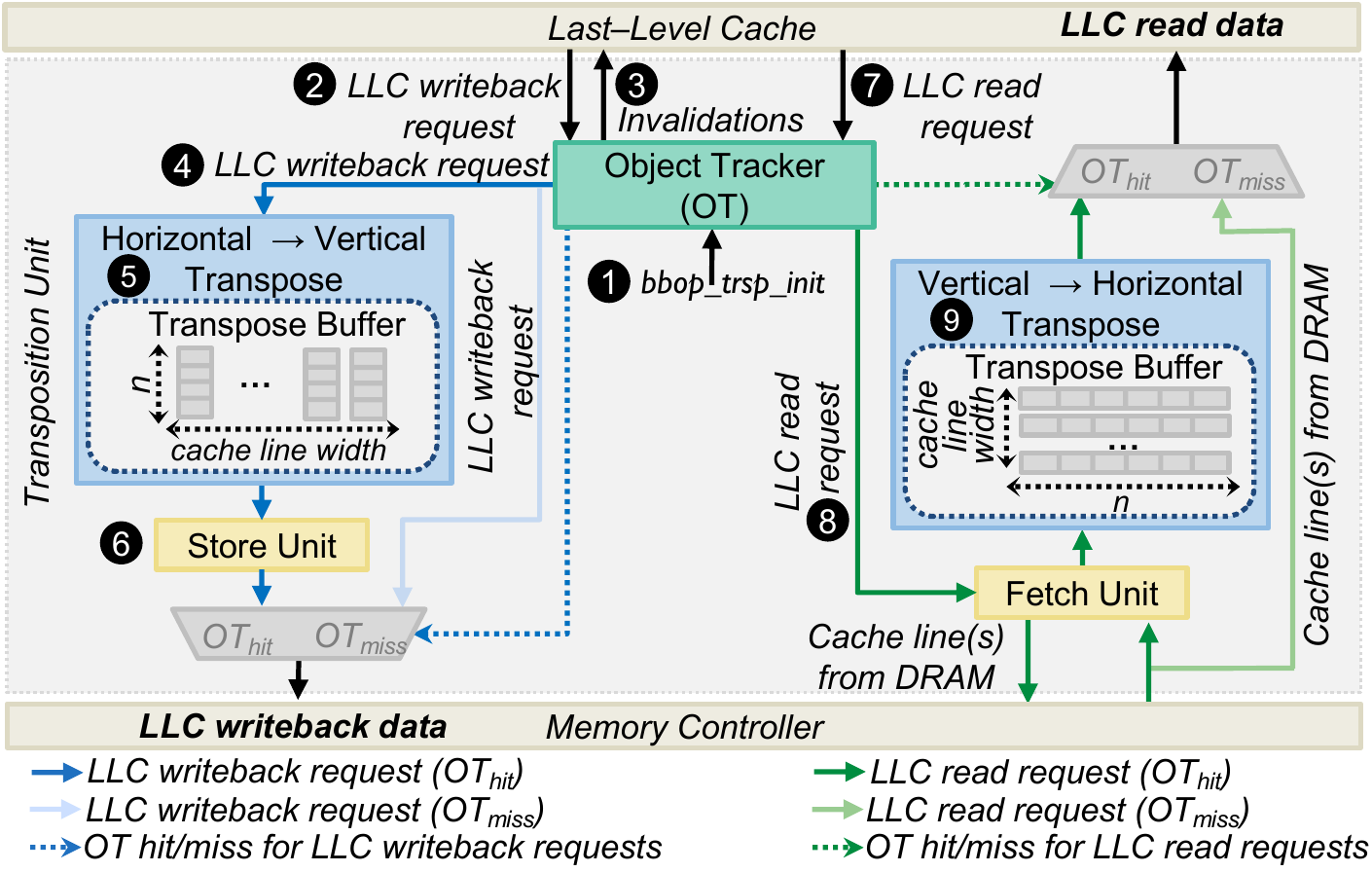}
    \caption{Major components of the \omi{data} transposition unit. 
    }
    \label{fig_transposing}
\end{figure}

\sgii{\cmr{Fig.}~\ref{fig_transposing} shows the key components of the transposition unit.} \nasrev{\sgii{The transposition unit keeps track of the memory objects that are used by \mech operations in a small cache in the transposition unit, called the \emph{Object Tracker}. To add an entry to the Object Tracker} when allocating a memory object used by \mech, the programmer adds an \omi{initialization} instruction called \texttt{bbop\_trsp\_init} (\cmr{\cref{sec:bbop}}) \sgii{\emph{immediately}} after the \texttt{malloc} that allocates the memory object \omiv{(\ding{182} in \cmr{Fig.}~\ref{fig_transposing})}.
\sgii{Assuming a system that employs lazy allocation, the}
\texttt{bbop\_trsp\_init} instruction informs the operating system (OS) that the memory object is a \emph{SIMDRAM object}. 
This allows the OS to perform virtual-to-physical memory mapping optimizations \sgii{for the object before the allocation starts}
(e.g., mapping the arguments of an operation to the same row/column in the physical memory). 
\sgii{When the \mech object's physical memory is allocated, the OS inserts} the base physical address, \sgii{the total size of the allocated data, and the \omiii{size of each element} in the object \omi{(provided by \texttt{bbop\_trsp\_init})} into the Object Tracker}.} 
\omi{As the initially-allocated data is placed in the CPU cache, the data starts in a horizontal layout until it is evicted from the cache.}

\omi{\mech stores \mech objects in DRAM using a vertical layout, since this is the layout used for in-DRAM computation (\cmr{\cref{sec:integ-overview}}). 
Since a vertically-laid-out $n$-bit element spans $n$ different cache lines in DRAM (with each cache line in a different DRAM row), \mech partitions \mech objects into \emph{\mech object slices}\omiii{, each of which is} $n$ cache lines in size. 
Thus, a \mech object slice in DRAM contains the vertically-laid-out bits of as many elements as bits in a cache line (e.g., 512 in a \SI{64}{\byte} cache line). 
Cache line $i$ ($0 \le i < n$) of an object slice contains bit~$i$ of \emph{all} elements stored in the slice. 
Whenever any one data element within a slice is requested by the CPU, the entire \mech object slice is brought into the LLC. 
Similarly, whenever a cache line from a \mech object is written back from the LLC to DRAM (i.e., it is evicted or flushed), \emph{all} $n-1$ remaining cache lines of the same \mech object slice are written back as well.\omiii{\footnote{\label{footnote:dbi}\omiii{The Dirty-Block Index~\cite{seshadri2014dirty} could be adapted for this purpose.}}} 
The use of object slices ensures correctness and simplifies the transposition unit.}

\sgii{\omi{Whenever the LLC writes back a cache line} to DRAM \omiii{(\circled{2} in \cmr{Fig.}~\ref{fig_transposing})}, 
the transposition unit checks the Object Tracker to see whether the cache line belongs to a \mech object. If the LLC request misses in the Object Tracker, the cache line does not belong to any \mech object, and the writeback request is forwarded to the memory controller \omi{as in a conventional system}.
If the LLC request hits in the Object Tracker, the cache line belongs to a \mech object, and \omi{thus} must be transposed \cmr{from the horizontal layout to the vertical layout}. An Object Tracker hit triggers two actions.}

\omi{First, the Object Tracker issues invalidation requests to \emph{all} $n-1$ remaining cache lines of the same \mech object slice \omiii{(\circled{3} in \cmr{Fig.}~\ref{fig_transposing})}.\textsuperscript{\ref{footnote:dbi}}
\omiii{We extend the LLC to support a special invalidation request type, which sends both dirty \emph{and} unmodified cache lines to the transposition unit (unlike a regular invalidation request, which simply invalidates unmodified cache lines).}
\omiii{The Object Tracker issues these invalidation requests for the remaining cache lines, ensuring that \emph{all}} cache lines of the object slice arrive at the transposition unit to perform the horizontal-to-vertical transpos\omiv{ition} correctly.}

\sgii{Second, the writeback request is forwarded \omiv{(\circled{4} in \cmr{Fig.}~\ref{fig_transposing})} to a horizontal-to-vertical transpose buffer, which performs the bit-by-bit transposition. We design the transpose buffer \omiii{(\circled{5})} such that it can transpose all bits of a horizontally-laid-out cache line in a single cycle. As the other cache lines belonging from the slice are evicted (as a result of the Object Tracker's invalidation requests) and arrive at the transposition unit, they too are forwarded to the transpose buffer, and their bits are transposed.  Each horizontally-laid-out cache line maps to a specific set of bit columns in the vertically-laid-out cache line, which is determined using the physical address of the horizontally-laid-out cache line. Once all $n$ cache lines in the \mech object slice have been transposed, the Store Unit generates DRAM write requests for each vertically-laid-out cache line, and sends the requests to the memory controller (\circled{6}).}

\sgii{When a program wants to read data that belongs to a \mech object, 
\omi{and} the data is not in the CPU caches, the LLC issues a read request to DRAM \omiv{(\circled{7} in \cmr{Fig.}~\ref{fig_transposing})}. If the address of the read request does not hit in the Object Tracker, the request is forwarded to the memory controller, 
\omi{as in a conventional system}. 
If the address of the read request hits in the Object Tracker, the read request is part of a \mech object, and the Object Tracker sends a signal \omi{(\circled{8})} to the Fetch Unit. The Fetch Unit generates the read requests for \emph{all} of the vertically-laid-out cache lines that belong to the same \mech object slice as the requested data, and sends these requests to the memory controller. When the request responses for 
\omi{the object slice's cache lines} arrive, the Fetch Unit sends the cache lines to a vertical-to-horizontal transpose buffer \omiii{(\circled{9})}, which can transpose all bits of one vertically-laid-out cache line into the horizontally-laid-out cache lines \omiii{in one cycle.} 
The horizontally-laid-out cache lines are then 
\omi{inserted into} the LLC. The \omi{$n-1$} cache lines that were \emph{not} part of the original memory request, \omi{but belong to the same object slice,} are inserted \omi{into} the LLC in a manner similar to \omiii{conventional} prefetch requests~\omiv{\cite{srinath2007feedback}}.}

\subsection{ISA Extensions and Programming Interface} 
\label{sec:bbop}

\nasrev{The lack of an efficient and expressive programmer/system interface can negatively impact the performance and usability of the \mech framework. 
This would \revonur{put} data transposition \revonur{on the critical path of} \mech computation, which would \revonur{cause large} performance overheads. To address such issues and to enable the programmer/system to efficiently communicate with \mech, we extend the ISA with specialized \mech instructions. The main goal of the \mech ISA extensions is to let the \mech control unit know (1)~what \mech operations need to be performed and when, and (2)~what the \mech memory objects are and when to transpose them.} 

\nasrev{Table~\ref{table_isa_format} shows the \omiii{CPU} ISA \omiii{extensions} that the \mech framework exposes to the programmer.
There are three types of instructions: 
(1)~\sgii{\mech object} initialization instructions, 
(2)~instructions to perform different \mech operations, and 
(3)~predication instructions.
We discuss \texttt{bbop\_trsp\_init}, our only \mech object initialization instruction, in \cmr{\cref{sec:transposing}}.
The \omiii{CPU} ISA \omiii{extensions} for performing SIMDRAM \sgii{operations} can be further divided into two categories:
(1)~operations with one input operand (e.g., bitcount, ReLU), 
and (2)~operations with two input operands (e.g., addition, division, equal, maximum). \mech uses an array-based computation \omiii{model,} and \texttt{src} (i.e., \texttt{src} in 1-input operations and \texttt{src\_1, src\_2} in 2-input operations) and \texttt{dst} in these instructions represent source and destination \emph{arrays}.
\texttt{bbop\_op} represents the opcode of the \mech operation, while \texttt{size} and \texttt{n} represent the number of elements in the source and destination arrays, and the number of bits in each array element, respectively. To enable predication, \mech uses the \texttt{bbop\_if\_else} instruction in which, in addition to two source and one destination arrays, \texttt{select} represents the 
\omi{predicate} array \omi{(i.e., the predicate\omiii{, or mask,} bits)}.} 

\begin{table}[ht]
\tempcommand{0.8}
\caption{SIMDRAM ISA \sgii{extensions}.}
\label{table_isa_format}
\resizebox{\linewidth}{!}{
    \begin{tabular}{@{}cl@{}}
    \toprule
    \textbf{Type}                & \multicolumn{1}{c}{\textbf{ISA Format}}    \\ \midrule
    Initialization & \texttt{bbop\_trsp\_init address, \omi{size}, n}                         \\
    1-Input Operation           & \texttt{bbop\_op dst, src, \omi{size}, n}                                \\
    2-Input Operation           & \texttt{bbop\_op dst, src\_1, src\_2, size, n}               \\
    Predication                  & \texttt{bbop\_if\_else dst, src\_1, src\_2, select, size, n} \\ \bottomrule
    \end{tabular}
}
\end{table}

\nasrev{Listing~\ref{lst:simdraprog} shows how \omiii{\mech's CPU} ISA \omiii{extensions} can be used to perform in-DRAM computation, 
\omi{with an example code that performs element-wise addition or subtraction of two arrays (\texttt{A} and \texttt{B}) depending on the comparison of each element of \texttt{A} to the corresponding element of a third array (\texttt{pred}).} 
Listing~\ref{sublst:codea} shows the \omiv{original} C code for the computation, while Listing~\ref{sublst:codeb} shows the equivalent code using \mech operations. 
The lines that perform the same operations are highlighted using the same colors in both 
\omi{C code and SIMDRAM code}. 
The if-then-else condition in C code is performed in \mech using a predication instruction (i.e., \texttt{bbop\_if\_else} 
\omi{on} line \omiv{16} in Listing~\ref{sublst:codeb}). 
\mech treats the if-then-else condition as a multiplexer. Accordingly, \texttt{bbop\_if\_else} takes two source arrays and a 
\omi{predicate} array as inputs, where the 
\omi{predicate} is used to choose which source array should be selected as the output \omi{at the corresponding index}. 
To this end, we first 
\omi{allocate} 
\omi{two arrays to hold the addition and subtraction results} (i.e., arrays \texttt{D} and \texttt{E} 
\omi{on} line 10 in Listing~\ref{sublst:codeb}), and then populate them using \texttt{bbop\_add} and \texttt{bbop\_sub} (lines \omiv{13} and \omiv{14} in Listing~\ref{sublst:codeb}), respectively. 
We then 
\omi{allocate the predicate array} (i.e., array \texttt{F} on line \omiv{11} in Listing~\ref{sublst:codeb}) and populate \omi{it using \texttt{bbop\_greater}} (line \omiv{15} in Listing~\ref{sublst:codeb}). 
The addition, subtraction, and 
\omi{predicate} arrays form the three inputs \omiii{(arrays \texttt{D}, \texttt{E}, \texttt{F})} to the \texttt{bbop\_if\_else} instruction \omiii{(line \omiv{16} \omiii{ in Listing~\ref{sublst:codeb}})}, which stores the outcome \omi{of the predicated execution} to the destination array (i.e., array \texttt{C} in Listing~\ref{sublst:codeb}).}

\begin{figure}[h]
    \setcaptiontype{lstlisting}
    \begin{minipage}{.45\textwidth}
	    \begin{lstlisting}[style=myC]
int size = 65536;
int elm_size = sizeof(uint8_t);
uint8_t *A, *B, *C = (uint8_t*)malloc(size*elm_size);
uint8_t *pred = (uint8_t*)malloc(size*elm_size);
...
for(int i = 0; i < size; ++i) {
%\HilightGreen%    bool cond = A[i] > pred[i];
%\HilightPink%    if (cond)
%\HilightBlue%        C[i] = A[i] + B[i];
%\HilightPink%    else
%\HilightYellow%        C[i] = A[i] - B[i];
}
\end{lstlisting}
        \vspace{-23pt}
        \subcaption{C code for vector add/sub with \omi{predicated} execution}
        \label{sublst:codea}
    \end{minipage}
   ~
    \begin{minipage}{.45\textwidth}
	    \begin{lstlisting}[ style=myC]
int size = 65536;
int elm_size = sizeof(uint8_t);
uint8_t *A, *B, *C = (uint8_t*)malloc(size*elm_size);

bbop_trsp_init(A,size,elm_size);
bbop_trsp_init(B,size,elm_size);
bbop_trsp_init(C,size,elm_size);
uint8_t *pred = (uint8_t*)malloc(size*elm_size);
// D, E, F store intermediate data
uint8_t *D, *E = (uint8_t*)malloc(size*elm_size);
bool *F = (bool*)malloc(size*sizeof(bool));
...
%\HilightBlue%bbop_add(D, A, B, size, elm_size);
%\HilightYellow%bbop_sub(E, A, B, size, elm_size);
%\HilightGreen%bbop_greater(F, A, pred, size, elm_size);
%\HilightPink%bbop_if_else(C, D, E, F, size, elm_size);
\end{lstlisting}
       \vspace{-23pt}
       \subcaption{Equivalent code using SIMDRAM operations}
       \label{sublst:codeb}
   \end{minipage}
    \caption{\geraldorevi{Example code using SIMDRAM instructions. \gfrev{or operations?}}}
    \label{lst:simdraprog}
    \vspace{-10pt}

\end{figure}


\textfromsl{In this work, we assume that the programmer manually \revsgii{rewrites} the code to use SIMDRAM operations. We follow this approach when evaluating  real-world applications in \cmr{\cref{sec_real_world_kernels}}. 
\geraldorevi{We envision two programming models for SIMDRAM. In the first programming model, SIMDRAM operations \revonurii{are} encapsulated within userspace library routines to ease programmability.} With this approach, the programmer can optimize the SIMDRAM-based code to \revonurii{make} the most out of the underlying in-DRAM computing mechanism. 
\geraldorevi{In the second programming model, SIMDRAM operations \revonurii{are} transparently inserted within the application's binary using compiler assistance.} Since SIMDRAM is a SIMD-like compute engine, we expect that the compiler can generate SIMDRAM code without \revonurii{programmer} intervention in at least two ways. First, it can leverage auto-vectorization routines already present in modern compilers~\revdelrefr{\cite{AutoVect51:online, Autovect33:online}}\revdelrefa{\hl{[39, 76]}} to generate SIMDRAM code, by setting the width of the SIMD lanes equivalent to a DRAM row. For example, in LLVM~\revdelrefr{\cite{lattner2004llvm}}\revdelrefa{\hl{[60]}}, the width of the SIMD units can be defined using the "\texttt{-force-vector-width}" flag~\revdelrefr{\cite{AutoVect51:online}}\revdelrefa{\hl{[76]}}. A SIMDRAM-based compiler back-end can convert the LLVM intermediate representation instructions into \emph{bbop} instructions.
Second, the compiler can compose groups of existing SIMD instructions generated by the compiler (e.g., AVX2 instructions~\revdelrefr{\cite{firasta2008intel}}\revdelrefa{\hl{[29]}}) into blocks that match the size of a DRAM row, and then convert such instructions into a single SIMDRAM operation. \omi{Prior work~\revdelrefr{\cite{ahmed2019compiler} uses a similar approach for 3D-stacked PIM}}\revdelrefa{\hl{[4]}}. We leave the design of a compiler for SIMDRAM for future work.}


SIMDRAM instructions can be implemented by extending the ISA of the host CPU. This is \color{black}possible since there is enough unused opcode space to support the extra opcodes that SIMDRAM requires. To illustrate, prior works~\revdelrefr{\cite{lopes2013isa,lopes2015shrink}}\revdelrefa{\hl{[75,76]}} show that there are 389 unused operation codes considering only the AVX and SSE \revonur{extensions} for the x86 ISA. Extending the instruction set 
is a common approach 
\omi{to interface a CPU with} PIM architectures~\revdelrefr{\cite{ahn2015scalable, seshadri2017ambit}}\revdelrefa{\hl{[4,93]}}. 





\subsection{Handling Page Faults, Address Translation, \omi{Coherence,} and Interrupts}
\label{sec:pagefaults}

\nasirevi{
SIMDRAM handles 
\omi{four} key system mechanisms \omi{as follows}:

\geraldorevi{\begin{itemize}[itemsep=0pt, topsep=0pt, leftmargin=*]
    \item  \textit{Page Faults:} \revonur{W}e assume that the pages that are touched during in-DRAM computation are already present and pinned in DRAM. In case the \revonur{required} data is not present in DRAM, we rely on the conventional page fault handling mechanism to bring the required pages into DRAM. 
    
    \item \textit{Address Translation:} Virtual memory and address translation \sr{
    \omi{are} challenging} for \nasirevi{many} PIM architectures~\revdelrefr{\cite{ghose2018enabling, PEI, picorel2017near}}\revdelrefa{\hl{[5,37,86]}}. SIMDRAM is relieved of such challenge as it operates directly on physical addresses. When the CPU issues a SIMDRAM instruction, the instruction's \sr{virtual} memory addresses are translated \sr{into their corresponding physical addresses using} the same translation lookaside buffer (TLB) lookup mechanisms used by regular load/store operations.
    
    \item \omi{\textit{Coherence:} Input arrays to \mech may be generated or modified by the CPU, and the data updates may reside only in the cache (e.g., because the updates have not yet been written back to DRAM). To ensure that \mech does not operate on stale data, programmers are responsible for flushing cache lines~\cite{guide2016intel, manual2010arm} modified by the CPU. \cmr{\omiv{SIMDRAM can leverage coherence optimizations tailored to PIM to improve overall performance~\cite{lazypim,boroumand2019conda}}.}}
    
    \item  \textit{Interrupts:} Two cases where an interrupt could affect the execution of a SIMDRAM operation are (1)~on an application context switch, and (2)~on a page fault. In case of a context switch, the control unit's context needs to be saved and \nasirevi{then} restored \nasirevi{later} when the application resume\revonur{s} execution. \revonur{W}e do not expect to encounter a page fault during the execution of a SIMDRAM operation since\revonur{,} as previously mentioned, pages touched by SIMDRAM operations are \revonur{\cmr{expected to be loaded into and} pinned in DRAM.}
\end{itemize}}} 

\subsection{Handling Limited Subarray Size}
\label{sec:limited}

\textfromsl{\geraldorevi{SIMDRAM  
operates on data placed within the same subarray. However, a single subarray stores \revonur{only} 
\omi{several} megabytes of data. For example, a subarray with 1024 rows and a row size of \SI{8}{\kilo\byte} can only store \SI{8}{\mega\byte} of data. Therefore, SIMDRAM needs to use a mechanism that can efficiently move data within DRAM (e.g., across DRAM banks and subarrays). SIMDRAM can exploit (1)~RowClone Pipelined Serial Mode (PSM)~\revdelrefr{\cite{seshadri2013rowclone}}\revdelrefa{\hl{[92]}} to copy data between two banks by using the internal DRAM bus, or (2)~Low-Cost Inter-Linked Subarrays (LISA)~\revdelrefr{\cite{chang2016low}}\revdelrefa{\hl{[18]}} to copy rows between two subarrays within the same bank. We evaluate the performance overheads of using both mechanisms in~\cref{sec:eval:datamovement}}.} 
\omi{Other mechanisms \omiii{for} fast in-DRAM data movement~\cite{nom20,wang2020figaro} can also enhance \mech's capability.}

\subsection{Security Implications}
\label{sec:security}

SIMDRAM and other similar in-DRAM computation mechanisms that use dedicated DRAM rows to perform computation may 
\omi{increase} vulnerabilit\omiii{y} \omi{to} RowHammer attacks~\cite{kim2014flipping,revisitrh,frigo2020trrespass,mutlu2017rowhammer,mutlu2019rowhammer}\omii{.}  
\omii{W}e believe, and the literature suggests, that there 
\omi{should be} robust and scalable solutions to RowHammer, orthogonally to our work (e.g., \geraldorevii{BlockHammer~\cite{yaglikci2020blockhammerarxiv},} PARA~\cite{para}, TWiCe~\cite{lee2019twice}, \omi{Graphene~\cite{park2020graphene}}). 
Exploring RowHammer prevention and mitigation mechanisms 
\omi{in conjunction with} SIMDRAM (or other \cmr{PIM} 
approaches) requires special attention and research, which we leave for future work.

\subsection{SIMDRAM Limitations}
\label{sec:limitations}

\sgii{We note three key limitations of the current version of \revonur{the} SIMDRAM framework:}
\geraldorevi{
\begin{itemize}[itemsep=0pt, topsep=0pt, leftmargin=*]
    \item  \textit{Floating-Point Operations:} SIMDRAM supports \revonur{only} integer and fixed-\revonur{point} operations. Enabling floating-point operations in-DRAM while maintaining low area overheads is a challenge. For example, for floating-point addition, the IEEE 754 FP32 format~\cite{ieee754} requires \omiii{shifting} the mantissa by the difference of the exponents of elements. Since each bitline stores a data element in SIMDRAM, \revonur{shifting} \revonur{the value stored in one bitline} without compromising the values stored in other bitlines \revonur{at low cost is currently infeasible}.  
    
    \item \textit{Operations That Require Shuffling Data Across Bitlines:} Different from prior works (e.g., \revonur{DRISA}~\revdelrefr{\cite{li2017drisa}}\revdelrefa{\hl{[72]}}), SIMDRAM does \emph{not} add any extra circuitry to perform \sr{bit-shift} operations. Instead, SIMDRAM stores data in a vertical layout and \omi{can} perform 
    \omi{explicit} \sr{bit-shift} 
    operations \omi{(if needed)} by orchestrating row copies. 
    Even though this approach enables \mech to implement a large range of operations, it is not possible to perform shuffling and reduction operations \emph{across bitlines} without \sr{the inclusion of dedicated \omi{bit-shifting}} circuitry. This is due to the lack of physical connections across bitlines\sr{, which can} be solved by \sr{building} a \sr{bit-shift} engine \revonur{near} the sense amplifiers. 
    
    \item  \textit{Synchronization Between Concurrent In-DRAM Operations:} SIMDRAM can be easily modified to enable concurrent \revonur{execution of \emph{distinct operations} across different subarrays in DRAM}. However, this would require the implementation of software or hardware synchronization primitives to orchestrate the computation of a single task across different subarrays. 
    \omi{Ideas that are similar to SynCron~\cite{giannoula2021SynCron} can be beneficial.}
    \end{itemize}    }

\section{Methodology}
\label{methodology}

We implement SIMDRAM using the gem5 simulator~\cite{gem5} and compare it to a \omi{real} multicore CPU (Intel Skylake~\cite{intelskylake}), a \omii{real} high-end GPU (\omi{NVIDIA Titan V}~\cite{TitanV}), and a state-of-the-art \omi{processing-using-DRAM} mechanism (Ambit~\cite{seshadri2017ambit}). In all our evaluations, the CPU code is optimized to leverage AVX-512 instructions~\cite{firasta2008intel}. \geraldorevi{Table~\ref{table_parameters} shows the system parameters we use in our evaluations.} To measure CPU performance, we implement a set of timers in \texttt{sys/time.h}~\cite{systime}\omii{. T}o measure CPU energy consumption, we use Intel RAPL~\cite{hahnel2012measuring}. To measure GPU performance, we implement a set of timers using the \texttt{cudaEvents} API~\cite{cheng2014professional}. We capture GPU kernel execution time \sr{that excludes} data initialization/transfer time. To measure GPU energy consumption, we use the \texttt{nvml} API~\cite{NVIDIAMa14}. We report the average of five runs for each CPU/GPU data point, each with a \omii{warmup} phase to avoid cold cache effects. \omii{W}e implement Ambit \omii{on} gem5 \omii{and} validate our implementation rigorously with the results \omi{reported in \cite{seshadri2017ambit}}. \omii{We \omiii{use} \omiii{the same} vertical data layout in our Ambit \omiii{and SIMDRAM implementations}, \omiii{which} enables us to (1) evaluate all 16 SIMDRAM operations in Ambit using their equivalent AND/OR/NOT-based implementation\omiii{s}, and (2) highlight the benefits of Step 1 in the \mech framework (i.e., using an optimized MAJ/NOT-based implementation of the operations).} 
\omvuii{Our synthetic throughput analysis \omviii{(\cref{sec_performance})} uses 64M-element input
arrays.}

\begin{table}[ht]
   \caption{\omiii{Evaluated s}ystem configuration\omiii{s}.}
   \centering
   \footnotesize
   \tempcommand{1.3}
   \renewcommand{\arraystretch}{0.7}
   \resizebox{\columnwidth}{!}{
   \begin{tabular}{@{} c l @{}}
   \toprule
   \multirow{5}{*}{\shortstack{\textbf{Intel}\\ \textbf{Skylake CPU\omii{~\cite{intelskylake}}}}} & x86~\cite{guide2016intel}, 16~cores, 8-wide, out-of-order, \SI{4}{\giga\hertz};  \\
                                                                           & \emph{L1 Data + Inst. Private Cache:} \SI{32}{\kilo\byte}, 8-way, \SI{64}{\byte} line; \\
                                                                           & \emph{L2 Private Cache:} \SI{256}{\kilo\byte}, 4-way, \SI{64}{\byte} line; \\
                                                                           & \emph{L3 Shared Cache:} \SI{8}{\mega\byte}, 16-way, \SI{64}{\byte} line; \\
                                                                           & \emph{Main Memory:} \SI{32}{\giga\byte} DDR4-2400, 4~channels, 4~ranks \\
   \midrule
   \multirow{3}{*}{\shortstack{\textbf{\omi{NVIDIA}}\\ \textbf{\omi{Titan V} GPU\omii{~\cite{TitanV}}}}} & 6 graphics processing clusters, 5120 CUDA Cores;\\ 
                                                                            & 80 streaming multiprocessors, \SI{1.2}{\giga\hertz} base clock; \\
                                                                            & \emph{L2 Cache:} \SI{4.5}{\mega\byte} L2 Cache; \emph{Main Memory:} \SI{12}{\giga\byte} HBM~\omii{\cite{HBM,lee2016simultaneous}} \\
   \midrule
   \multirow{5}{*}{\shortstack{\omi{\textbf{Ambit~\cite{seshadri2017ambit}}}\\ \textbf{\omi{and SIMDRAM}}}} &  gem5 system emulation;  x86~\cite{guide2016intel}, 1-core, out-of-order, \SI{4}{\giga\hertz};\\
                                                                             & \emph{L1 Data + Inst. Cache:} \SI{32}{\kilo\byte}, 8-way, \SI{64}{\byte} line;\\
                                                                             & \emph{L2 Cache:} \SI{256}{\kilo\byte}, 4-way, \SI{64}{\byte} line; \\
                                                                             & \emph{Memory Controller:}  \SI{8}{\kilo\byte} row size, FR-FCFS~\cite{mutlu2007stall,zuravleff1997controller} scheduling\\
                                                                             & \emph{Main Memory:}  DDR4-2400, 1~channel, 1~rank, 16~banks \\
   \bottomrule
   \end{tabular}
   }
   \label{table_parameters}
\end{table}

We evaluate three different configurations of SIMDRAM \omii{where} 1 \geraldorevii{(\emph{SIMDRAM:1})}, 4 \geraldorevii{(\emph{SIMDRAM:4})}, and 16 \geraldorevii{(\emph{SIMDRAM:16})} banks out of all the banks in one channel (16 banks in our evaluations) have SIMDRAM computation capability. In the SIMDRAM 1-bank configuration, our mechanism exploits 65536 (\omi{i.e., size of an} \SI{8}{\kilo\byte} row buffer) SIMD lanes. Conventional DRAM architectures exploit bank-level parallelism (BLP) to maximize DRAM throughput~\omii{\cite{mutlu2008parallelism,kim2010thread,kim2012case,kim2016ramulator,lee2009improving}}. The memory controller can issue commands to different banks (one-per-cycle) on the same channel such that banks can operate in parallel. In SIMDRAM, banks in the same channel can operate in parallel, just like conventional banks. Therefore, to enable the required parallelism, SIMDRAM requires no more modifications. Accordingly, the number of available SIMD lanes\omii{, i.e., SIMDRAM's computation capability,} increases by exploiting BLP in SIMDRAM configurations (i.e., the number of available SIMD lanes in the \omii{16}-bank configuration is \omii{16} $\times$ 65536).\omii{\footnote{\omii{SIMDRAM computation capability can be further increased by enabling and exploiting subarray-level parallelism in each bank~\omiii{\cite{kim2012case,chang2016low,chang2014improving,kang2014co}}.}}} 

\section{Evaluation}
\label{evaluation}
\label{sec_evaluation}

\ifasploscr

\omii{We demonstrate the advantages of the \mech framework by evaluating: (1)~SIMDRAM's throughput and energy consumption \omii{for a wide range of operations}; (2)~SIMDRAM's performance benefits \omiii{on} real-world applications; \cmr{and} (3)~SIMDRAM's performance and energy benefits \omiii{over} a closely\omiii{-}related processing-using-\omiii{cache} architecture~\cite{dualitycache}. Finally, we evaluate \cmr{the area cost of SIMDRAM}.}
\cmr{The \omvuii{definitive} version of the paper~\cite{extendedsimdram} \omvi{demonstrates}
(1)~how SIMDRAM's triple-row-activation (TRA) operations \omvuii{are more }scalable and variation-tolerant than quintuple-row-activations (QRAs) used by prior works~\cite{ali2019memory, angizi2019graphide}; and 
(2)~the \omvi{large} performance \omvi{benefits} of \mech even \omvi{in the presence of} worst-case data movement and data transposition overheads.}

\else

\omii{We demonstrate the advantages of the \mech framework by evaluating: (1)~SIMDRAM's throughput and energy consumption \omii{for a wide range of operations}; (2)~SIMDRAM's performance benefits \omiii{on} real-world applications; (3)~SIMDRAM's performance and energy benefits \omiii{over} a closely\omiii{-}related processing-using-\omiii{cache} architecture~\cite{dualitycache}; and (4)~the reliability of SIMDRAM operations. Finally, we evaluate three key overheads in SIMDRAM\omii{:} in-DRAM data movement, data transposition, and area cost.}

\fi


\subsection{Throughput Analysis}
\label{sec_performance}

\begin{figure}[ht]
   \centering
   \includegraphics[width=1.0\linewidth]{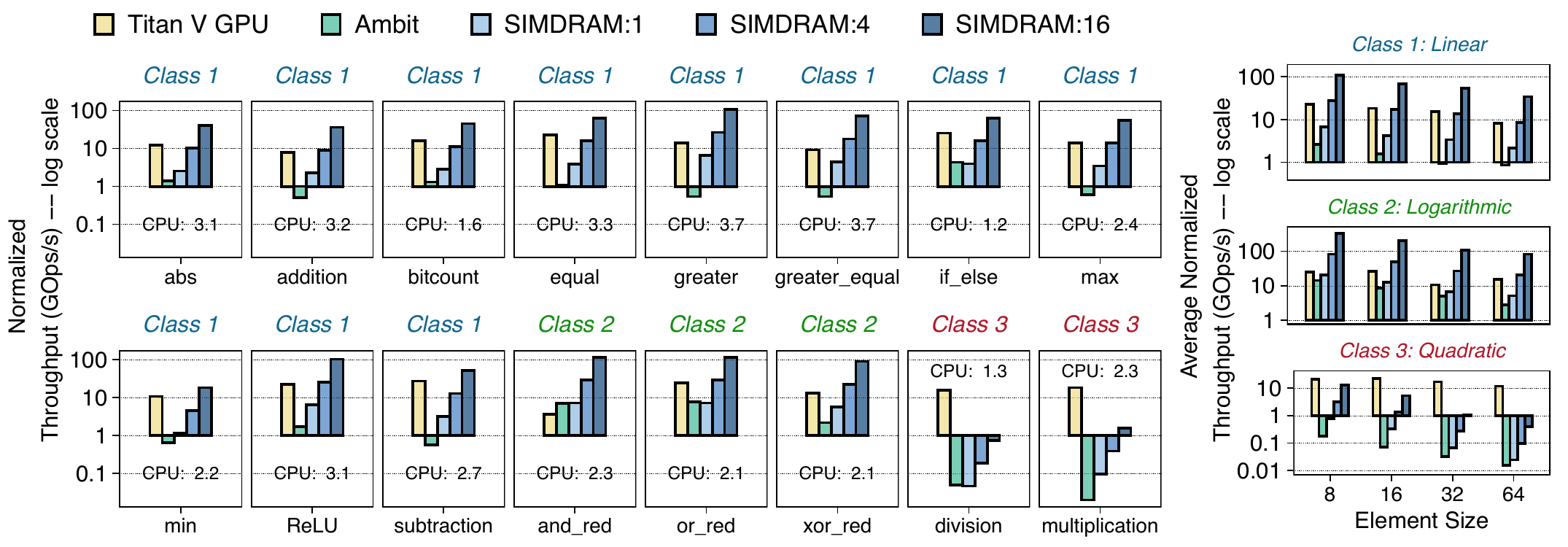}
   \caption{\omvi{Normalized throughput of 16  operations. \omvuii{SIMDRAM:\emph{X} uses \emph{X} DRAM banks for computation.}}}
 
   \label{fig_throughput}
\end{figure}

\cmr{Fig.}~\ref{fig_throughput} (\omi{left}) shows the \omii{normalized} throughput of all \omii{16} \mech operations \omi{(\cmr{\cref{sec:supported:ops}}) compared to \omii{those on CPU, GPU, and Ambit} (normalized to the multicore CPU throughput), for 
 a\omii{n element size} of 32~bits. \omv{We provide the} \omiii{absolute throughput of the baseline CPU (in GOps/s) in each graph.} 
 We classify each operation based on how the latency of the operation scales with respect to \omii{element size} $n$\omii{.}\footnote{\omdef{Appendix~\ref{apdx:op-class} \omiii{discusses the scalability} of each operation\omii{.}}}
 Class~1\omii{, 2, and 3 operations} scale linearly\omii{,} logarithmically, and quadratically with $n$\omii{, respectively}.} 
\omi{\cmr{Fig.}~\ref{fig_throughput} (right) shows how the average throughput across all operations of the same class scales relative to \omii{element size}.
 We evaluate element sizes \revonur{of} 8, 16, 32, 64 bits. We normalize the figure to the average throughput on a CPU}. 
 
\omi{W}e \omi{make} \omi{four} observations \omi{from \cmr{Fig.}~\ref{fig_throughput}}. \omi{First, we observe that SIMDRAM outperforms the three state-of-the-art baseline systems i.e., CPU/GPU/Ambit. \omii{Compared to CPU/GPU, SIMDRAM\omiii{'s throughput is} 
5.5$\times$/0.4$\times$, 
22.0$\times$/1.5$\times$, and 
88.0$\times$/5.8$\times$ \omiii{that of \omiv{the} CPU/GPU,} 
\omii{averaged across all 16 SIMDRAM operations} for 1, 4, and 16~banks, respectively. To ensure fairness, we compare Ambit, which uses a single DRAM bank in our evaluations, only against \emph{SIMDRAM:1}.\footnote{\omiii{Ambit's throughput scales proportionally to bank count, just like SIMDRAM's.}} Our evaluations show that \emph{SIMDRAM:1} outperforms Ambit by 2.0$\times$, averaged across all 16 SIMDRAM operations.}} \omi{Second, SIMDRAM outperforms the GPU baseline when we use more than four DRAM banks for \emph{all} the linear and logarithmic operations. \emph{SIMDRAM:16} provides 5.7$\times$ (9.3$\times$) \omiii{the throughput of} the GPU across all linear (logarithmic) operations, on average. \emph{SIMDRAM:16}\omiii{'s throughput is} 83$\times$ (189$\times$) and 45.2$\times$ (19.9$\times$) \omiv{that of CPU and Ambit, respectively}, averaged across all linear (logarithmic) operations.} \omi{Third}, we observe that both  \omi{the multicore CPU} baseline and GPU outperform \geraldorevii{\emph{SIMDRAM:1}, \emph{SIMDRAM:4}, and \emph{SIMDRAM:16}} \omi{\emph{only} for the} division and multiplication operations. This is due to the quadratic nature of our bit-serial implementation of \omi{these two} operations. \omi{Fourth}, as expected, we observe a drop in the throughput for \omii{\emph{all}} operations \omii{with} increasing element size, since the latency \omii{of} each operation increases \omi{with} element size. \omi{W}e conclude that SIMDRAM \omii{significantly} \omi{outperforms all three state-of-the-art baselines} for a wide range of operations.

\subsection{Energy Analysis}
\label{sec_energy}

\begin{figure}[h]
   \centering
   \includegraphics[width=1.0\linewidth]{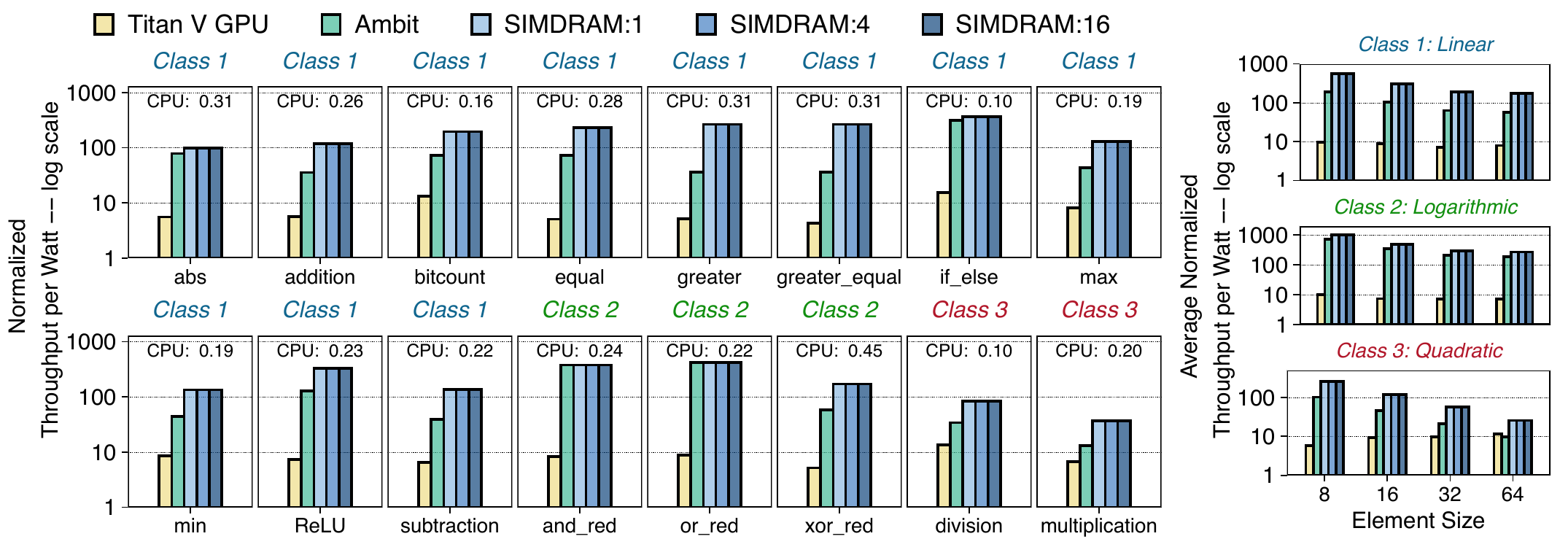}%
   \caption{\omvi{Normalized  energy efficiency of 16 operations.}
   }
   \label{fig_energy}
\end{figure}

We use \omi{CACTI}~\cite{cacti} to evaluate SIMDRAM's \omi{energy} consumption. Prior work~\cite{seshadri2017ambit} shows that each additional simultaneous row activation increases energy consumption by 22\%. We \omii{use} this observation in evaluating the energy consumption of SIMDRAM, which requires TRAs. \cmr{Fig.}~\ref{fig_energy} compares the energy efficiency (\omii{\omii{T}hroughput} per Watt) of SIMDRAM against the GPU and Ambit baselines\omi{, normalized to the CPU baseline}. \omiv{We provide the absolute Throughput per Watt of the \omv{baseline} CPU in each graph}. We make \omii{four} observations. First, SIMDRAM significantly increases energy efficiency for \emph{all} operations \sgii{over \omii{\emph{all} three} baselines}. 
\omi{SIMDRAM’s energy efficiency is 257$\times$, 31$\times$, and 2.6$\times$ that of CPU, GPU, and Ambit, respectively, averaged across all 16 operations.} The energy savings in SIMDRAM directly result from (1)~avoiding the costly off-chip round-trips to load\omiii{/store} data from\omiii{/to} memory, (2)~exploiting the abundant memory bandwidth within the memory device, reducing execution time\omii{, and (3)~reducing the number of TRAs required to compute a given operation by leveraging an optimized majority-based implementation of the operation}. Second, similar to \omi{our results on} throughput \omi{(\cmr{\cref{sec_performance}})}, the energy efficiency \omi{of} SIMDRAM reduces \omi{as} element size \omi{increases}. \omi{However, the energy efficiency of the CPU \omiii{or} GPU does not}. This is because (1)~for \omii{all SIMDRAM operations}, the number of TRAs increases with element size; and (2) \omii{CPU and GPU can} fully utiliz\omii{e} their \omii{wider} arithmetic units with larger (i.e., 32\omi{-} and 64\omi{-}bit) element sizes. 
Third, even though \omii{SIMDRAM} multiplication and division operations \omi{scale} poorly \omii{with} element size, the SIMDRAM implementations of these operations are significantly more energy-efficient compared to the CPU and GPU baselines,  
\geraldorevi{making SIMDRAM \revonur{a} competitive \revonur{candidate} \omiii{even} for multiplication and division operations.} \omii{Fourth, since \emph{both} SIMDRAM's throughput and power consumption increase proportionally to the number of banks, the \omii{T}hroughput per Watt for SIMDRAM 1-, 4-, and 16-bank configurations is the same.} \omi{We conclude that SIMDRAM is more energy\omii{-}efficient than all three state-of-the-art baselines for a wide range of operations. }

\subsection{Effect on Real-World Kernels}
\label{sec_real_world_kernels}

We evaluate SIMDRAM with a set of kernels that represent the behavior of \omi{selected} important real-world applications from different domains. The evaluated kernels \omi{come from} databases (TPC-H query 1~\cite{tpch}, BitWeaving~\cite{li2013bitweaving}), convolutional neural networks (LeNET-5~\cite{lecun2015lenet}, VGG-13~\cite{simonyan2014very}, VGG-16~\cite{simonyan2014very}), \omi{classification algorithms} (k-nearest neighbors~\cite{lee1991handwritten}), and  image processing (brightness\omii{~\cite{gonzales2002digital}}). These kernels rely on many of the basic operations we evaluate in \cmr{\cref{sec_performance}}. \omdef{We provide a brief description of each kernel and the SIMDRAM operations \omdefii{it utilizes} in Appendix~\ref{appendix}}. 

\cmr{Fig.}~\ref{fig_speedup_real_world} shows the \omi{performance of \mech and our baseline configurations} \omiii{for} each kernel, normalized to \omii{that} of the multicore CPU. 
We make \omi{four} observations. 
\omi{First, \emph{SIMDRAM:16} greatly outperforms the CPU \omii{and} GPU baselines, \omiv{providing 21$\times$ \omii{and} 2.1$\times$ 
the performance of the CPU and GPU, respectively, on average across all \omii{seven} kernels.
\omii{\omiv{SIMDRAM has a maximum performance of 65$\times$ \omii{and} 5.4$\times$ that of the CPU and GPU, respectively (for the BitWeaving kernel in both cases).}}}
\omii{Similarly, \emph{SIMDRAM:1} \omiv{provides 2.5$\times$ the performance of Ambit (which also uses a single bank for in-DRAM computation)}, on average across all seven kernels, with a maximum \omiv{of 4.8$\times$ the performance of Ambit} for the TPC-H kernel.}} 
\omi{Second, even with a single DRAM bank, SIMDRAM \emph{always} outperforms the CPU baseline, \omiv{providing 2.9$\times$ the performance of the CPU on average} across all kernels.} \omi{Third, 
\emph{SIMDRAM:4} provides 2$\times$ and 1.1$\times$ \omiv{the performance of} the GPU \omii{baseline} for \omiv{the BitWeaving and brightness kernels, respectively}.}  Fourth, \omii{despite GPU's higher multiplication throughput compared to \mech (\cmr{\cref{sec_performance}}), \emph{SIMDRAM\omii{:16}} outperforms the GPU baseline \omiii{even} for kernels 
that heavily rely on multiplication \omdef{(Appendix~\ref{appendix})} (e.g., by 1.03$\times$ and 2.5$\times$ for kNN and TPC-H kernels, respectively). This speedup is a direct result of exploiting the high \omii{in-DRAM} bandwidth in \mech to avoid the memory bottleneck in GPU caused by the large amounts of intermediate data generated in such kernels.} \revonur{We conclude that} SIMDRAM is \omi{an effective and efficient} substrate to accelerate many common\revonuri{ly}-used real-world applications.

\begin{figure}[h]
   \centering
   \includegraphics[width=\linewidth]{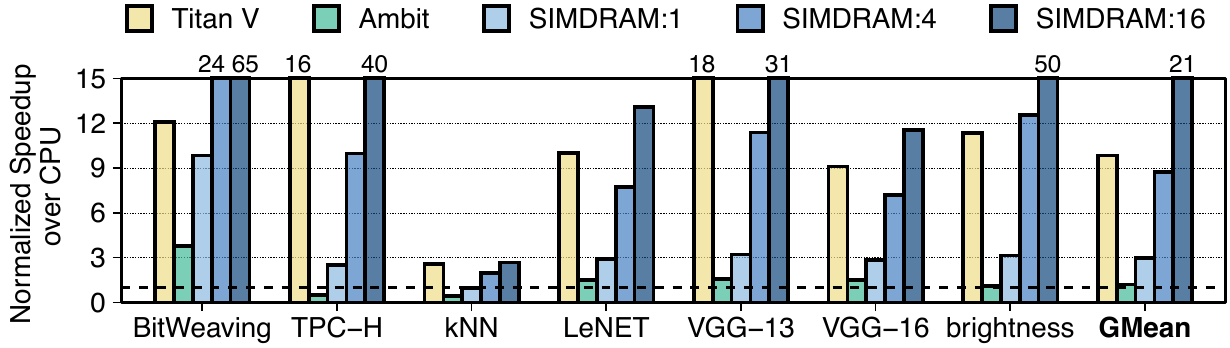}%
   \caption{\omvi{\omii{Normalized s}peedup of real-world kernels.} 
   }
   \label{fig_speedup_real_world}
\end{figure}

\subsection{\omi{Comparison to DualityCache}}
\label{sec:eva:dualitycache}

\nasirevi{We compare SIMDRAM to 
\omi{DualityCache~\cite{dualitycache}, a closely-related 
\omiii{processing-using-cache} architecture.}
DualityCache is an in-cache computing framework that performs computation using discrete logic elements (e.g., logic gates, latches, muxes) that are added to the SRAM peripheral circuitry. \revonur{In-cache computing approaches \revonuri{(such as DualityCache)} need  data to be brought into the cache first\revonuri{, which} requires extra data movement (and even more \omii{if} the working set of the application does not fit in the cache) compared to in-memory computing approaches (like SIMDRAM).}}

\begin{figure}[h]
   \centering
   \includegraphics[width=\linewidth]{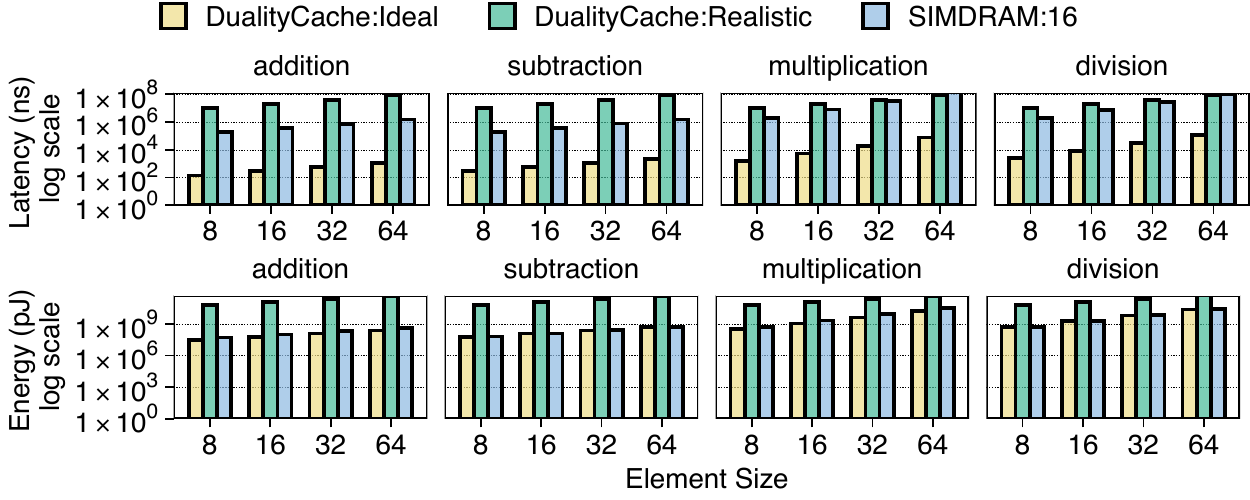}%
   \caption{\omvi{Latency and energy to execute 64M operations.}
   }
   \label{fig_dualitycache}
\end{figure}

\cmr{Fig.}~\ref{fig_dualitycache} \omii{(top)} compares the \omii{latency} of SIMDRAM against DualityCache~\cite{dualitycache} for the subset of operations that \emph{both} SIMDRAM and DualityCache support \revonur{(i.e., addition, subtraction, multiplication, and division)}. In this experiment, we study three different configurations. \revonuri{First,} \emph{\geraldorevii{DualityCache:\omiii{Ideal}}} has \revonuri{\emph{all}} data \revonuri{required} for DualityCache residing in the cache. Therefore, results for \emph{\geraldorevii{DualityCache:\omiii{Ideal}}} do \omiii{\emph{not}} include the overhead of moving data from DRAM to the cache\revonur{, making it an unrealistic configuration that needs the data to already reside and fit in the cache}. \revonuri{Second,} \emph{\geraldorevii{DualityCache:\omiii{Realistic}}} includes the overhead of data movement from DRAM to the cache. Both DualityCache configurations compute \omiv{on} an input array of \SI{45}{\mega\byte}. \omii{\revonur{Third}, \omii{\emph{SIMDRAM:16}.}}
\omiv{For all three configurations, we use the same cache size (\SI{35}{\mega\byte}) as the original DualityCache work~\cite{dualitycache} to provide a fair comparison}. 
As shown in the figure, \revonur{SIMDRAM \omii{greatly outperforms} \revonuri{DualityCache} when data movement is \revonurii{realistically} taken into account.} \geraldorevii{\emph{SIMDRAM:16}} \omii{outperforms} \emph{\geraldorevii{DualityCache:\omiii{Realistic}}} for all four operations \revonuri{(by 52.9$\times$, 52.\omii{4}$\times$, 1.\omii{8}$\times$, and 2.\omii{1}$\times$ for addition, subtraction, multiplication, and division} respectively, on average across all element sizes). \nasirevi{\revonur{SIMDRAM\omii{'s performance improvement} comes at a much lower area overhead compared to DualityCache. DualityCache \revsgii{(including its peripherals, transpose memory unit, controller, miss status holding registers, and crossbar network)} \omiv{has an area overhead of 3.5\% in a high-end CPU, whereas SIMDRAM has an area overhead of only 0.2\%~\omv{(\cmr{\cref{sec_area_overhead}})}.}
As a result, SIMDRAM can \revonuri{actually fit} \revsgii{a} significantly higher number of SIMD lanes in a given area compared to DualityCache. 
\revsgii{Therefore, SIMDRAM's \omii{performance improvement \omiv{per unit area} would be much larger \omiii{than \omv{that} we observe in \cmr{Fig.}~\ref{fig_dualitycache}.}}}}
\revonuri{We conclude that SIMDRAM achieve\omii{s} higher performance at lower area \omix{cost over} DualityCache\omii{, when we consider DRAM-to-cache data movement}.}}

\omi{\cmr{Fig.}~\ref{fig_dualitycache} \omii{(bottom)} shows} the energy consumption of \emph{DualityCache:\omiii{Realistic}}, \emph{DualityCache:\omiii{Ideal}}, and \emph{SIMDRAM:16} when performing \omi{64M addition, subtraction, multiplication, and division operations}. We make two observations. First, compared to \emph{\omii{Duality}Cache:\omiii{Ideal}}, \emph{SIMDRAM:16} increases average energy consumption by 60\%. This \omii{is} because while the energy per bit to \omii{perform} computation in DRAM (\SI{13.3}{\nano\joule}/bit~\cite{micropower,vogelsang2010understanding}) is smaller than the energy per bit to \omii{perform} computation in the cache (\SI{60.1}{\nano\joule}/bit~\cite{eckert2018neural}), the DualityCache implementation of \omiii{each} operation requires fewer iterations than \omiv{its} equivalent SIMDRAM implementation. Second, \emph{SIMDRAM:16} reduces average energy by 600$\times$ over \emph{\omii{Duality}Cache:\omiii{Realistic}} \omii{because} \emph{DualityCache\omii{:\omiii{Realistic}}} needs to load \emph{all} input data from DRAM, \omiv{incurring} high energy overhead (a DRAM access consumes 650$\times$ \omiii{the energy-per-bit of} a DualityCache operation~\omii{\cite{eckert2018neural, dualitycache}}). \omii{In contrast,} SIMDRAM operates on data that is already present in DRAM, eliminating any data movement \omii{overhead}. \omii{We conclude that SIMDRAM \omiii{is} much more efficient than DualityCache, when cache-to-DRAM data movement is \omiv{realistically} considered.}

\ifasploscr

\else

\subsection{Reliability}
\label{sec_reliability}

We use SPICE simulations to test the reliability of SIMDRAM for different technology nodes and \revonur{varying} \revonuri{amounts} of process variation. At the core of SIMDRAM, \omiii{there are} two back-to-back triple-row activations (TRAs). Table~\revdelrefr{\ref{table:variation}}\revdelrefa{\hl{3}} shows the \revonur{characteristics} of TRA and two back-to-back TRAs (TRAb2b) for the \omii{45, 32, and } \SI{22}{\nano\metre} technology nodes. We compare these with the reliability of quintuple-row activations (QRA\revonur{s}), used by prior works~\revdelrefr{\cite{ali2019memory, angizi2019graphide}}\revdelrefa{\hl{[7,9]}} to implement bit-serial addition. We use the reference \SI{55}{\nano\metre} DRAM model from Rambus~\revdelrefr{\cite{rambus_model}}\revdelrefa{\hl{[89]}} and scale it based on the ITRS roadmap~\revdelrefr{\cite{itrs_model,vogelsang2010understanding}}\revdelrefa{\hl{[48,104]}} to model smaller technology nodes 
following the PTM transistor models~\revdelrefr{\cite{ptm_model}}\revdelrefa{[83]}.  
\geraldorevi{The goal of our analysis is to understand the reliability trends for TRA and QRA operations \omii{with} technology \omii{scaling}.} For each technology \revonur{node} and \revonurii{process} variation \revonuri{amount}, we run Monte-Carlo simulations for $10^4$ iterations. 

\begin{table}[h]
\footnotesize
\tempcommand{0.8}
\renewcommand{\arraystretch}{1.1}
\centering
\caption{\cmr{Process variation's effect on TRA/QRA failure rates.}}
\begin{tabular}{ |c||c||c|c|c|c|c| } 
\hline
& \textbf{Variation (\%)} & \textbf{$\pm$ 0} & \textbf{$\pm$ 5} & \textbf{$\pm$ 10} & \textbf{$\pm$ 20}\\
\hhline{|=#=#=|=|=|=|}
\multirow{3}{*}{\textbf{45~nm}} & TRA Failure (\%)& 0 & 0 & 0.02 & 3.01 \\ 
& TRAb2b Failure (\%)& 0 & 0 & 0.04 & 5.93\\ 
& QRA Failure (\%) & 0 & 0 & 0.35 & 6.54\\ 
\hline
\multirow{3}{*}{\textbf{32~nm}} & TRA Failure (\%) & 0 & 0 & 0.35 & 3.90 \\ 
& TRAb2b Failure (\%) & 0 & 0 & 0.69 & 7.64\\ 
& QRA Failure (\%) & 0 & 0.42 & 6.33 & 11.52\\
\hline
\multirow{3}{*}{\revDMicro{\textbf{22~nm}}} & \revDMicro{TRA Failure (\%)} & \revDMicro{0} & \revDMicro{0} & \revDMicro{0.42} & \revDMicro{4.50} \\ 
& \revDMicro{TRAb2b Failure (\%)} & \revDMicro{0} & \revDMicro{0} & \revDMicro{0.84} & \revDMicro{8.83}\\ 
& \revDMicro{QRA Failure (\%)} & \omii{error} & \omii{error} & \omii{error} & \omii{error}\\
\hline
\end{tabular}
\label{table:variation}
\end{table}

We make \geraldorevi{four} observations. First, for all \geraldorevii{process} variation ranges
, TRA and TRAb2b perform more reliably than QRA. Specifically, TRA and TRAb2b perform without errors \omii{for} 5\% variation. Second, while moving from \SI{45}{\nano\metre} to \SI{32}{\nano\metre}, we observe that the error rate of QRA \omi{increases} faster than \omiii{than that of} TRA, making \omi{QRA} less reliable as the technology \omii{node} size \omii{reduces}. Third, for TRA and TRAb2b in \SI{22}{\nano\metre}, we observe a similar trend of increased error rate while still having zero error rate for 5\% process variation. In our simulations, QRA does not perform correctly 
\omii{in} the projected \SI{22}{\nano\metre} DRAM. For example, \texttt{MAJ(11100)} \omiii{always} leads to the incorrect outcome of \texttt{\omii{`}0'}. This is because charge sharing between five capacitors in QRA does not lead to enough voltage on the bitline for the sense amplifier to pull up the bitline to the value \texttt{\omii{`}1'}. We believe that proposals based on QRA require changes to the circuit elements (e.g., transistors in the sense amplifier) to enable \omii{correct} operation in \omiii{the} \SI{22}{\nano\metre} technology \omii{node}. 
\geraldorevi{Fourth, a TRA can fail depending on the amount of manufacturing process variation. We observe that a TRA starts to fail \revonur{when} process variation \revonur{is} larger than 10\%, for all \omii{technology nodes}. Since SIMDRAM operations are executed within a DRAM module, it is quite challenging to leverage existing \revonur{in-}DRAM \revonur{or in-memory-controller} error
correction mechanisms~\omiii{\cite{patel19, patel20, yixin14, meza2015revisiting}}. The same problem exists for \omii{other} \omi{processing-using-DRAM} mechanisms~\omiv{\cite{seshadri2017ambit, seshadri2013rowclone, li2017drisa, deng2018dracc, xin2020elp2im, ali2019memory,angizi2019graphide, li2018scope,seshadri2019dram, seshadri2017simple,seshadri2015fast, seshadri2016processing,subramaniyan2017parallel,dlugosch2014efficient}}.} 
\omii{We conclude that the TRA operations SIMDRAM relies on are much more scalable and variation-tolerant than QRA operations some prior works rely on.} \omi{We leave a study of
reliability solutions for future work.}

\fi 

\ifasploscr

\else

\subsection{Data Movement Overhead}
\label{sec:eval:datamovement}

\omi{There may be cases where the output of a \mech operation that is used as an input to a subsequent operation does not reside in the same subarray as other inputs.}  
\omii{For example, consider the computation $C = OP(A, B)$. If the output of the SIMDRAM operation $OP$ is an input to a subsequent SIMDRAM operation, \revonur{$C$} needs to move to the same subarray as the other inputs of the subsequent operation, before the operation can start.} 
\geraldorevi{\cmr{Fig.}~\ref{fig_data_movement} shows the \omii{distribution of the worst-case latency} overhead of moving the output of each of our \omii{16 SIMDRAM operations with 8-, 16-, 32-, and 64-bit element sizes \omii{in \emph{SIMDRAM:1}}} \omii{to a different subarray within the same bank, i.e., \emph{intra-bank} (using LISA~\revdelrefr{\cite{chang2016low}}\revdelrefa{\hl{[18]}}) or a different bank, i.e., \emph{inter-bank} (using RowClone PSM~\revdelrefr{\cite{seshadri2013rowclone}}\revdelrefa{\hl{[92]}})}}. 
We make two observations. First, \omii{intra-bank data movement \omii{(\cmr{Fig.}~\ref{fig_data_movement}, left)} results in \omii{only} 0.39\% latency overhead, averaged across all 16 \omii{SIMDRAM operations and four different element sizes}} (max. 1.52\% for 8-bit reduction, min. 0.001\% for 64-bit multiplication). 
Second, 
\omii{inter-bank data movement} \omii{(\cmr{Fig.}~\ref{fig_data_movement}, right) results in 17.5\% \revonur{\omii{latency}} overhead, averaged across all \omii{16 SIMDRAM operations and four different element sizes}} (max. 68.7\% for 8-bit reduction, min. 0.03\% for 64-bit multiplication). 
\omiv{We observe that the latency overhead of moving data, as a fraction of the total \omv{computation} latency \emph{decreases} \omiii{with} element size, \omii{because} the \omv{computation} latency of \omii{each} SIMDRAM operation increases \omii{with element size}.} 
\revonur{We conclude that while efficient data movement is a challenge in processing-in-memory architectures that rely on moving and aligning operands, the performance overhead of data movement in SIMDRAM stays within \revonuri{an} acceptable range} \omii{even under worst-case assumptions.}

\begin{figure}[h]
   \centering
   \includegraphics[width=\linewidth]{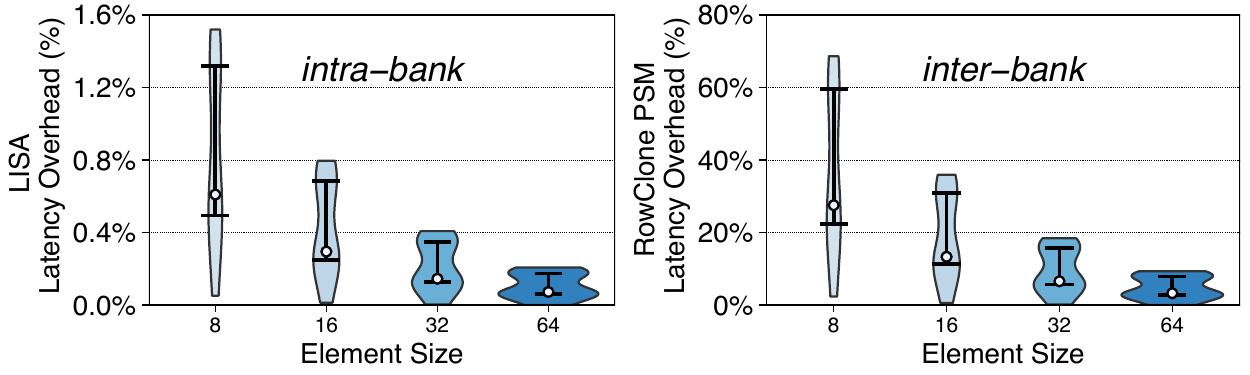}%
   \caption{\omii{Latency overhead distribution of \omiii{worst-case} intra-bank (left) and inter-bank (right) data movement} \cmr{for \emph{SIMDRAM:1}. Error bars depict the 25th and 75th percentiles.}} 
   \label{fig_data_movement}
\end{figure}

\fi 

\ifasploscr

\else

\subsection{Data Transposi\omii{tion} Overhead}
\label{sec:eval:transposing}

\nasirevi{
\revonuriii{\omii{T}ransposition} of the data in one subarray can overlap with in-DRAM computation in another subarray. As a result, if the data required for in-DRAM computation spans over multiple subarrays, only the transposition of the data in the first subarray is on the critical path of \omii{SIMDRAM} execution. The data in each remaining subarray is then transposed simultaneously with the in-DRAM computation in the previous subarray.} 

\nasirevi{To better understand the overhead of transposing data, we evaluate the \omiii{worst-case} \geraldorevii{latency of data transposition\omiii{, which is when SIMDRAM's data initially resides in the cache in a horizontal layout.  Before the computation of the SIMDRAM operation can start, this data needs to be transposed to a vertical layout and transferred to DRAM, incurring additional latency.} 
\cmr{Fig.}~\ref{fig_transposing_latency_transpose} shows \omiii{this worst-case data} \geraldorevii{transposition latency and the \omii{distribution of latency overhead of} data transposi\omii{tion in \emph{SIMDRAM:1}} across all \omii{16} SIMDRAM operations,} as a function of \omi{element \omii{size}}.  We make \geraldorevii{three} observations. \omii{First}, \omii{in \emph{SIMDRAM:1} (\emph{SIMDRAM:16}),} \omii{data} transposition incurs \geraldorevii{7.1}\% \omii{(44.6\%)} \omii{latency overhead} across all SIMDRAM operations (min.\ \geraldorevii{0.03\% \omii{(0.55\%)} for 64-bit multiplication, max.\ 38.9\% \omii{(91.1\%)} for 8-bit \revonuriii{AND-reduction} and {OR-reduction}}). \omii{\omiii{As shown in \cmr{\cref{sec_performance}},} for all the evaluated element sizes, \emph{SIMDRAM:1} (\emph{SIMDRAM:16}) outperforms the CPU \omii{and GPU} baselines by 5.5$\times$ and \omii{0.4$\times$} \omii{(88.0$\times$ and 5.8$\times$) on average across all 16 SIMDRAM operations, respectively.} Even \revonuriii{when we include the data} transposition overhead, \emph{SIMDRAM:1} (\emph{SIMDRAM:16}) still outperforms both the CPU and GPU baselines by  
\geraldorevii{4.0$\times$ and \omii{0.24$\times$} \omii{(20.0$\times$ and 1.4$\times$)}} on average across all 16 SIMDRAM operations. }
Our analysis \revonuriii{for kernels that represent the behavior of real-world applications} (\cmr{\cref{sec_real_world_kernels}}) \omiii{\emph{already includes}} the  data transposition \omii{overhead}. 
\omii{Second}, the data transposi\omii{tion} latency \omii{significantly} increases  \omii{with element} size (\omiii{by} 9.7$\times$ from 8-bit \omii{elements} to 64-bit \omii{elements}). \geraldorevii{\omii{T}he number of cache lines that need to be transposed increases linearly \omii{with element size}, which, in turn, increases the total transposi\omii{tion} latency. Third, even though the \omii{transposition} latency increases with element size, the \omii{transposition} overhead \omi{as a fraction of the total latency} \emph{decreases} \omiii{with} element size, \omii{because} the latency of \omii{each} SIMDRAM operation also increases \omii{with element size}. \omii{Since 
the transposition of data in each subarray is overlapped with the computation in another subarray,} 
the increase in \omii{transposition} latency is amortized over an even higher increase in the SIMDRAM operation latency.} We conclude that SIMDRAM \revonuriii{can} efficiently perform in-DRAM computation even when \omiii{worst-case} data transposition overhead is taken into account.}
}

\begin{figure}[h]
   \centering
   \includegraphics[width=\linewidth]{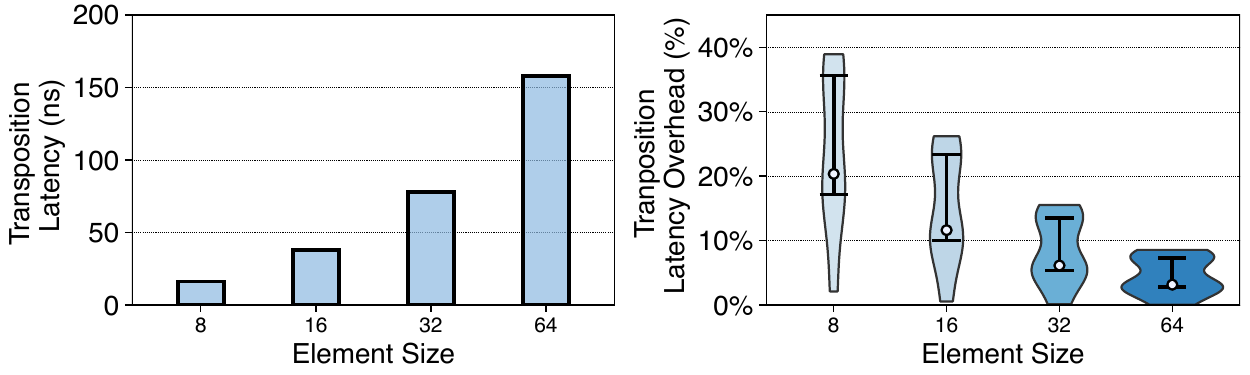}%
   \caption{\omiii{Worst-case latency} \omii{(left)} and \omiii{worst-case} \omii{latency overhead distribution (\omii{right}) of \omiii{data} transposition in 16 SIMDRAM operations for \emph{SIMDRAM:1}. \omii{Error bars depict the 25th and 75th percentiles, and a \omiv{bubble} depicts the 50th percentile.}}}
  \label{fig_transposing_latency_transpose}
 \end{figure}

\fi 

\subsection{Area Overhead}
\label{sec_area_overhead}

We use \revonur{CACTI}~\cite{cacti} to evaluate the area overhead of the primary components in the SIMDRAM design using a \SI{22}{\nano\meter} technology node. SIMDRAM does not \omi{introduce any modifications} to DRAM circuitry \revonur{other} than those proposed by Ambit\geraldorevii{,} \geraldorevi{which \revonur{has} an area overhead of $<$1\% \revonur{in a commodity} DRAM chip~\revdelrefr{\cite{seshadri2017ambit}}\revdelrefa{\hl{[93]}}}. \revonur{Therefore, SIMDRAM's area overhead over Ambit is only}
two structures in the memory controller: the control and transposition units. 

\textbf{Control Unit Area Overhead.} \geraldorevii{The main component\omi{s} in \omi{the} SIMDRAM control unit are the (1)~\emph{bbop} FIFO, (2)~\uprog{} Scratchpad, (3)~\uop{} Memory.} We size the \emph{bbop} FIFO and \uprog{} Scratchpad to \SI{2}{\kilo\byte} each. The size of the \emph{bbop} FIFO is enough to hold up to 1024 \emph{bbop} \omii{instructions}, which we observe is more than enough for our real-world applications. \omi{T}he size of the \uprog{} Scratchpad is large enough to store the \uprog{}s for all 16 SIMDRAM \omii{operations} that we evaluate in the paper (16~\uprog{}s $\times$ \SI{128}{\byte} max per \uprog{}). We use a \SI{128}{\byte} scratchpad for the \uop{} Memory.\textsuperscript{\ref{footnote:upgram}} 
We estimate that the \omi{SIMDRAM} control unit area is \SI{0.04}{\milli\meter\squared}. 

 \geraldorevii{\textbf{Transposition Unit Area Overhead.} The primary components in the transposition unit are (1)~the Object Tracker and (2)~two transposition buffers. 
 We use an \SI{8}{\kilo\byte} fully-associative cache with a 64-bit cache line size for the Object Tracker. This is enough to store 1024~entries in the Object Tracker, where each entry holds the base physical address of a SIMDRAM object (19~bits), the total size of the allocated data (32~bits), and the \omii{size} of each element in the object (6~bits). Each transposition buffer is \SI{4}{\kilo\byte}, to transpose up to a 64-bit SIMDRAM object (64-bit $\times$ \SI{64}{\byte}).}  
 \omi{We estimate the transposition unit area is \SI{0.06}{\milli\meter\squared}.} 
\omi{Considering the area of the control and transposition units,
SIMDRAM \omii{has} an area overhead of \omii{only} 0.2\% compared to the die area of an Intel Xeon E5-2697 v3 CPU~\cite{dualitycache}.} 
\omi{W}e conclude that SIMDRAM has low area cost. 

\section{Related Work}
\label{relatedWork}

To our knowledge, SIMDRAM is the first end-to-end framework that supports in-DRAM computation flexibly and transparently \omi{to the user}. We highlight SIMDRAM’s key contributions by contrasting it with state-of-the-art \omii{processing-in-memory} designs.

\noindent\textbf{Processing-\omiii{near}-Memory \omiii{(PnM)} \sgii{\omii{w}ithin} 3D-Stacked Memories.} Many recent works (e.g.,~\cite{ahn2015scalable,
nai2017graphpim,
boroumand2018google, 
lazypim, 
top-pim, 
gao2016hrl, 
santos2017operand, 
NIM, 
gu2020ipim, 
lenjani2020fulcrum, 
gokhale1995processing, 
fernandez2020natsa, 
cali2020genasm, 
boroumand2019conda, 
giannoula2021SynCron,
santos2018processing, 
PEI,
hsieh2016accelerating,
hsieh2016transparent,
kim2018grim,
drumond2017mondrian,
gao2017tetris,
Kim2016,
farmahini2015nda,
loh2013processing,
devaux2019true,
picorel2017near}) explore adding logic directly to the logic layer \sgii{of} 3D-stacked memories (e.g., \sgii{High-Bandwidth Memory}~\revdelrefr{\cite{HBM, lee2016simultaneous}}\revdelrefa{\hl{[62\revonurii{, 63}]}}, \sgii{Hybrid Memory Cube}~\revdelrefr{\cite{HMC2}}\revdelrefa{\hl{[49]}}). \omii{T}he implementation of SIMDRAM is considerably simpler, and relies on minimal modifications to \omii{commodity} DRAM \omii{chips}.

\noindent \textbf{Processing-\omiii{using}-Memory (\omiii{PuM}).} Prior works propose mechanisms wherein the memory arrays themselves perform various operations in bulk~\omii{\cite{seshadri2017ambit, seshadri2013rowclone, li2017drisa, deng2018dracc, xin2020elp2im, ali2019memory,angizi2019graphide, li2018scope,seshadri2019dram, seshadri2017simple,seshadri2015fast, seshadri2016processing,dlugosch2014efficient,subramaniyan2017parallel}}. SIMDRAM supports a much \sgii{wider} range of operations (compared to \revdelrefr{\cite{xin2020elp2im,li2017drisa,deng2018dracc,seshadri2017ambit, seshadri2013rowclone, ali2019memory,angizi2019graphide, li2018scope}}\revdelrefa{\hl{[8, 10, 23, \color{blue}72\color{black}, 73, 97, 98, 111]}}), at lower computational cost (compared to \revdelrefr{\cite{xin2020elp2im,seshadri2017ambit}}\revdelrefa{\hl{[98, 111]}}), \sgii{at lower} area overhead (compared to \revdelrefr{\cite{li2017drisa}}\revdelrefa{\hl{[73]}}), and \sgii{with more} reliable execution (compared to \revdelrefr{\cite{ali2019memory,angizi2019graphide}}\revdelrefa{\hl{[8, 10]}}).

\noindent \textbf{Processing-\omiii{in}-Cache.} \omi{Recent works~\cite{aga2017compute, eckert2018neural, dualitycache} propose} in-SRAM accelerators that take advantage of the SRAM bitline structures to perform bit-serial computation \omi{in caches}.  SIMDRAM shares similarities with these approaches, but offers a significantly lower cost per bit by exploiting the high density and low \omii{cost of} DRAM technology. \omi{We show the large performance and energy advantages of SIMDRAM compared to \omii{DualityCache}~\cite{dualitycache} in \cmr{\cref{sec:eva:dualitycache}}.}

\noindent \textbf{Frameworks for \omii{PIM}.
} 
\omi{F}ew prior works tackle the challenge of providing end-to-end support for \omii{PIM}. \omi{W}e describe these frameworks and their limitations for in-DRAM computing. \sgii{DualityCache~\cite{dualitycache} is} an end-to-end framework for in-cache computing. DualityCache utilizes \omiii{the} \sgii{CUDA}/OpenAcc programming languages~\omiii{\cite{cheng2014professional, OpenACCA1:online}} to generate code for an in-cache mechanism that execute\geraldorevi{s} a fixed set of operations in a \revsgii{single\omii{-}instruction multiple\omii{-}thread (SIMT)} manner. Like SIMDRAM, DualityCache stores data in a vertical layout through the \revonurii{bitlines} of the SRAM array. It \revonurii{treats} each \revonurii{bitline} as an independent execution thread and utilizes a crossbar network to allow inter-thread communication across \revonurii{bitlines}. Despite its benefits, employing DualityCache \revonurii{in DRAM} is not straightforward for two reasons. 
\omii{First, extending the DRAM subarray with the crossbar network utilized by DualityCache \revonurii{in SRAM} to allow inter-thread communication \omi{would} impose \omiii{a} prohibitive area overhead \omii{in DRAM} (9$\times$ the DRAM subarray area).} 
Second, as an in-cache \revonurii{computing} solution, DualityCache does \revonurii{\emph{not}} account for the limitations of in-DRAM computing, \omiii{i.e.,} \omiv{DRAM
operations that destroy input data, limited number of DRAM rows
that are capable of processing-using-DRAM, and the need to avoid
costly in-DRAM copies.} 
\omi{\omiv{We have already shown that SIMDRAM achieves higher performance at lower area overhead than DualityCache, \omii{when DRAM-to-cache data movement is \omiii{realistically} taken into account} (\cmr{\cref{sec:eva:dualitycache}}).}}


Two prior works propose frameworks targeting ReRAM devices. \revsgii{Hyper-AP~\revdelrefr{\cite{zha2020hyper}}\revdelrefa{\hl{[112]}} is} a framework for \revsgii{associative processing} using ReRAM. Since Hyper-AP targets \sgii{associative processing}, the proposed framework is \emph{fundamentally} different from \omii{SIMDRAM}. \revsgii{IMP~\revdelrefr{\cite{fujiki2018memory}}\revdelrefa{\hl{[31]}} is} a framework for in-situ ReRAM operations. Like DualityCache, the IMP framework depends on particular structures of the ReRAM array (\revonurii{such} as \revonurii{analog-to-digital/digital-to-analog converters}) to \omi{perform} computation \sgii{and, thus,} \revonurii{is not} applicable to an in-DRAM substrate \revonurii{that performs \revsgii{bulk bitwise} operations}. Moreover, \geraldorevi{DualityCache, Hyper-AP, and IMP} \omi{each} have a rigid ISA that enables \revonurii{only} a \revonurii{limited set} of in-memory operations \revonurii{(DualityCache supports 16 in-memory operations\revsgii{, while} both Hyper-AP and IMP support 12)}. In contrast, SIMDRAM is the first framework \geraldorevi{for \omiii{PuM}} that is flexible, \sgii{providing} a methodology that allows new operations to be \omi{integrated and} computed in memory as needed. \revonurii{In summary}, SIMDRAM fills the gap for a \revonurii{flexible} end-to-end framework \revonurii{that targets} \omii{processing-using-DRAM}. 

\section{Conclusion}
\label{conclusion}

\omii{We introduce \mech, a massively\omii{-}parallel general-purpose processing-using-DRAM framework that (1)~enables the efficient implementation of \omii{a wide variety of operations in DRAM,} in SIMD fashion, and (2)~provides a flexible mechanism to support the implementation of arbitrary user-defined operations. \mech introduces a \omii{new} three-\omiii{step} framework to enable efficient MAJ/NOT-based in-DRAM implementation for complex operations of different categories (e.g., arithmetic, relational, predication), and is applicable to a wide range of real-world applications.  We design the hardware and
ISA support for SIMDRAM framework to (1)~address key system
integration challenges, and (2)~allow programmers to employ new
SIMDRAM operations without hardware changes.} We 
\omii{experimentally demonstrate} that \mech provides significant \omiii{performance and energy benefits} over \omi{state-of-the-art CPU, GPU, and \omiii{PuM} systems. We hope that future work builds on our framework to \omiii{further} \omii{ease} \omiii{the} adoption \omiii{and improve the performance and efficiency} of processing-using-DRAM architectures and applications.}

\section*{Acknowledgments} 

\omii{We thank our shepherd Thomas Wenisch and the anonymous reviewers of MICRO 2020 \omiii{and ASPLOS 2021} for \omiii{their feedback}.} \omi{We thank the
SAFARI Research Group members for valuable feedback and the stimulating intellectual environment they provide. We acknowledge the generous gifts \omiii{of} our industrial partners\omiii{, especially} Google, Huawei, Intel, Microsoft, \omiii{and} VMware. \omii{This research was partially
supported by the Semiconductor Research Corporation.}}
\vfill\eject

 \balance 
\bibliographystyle{IEEEtranS}
\bibliography{refs}

\begin{figure*}[!b]
    \centering
    \includegraphics[width=\linewidth]{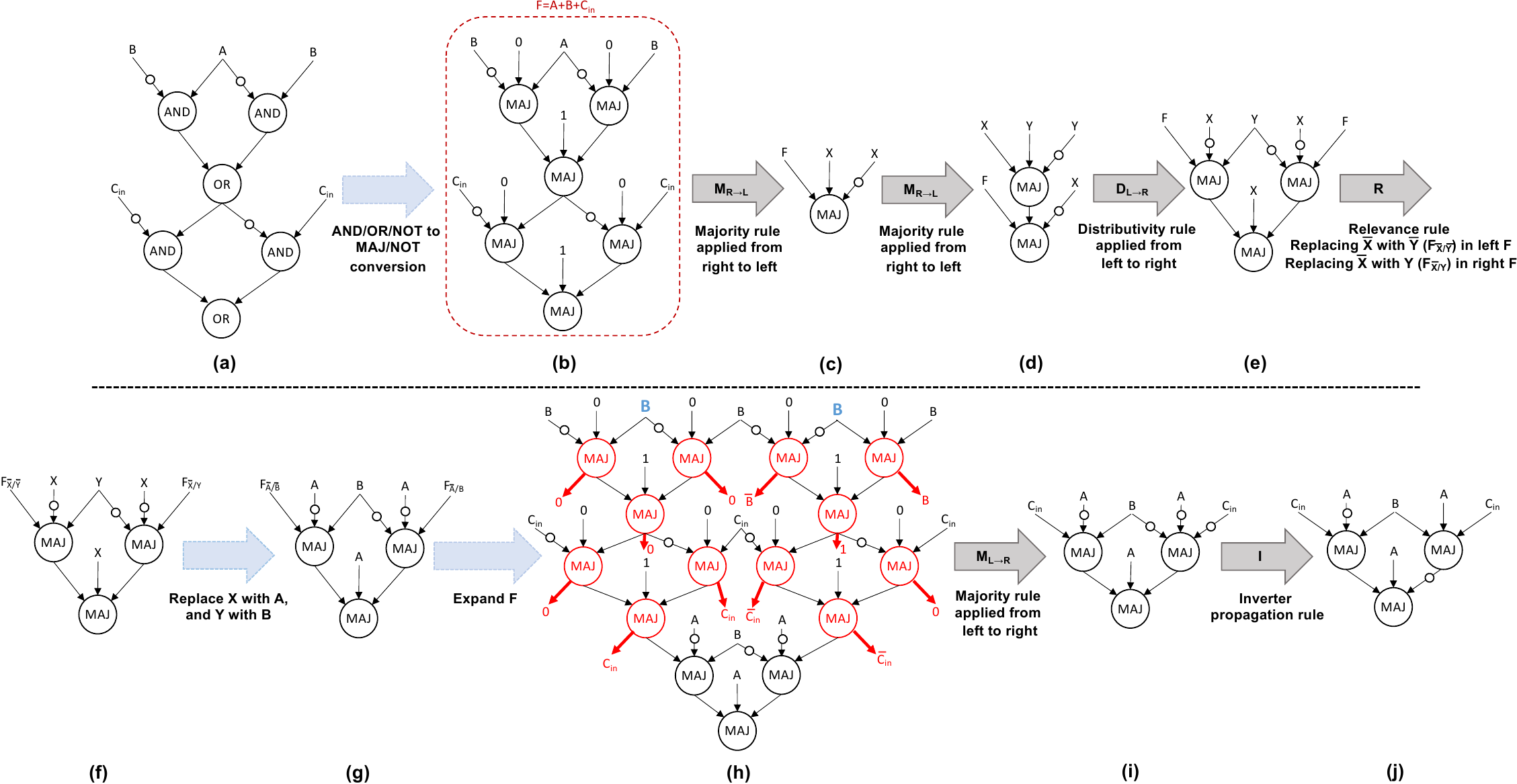}
    \caption{Synthesizing SIMDRAM circuit for a full addition.}
    \label{fig_add-maj}
\end{figure*}

 \newpage

\appendix
\noindent \textbf{\Large APPENDIX}
 \section{AIG-to-MIG Conversion}
 \label{apdx:aoi-to-mig}


The conversion from AND/OR/NOT representation of an operation to its MAJ/NOT representation \juang{relies} on a set of \juang{transformation} rules that are derived \juang{from the characteristics of the MAJ operation.} 
Table~\ref{table:rules} lists the set of transformation rules that we use 
to synthesize a circuit for a desired operation with MAJ and NOT gates. 
We use full \omdefi{addition} as a \juang{running} example to describe the process of synthesizing a \juang{MAJ/NOT-based} circuit\omdefi{,} starting from an AND/OR/NOT representation of the circuit \omdefi{and} using the transformation rules. 
\juang{We obtain MAJ/NOT-based circuits for other \mech operations following the same method. 
In a later step (\cref{sec:mi-aap}), we translate \omdefi{a} MAJ/NOT-based circuit to sequences of \omdefi{\aaps} operations.}

\begin{table}[h]
    \caption{\omdefi{MAJ/NOT} transformation Rules~\cite{epflmaj}.}
    \label{table:rules}
    \centering
    \footnotesize
    \tempcommand{1.3}
    \renewcommand{\arraystretch}{0.8}
    \resizebox{\columnwidth}{!}{
    \begin{tabular}{ll}
    \toprule
    \textbf{Commutativity (C)} & $M(x, y, z) = M(y, x, z) = M(z, y, x)$\\
    \cmidrule(rl){1-2}
    \multirow{2}{*}{\textbf{Majority (M)}} & if$(x = y): M(x, y, z) = x = y$ \\
    & if$(x = \overline{y}): M(x, y, z) = z$ \\
    \cmidrule(rl){1-2}
    \textbf{Associativity (A)} & $M(x, u, M(y, u, z)) = M(z, u, M(y, u, x))$\\
    \cmidrule(rl){1-2}
    \textbf{Distributivity (D)} & $M(x, y, M(u, v, z)) = M(M(x, y, u), M(x, y, v), z)$\\
    \cmidrule(rl){1-2}
    \textbf{Inverter Propagation (I)} & $\overline{M}(x, y, z) = M(\overline{x}, \overline{y}, \overline{z})$\\
    \cmidrule(rl){1-2}
    \textbf{Relevance (R)} & $M(x, y, z) = M(x, y, \mathbf{z}^{}_{x/\overline{y}})$\\
    \cmidrule(rl){1-2}
    \textbf{Complementary Associativity (CA)} & $M(x, u, M(y, \overline{u}, z)) = M(x, u, M(y, x, z))$\\
    \bottomrule
    \end{tabular}
    }
\end{table}



\omdefi{Fig.}~\ref{fig_add-maj}a shows the optimized AND/OR/\omdefi{Inverter} (i.e., AND/OR/NOT) \omdefi{Graph} (AOIG) representation of a full \omdefi{addition} (i.e., F = A + B + C$_{in}$). As shown in \omdefi{Fig.}~\ref{fig_add-maj}b, the naive way to transform the AOIG to a \omdefi{Majority/Inverter} (i.e., MAJ/NOT) Graph (MIG) representation, is to replace every AND and OR \omdefi{primitive} with a \omdefi{three-input} MAJ \omdefi{primitive} where the third input is 0 or 1, respectively. The resulting MIG is in fact Ambit's~\cite{seshadri2017ambit} representation of the full \omdefi{addition}. 
While the AOIG in \omdefii{Fig.~\ref{fig_add-maj}a} is optimized for AND/OR/NOT operations, the resulting MIG in Fig.~\ref{fig_add-maj}b can be further \omdefi{optimized} by exploiting the 
\juang{transformation rules of the MAJ \omdefi{primitive} (Table~\ref{table:rules}\omdefii{, replicated from~\cite{epflmaj}})}. \omdefi{The MIG optimization is performed in two key steps: (1) node reduction, and (2) MIG reshaping.}

\omdefi{\textbf{Node reduction.} In order to optimize the MIG in Fig.~\ref{fig_add-maj}b, the first step is to reduce the number of MAJ nodes in the MIG.} 
As shown in Table 1, rules \textbf{M} and \textbf{D} reduce the number of nodes in a MIG if applied from left to right (i.e., $\mathbf{M}^{}_{L\rightarrow{R}}$) and from right to left (i.e., $\mathbf{D}^{}_{R\rightarrow{L}}$), respectively. $\mathbf{M}^{}_{L\rightarrow{R}}$ replaces a MAJ \omdefi{node} with \omdefi{a single} \omdefi{value}, and $\mathbf{D}^{}_{R\rightarrow{L}}$ replaces three \omdefi{MAJ nodes} with two \omdefi{MAJ nodes in the MIG}. \omdefi{The node reduction step applies} $\mathbf{M}^{}_{L\rightarrow{R}}$ and $\mathbf{D}^{}_{R\rightarrow{L}}$ as many times as possible to reduce the the number of MAJ operations \omdefi{in the MIG. We can see in Fig.~\ref{fig_add-maj}b that none of the two rules are applicable \juang{in the particular case of the full addition MIG}.} \omdefii{Therefore, Fig~\ref{fig_add-maj}b remains unchanged after applying node reduction.}


\omdefi{\textbf{MIG reshaping.}} When no further node reduction is possible, we \emph{reshape} the MIG in an effort to enable more node reduction opportunities by repeatedly using two sets of rules: 
(1)~rules $\mathbf{M}^{}_{R\rightarrow{L}}$, $\mathbf{D}^{}_{L\rightarrow{R}}$, and \textbf{R} to temporarily inflate the MIG and create more node reduction opportunities with the \omdefi{help of the} new nodes, and (2)~rules \textbf{A} and \revAMicro{\textbf{CA}}, to exchange variables between adjacent nodes.  \revAMicro{Note that in this step, rules \textbf{M} and \textbf{D} are \omdefi{applied} in the reverse direction compared to \omdefi{the} previous step \omdefi{(i.e., node reduction step)} which results in increasing the number of nodes in the MIG.} \omdefi{We now describe the MIG reshaping process for the full addition example (Fig.~\ref{fig_add-maj}b). For simplicity,} we first assume that the entire MIG is represented as function \textbf{F} \omdefi{that computes the full addition of the input operands \textbf{A} and \textbf{B}. Then, we} apply rule $\mathbf{M}^{}_{R\rightarrow{L}}$ \omdefi{while} introducing variable \textbf{X} to the MIG \omdefi{(as \omdefii{$F = M (F, x, \overline{x})$)}} without impacting the functionality of the MIG (Fig. \ref{fig_add-maj}c). We then apply \omdefi{the same rule again, and replace \textbf{X} with a new MAJ node while introducing variable \textbf{Y} (Fig.~\ref{fig_add-maj}d). Next,}  
by applying rule $\mathbf{D}^{}_{L\rightarrow{R}}$, we introduce a new \omdefi{MAJ} node and distribute the function \textbf{F} across \omdefi{the} two MAJ nodes \omdefi{(Fig.~\ref{fig_add-maj}e)}. \omdefii{Now, by applying rule \textbf{R} to the function \textbf{F} on the left, variable ${\overline{\textbf{X}}}$ is replaced with variable $\overline{\textbf{Y}}$ in the function \textbf{F} on the left. Similarly, by applying rule \textbf{R} to the function \textbf{F} on the right, variable ${\overline{\textbf{X}}}$ is replaced with variable ${\textbf{Y}}$ in the function \textbf{F} on the right \omdefi{(Fig.~\ref{fig_add-maj}f)}.} 
At this point, since \omdefi{rule} $\mathbf{M}^{}_{R\rightarrow{L}}$ holds with any given two variables, we can safely replace \textbf{X} and \textbf{Y} with variables \textbf{A} and \textbf{B}, respectively (Figure~\ref{fig_add-maj}g). Next, we expand function \textbf{F} \omdefi{(Fig.~\ref{fig_add-maj}h)} and the variables replaced as a result of the previous rule are highlighted in blue. As shown in Fig.~\ref{fig_add-maj}h, the resulting graph after expanding function \textbf{F} has multiple node reduction opportunities using rule $\mathbf{M}^{}_{L\rightarrow{R}}$ and starting from the top of the graph. The nodes that can be eliminated using this rule are marked in red and the replacing value is indicated with a red arrow leaving the node. Fig.~\ref{fig_add-maj}i shows the same MIG after resolving all the node reductions. We next use rule \textbf{I} to remove all three NOT \omdefi{primitives} in the rightmost MAJ node. The final optimized MIG that is shown in Figure~\ref{fig_add-maj}j \omdefi{requires only 3 MAJ \omdefi{primitives} to perform the full addition operation}\omdefii{, as opposed to the 6 we started with (in Fig.~\ref{fig_add-maj}b)}. 

\omdefi{The node reduction step followed by the MIG reshaping step are repeated (for a predefined number of times) until we achieve an optimized MIG that requires minimal number of MAJ operations to perform the desired in-DRAM operation.} \revBMicro{The process of converting an operation to a MAJ-based implementation can be automated as suggested by prior work~\cite{epflmaj,soeken2016mig}.}

 \section{Row-to-Operand Allocation}
 \label{apdx:row-to-op}


\nasrev{Algorithm~\ref{alg_mapping} describes SIMDRAM’s row-to-operand allocation procedure. \sgii{To enable in-DRAM computation, our \omdef{allocation} algorithm copies (i.e., maps) input operands for each MAJ node in the MIG from D-group rows (where the operands normally reside) into compute rows.  However, due to the limited number of compute rows, the \omdef{allocation} algorithm cannot \omdef{allocate DRAM rows to} all input operands from all MAJ nodes at once.} 
To address this issue, \juangg{the} \omdef{allocation} algorithm divides the \omdef{allocation} process into \emph{phases}. Each phase \omdef{allocates as many compute rows to} operands as possible. For example, \sgii{because no rows are \omdef{allocated} yet, the initial phase (Phase~0)} has all six compute rows \sgii{available for \omdef{allocation} (i.e., the rows are vacant),} and can \omdef{allocate up to six input operands to the compute rows}. 
\sgii{A phase} is considered finished \sgii{when either
(1)~there are not enough vacant compute rows to \omdef{allocate} all input operands for the next logic primitive that needs to be computed, or
(2)~there are no more MAJ primitives left to process in the MIG.} 
\omdef{The phase information is used when generating the \uprog{} for the MIG in Task 2 of Step 2 of SIMDRAM framework (\cref{sec:framework:step2:generating}), where \uop{}s for all MAJ primitives in phase $i$ are generated prior to the MAJ primitives in phase $i+1$. Knowing that all the MAJ primitives in phase $i$ are performed before the next phase $i+1$ starts, \omi{the \omdef{allocation} algorithm can safely \omdef{\omdefii{reuse}}} \sgii{the compute rows for use} in phase $i+1$, without worrying about the output of a MAJ primitive being overwritten by a new row-to-operand allocation. 
}}

\nasrev{\sgii{We now describe the \omdef{row-to-operand allocation} algorithm in detail, using the MIG for full \omdef{addition} in \cmr{Fig.}~\ref{fig_output_mapping}a as an example of a MIG being traversed by the algorithm.}
The \omdef{allocation} algorithm starts \sgii{at} Phase~0. \sgii{Throughout its execution, the algorithm maintains 
(1)~the list of free compute rows \omdefii{(\omdefii{rows in the B-group of the subarray, shown in Fig.~\ref{fig_subarray}})} that are available for \omdef{allocation} (\emph{B\_rows} and \emph{B\_rows\_DCC} in Algorithm~\ref{alg_mapping}\omdefii{, initialized in lines 3--4}); and
(2)~}the list of \omdef{row-to-operand} \omdef{allocations} associated with each MAJ node, tagged with the phase number that the \omdef{allocations} were performed in (\emph{row\_operand\_allocation} in Algorithm~\ref{alg_mapping}). Once a \omdef{row-to-operand} \omdef{allocation} is performed, the algorithm removes the compute row used for the \omdef{allocation} from the list of the free compute rows, and adds the new \omdef{allocation} to the list of \omdef{row-to-operand allocations generated} in that phase for the corresponding MAJ node. The algorithm follows a simple procedure to \omdef{allocate compute rows to}  
the input operands of the MAJ nodes in the MIG. The algorithm does a topological traversal starting with the leftmost MAJ node in the highest level of the MIG (\sgii{e.g.}, Level 0 in \cmr{Fig.}~\ref{fig_output_mapping}a), and traverses all the MAJ nodes in each level, before moving to the next \omdef{lower} level of the graph. }

\algnewcommand{\algorithmicgoto}{\textbf{go to}}%
\algnewcommand{\Goto}[1]{\algorithmicgoto~\ref{#1}}%
\begin{algorithm}[!ht]
  \caption{\juang{\mech's \omdef{Row-to-Operand Allocation Algorithm}.}}
  \label{alg_mapping}
  \tiny
  \begin{algorithmic}[1]

    \State \textbf{Input:} MIG \texttt{G} = (\texttt{V}, \texttt{E})   \algorithmiccomment{Majority-Inverter Graph \texttt{G} nodes <vertex, edge>}
    \State \textbf{Output:} \texttt{row\_operand\_allocation} \algorithmiccomment{Allocation map of rows to operands}
    
    \vspace{0.5em}
    \State \texttt{B\_rows} $\gets$ \texttt{\{T0, T1, T2, T3\}}
    \State \texttt{B\_rows\_DCC} $\gets$  \texttt{\{DCC0, DCC1\}} 
    \State \texttt{phase} $\gets$ 0
    \State \texttt{row\_operand\_allocation\_map} $\gets \emptyset$
    \vspace{0.5em}
  
    \ForEach{level in \texttt{G}}
        \ForEach{\texttt{V} in \texttt{G}[level]}
            \ForEach{\texttt{input edge} in \texttt{E[V]}}
                \State Search for \texttt{input edge}'s parent
    \If{input edge has no parents}\tikzmark{top}
                    \If{\texttt{input edge} is negated}
                    \State Allocate row in \texttt{B\_rows\_DCC} to \texttt{input edge}
                        \State Remove allocated row from \texttt{B\_rows\_DCC}
                    \Else
                        \State Allocate row in \texttt{B\_rows} to  \texttt{input edge}
                        \State Remove allocated row from \texttt{B\_rows} 
                    \EndIf \tikzmark{bottom}
                \Else \tikzmark{topblue}
                    \If{ \texttt{input edge} is negated}
                        \State Map allocated parent row in \texttt{B\_rows\_DCC} to  \texttt{input edge}
                    \Else
                        \State Map allocated parent row in \texttt{B\_rows} to  \texttt{input edge}
                    \EndIf
                \EndIf \tikzmark{bottomblue}

            \If{\texttt{B\_rows} and \texttt{B\_rows\_DCC} are empty} \tikzmark{topgreen}
                \State \texttt{phase} $\gets$ \texttt{phase + 1}
                \State \texttt{B\_rows} $\gets$ \texttt{\{T0, T1, T2, T3\}}
                \State \texttt{B\_rows\_DCC} $\gets$ \texttt{\{DCC0, DCC1\}} 
            \EndIf \tikzmark{bottomgreen}
            
             \State \texttt{row\_operand\_allocation} $\gets$ (\texttt{input edge}, allocated row, \texttt{phase}) \tikzmark{right}
            \EndFor 
        \EndFor
    \EndFor
    
  \end{algorithmic}
  \AddNoteRed{top}{bottom}{right}{\normalsize Case 1}
    \AddNoteBlue{topblue}{bottomblue}{right}{\normalsize Case 2}
    \AddNoteGreen{topgreen}{bottomgreen}{right}{\normalsize Case 3}
  
\end{algorithm}

\nasrev{For each of the three 
input edges (i.e., operands) of any given MAJ node, the algorithm checks for the following three possible cases and performs the \omdef{allocation} accordingly: }

\noindent\nasrev{\sgii{\textbf{Case~1:}} if the edge is not connected to another MAJ node in \sgii{a} \omdef{higher} level of the graph \omdef{(line 11 in Algorithm~\ref{alg_mapping})}, i.e., the edge does not have a parent (e.g., \sgii{the three edges entering} the blue node in \cmr{Fig.}~\ref{fig_output_mapping}a), \sgii{and a compute row is available,} the 
input operand associated with the edge is considered \sgii{to be a source input, and is currently located} in the D-group rows of the subarray. As a result, the algorithm copies the input operand associated with the edge from \sgii{its D-group row} to the first available compute row. Note that if the edge \omdef{\emph{is}} complemented, i.e., the input operand is negated (e.g., the edge with operand A for the blue node in \cmr{Fig.}~\ref{fig_output_mapping}a), the algorithm \omdef{allocates the first available compute row with dual contact cells (DCC0 or DCC1) to}  the input operand of the edge (lines \omdef{12--14} in Algorithm~\ref{alg_mapping}). If the edge is \omdef{\emph{not}} complemented (e.g., the edge with operand B for the blue node in \cmr{Fig.}~\ref{fig_output_mapping}a), \omdef{a regular compute row is allocated to} the input operand 
(lines \omdef{15--17} in Algorithm~\ref{alg_mapping}). }

\noindent\nasrev{\sgii{\textbf{Case~2:}} if the edge is connected to another MAJ node in \sgii{a} higher level of the graph \omdef{(line 18 in Algorithm~\ref{alg_mapping})}, 
\omdef{the edge has a parent node and the value of the input operand associated with the edge equals the \omdefii{result} of the parent node, \omdefii{which}} is available in the compute rows that hold the result of the parent MAJ node. As a result, the algorithm maps the input operand of the edge to a compute row that holds the result of its parent node (lines \omdef{19--22} in Algorithm~\ref{alg_mapping}). 
}

\noindent\nasrev{\sgii{\textbf{Case~3:}} if there are no free compute rows available, the algorithm \omdef{considers} the phase as \emph{complete} and continues the \omdef{allocations} in the next phase (lines \omdef{23--26} in Algorithm~\ref{alg_mapping}). }

\nasrev{Once \omdef{DRAM rows are allocated to} all the edges connected to a MAJ node, the algorithm stores the \omdef{row-to-operand allocation} information of the three input operands of the MAJ node in \omdef{\emph{row\_operand\_allocation}} (line \omdef{27} in Algorithm~\ref{alg_mapping}) 
and associates this information with the MAJ node and the phase number that the \omdef{allocations} were performed in. The algorithm finishes once \omdef{DRAM rows are allocated to} \emph{all} the input operands of all the MAJ nodes in the MIG. \cmr{Fig.}~\ref{fig_output_mapping}b shows these \omdef{allocations} as the output of Task 1 for the full \omdef{addition} example. \sgii{The resulting} \emph{row\_operand\_allocation} is then used in \omdef{Task 2 of Step 2 of the SIMDRAM framework} (\cmr{\cref{sec:framework:step2:generating}}) to generate the \omiii{series} of \uop{}s to compute the operation that the MIG represents. } 

 \section{Scalability of Operations}
 \label{apdx:op-class}

\revGeraldo{\nastaran{Table~\ref{table_aaps_operations} lists the semantics and the total number of AAP/APs required for each of the 16 \mech operations that we evaluate in this paper (\cref{sec_evaluation}) for input element(s) of size $n$.}} \omdef{Each operation is classified based on how the latency of the operation scales with respect to the element size $n$. Class 1, 2, and 3 operations scale linearly, logarithmically,and quadratically with $n$, respectively.}

\begin{table}[h]
\caption{\sgii{Evaluated \mech operations (for $n$-bit data)}.}
\label{table_aaps_operations}
\centering
\tempcommand{1.2}
\resizebox{\linewidth}{!}{
\begin{tabular}{|l|l|l|l|l|}
\hline
\multicolumn{1}{|l|}{\textbf{Type}} & \multicolumn{1}{l|}{\textbf{Operation}} & \multicolumn{1}{l|}{\textbf{\# AAPs/APs}} & \multicolumn{1}{l|}{\textbf{Class}} & \multicolumn{1}{l|}{\textbf{Semantics}} \\ 
\hline
\hline
\multirow{10}{*}{Arithmetic}& abs                       & \(10n - 2\)         & Linear              & \(dst = (src > 0) ?\ src : -(src)\) \\ \cline{2-5} 
                            & addition                  & \(8n + 1\)          & Linear              & $dst = src_1 + src_2 $                             \\ \cline{2-5} 
                            & \multirow{2}{*}{bitcount} &  $\Omega = 8n - 8\log_2 (n+1) $       &   \multirow{2}{*}{Linear}   & \multirow{2}{*}{ $\sum_{i=0}^{n} src(i)$}                        \\ \cline{3-3}
                            &                &  $O = 8n $             &                         &                    \\ \cline{2-5} 
                            & division       & $8n^{2} + 12n$         &  Quadratic  &            $dst = \frac{src_1}{src_2} $                \\ \cline{2-5} 
                            & max            & $10n + 2$              &  Linear     & $dst = (src_1 > src_2)?\ src_1 : src_2 $\\ \cline{2-5} 
                            & min            & $10n + 2   $           &  Linear     & $dst = (src_1 < src_2)?\ src_1 : src_2 $   \\ \cline{2-5} 
                            & multiplication & $ 11n^{2} - 5n - 1 $   &  Quadratic  & $dst = src_1 \times  src_2  $                      \\ \cline{2-5} 
                            & ReLU           & $3n + ((n-1)\ mod\ 2)$ &  Linear     & $dst = (src \geq 0)?\ src : 0  $ \\ \cline{2-5} 
                            & subtraction    & $8n + 1  $             &  Linear     & $dst = src_1 - src_2   $                        \\ \hline \hline
Predication                 & if\_else       & $7n$                   &  Linear     & $dst = (sel)?\ src_1 : src_2   $                 \\ \hline \hline
\hline
\multirow{3}{*}{Reduction}  & and\_reduction &  $5\floor{\frac{n}{2}} + 2$   &  Logarithmic & $Y = src(1)  \wedge src(2) \wedge src(3)  $                                  \\ \cline{2-5} 
                            & or\_reduction  & $5\floor{\frac{n}{2}}  + 2$   &  Logarithmic & $ Y = src(1)  \vee src(2) \vee src(3)$             \\ \cline{2-5} 
                            & xor\_reduction & $6\floor{\frac{n}{2}}  + 1$   &  Logarithmic & $Y = src(1)  \oplus src(2) \oplus src(3)$                                          \\ \hline \hline
\multirow{3}{*}{Relational} & equal          & $4n + 3$                      &  Linear      & $dst = (src_1 == src_2) $                       \\ \cline{2-5} 
                            & greater        & $3n + 2      $                &  Linear      & $dst = (src_1 > src_2)       $       \\ \cline{2-5} 
                            & greater\_equal & $3n + 2      $                & Linear       & $dst = (src_1 \geq src_2)    $       \\ \hline
\end{tabular}
}

\end{table}
 
 \section{Evaluated Real-World Applications}
 \label{appendix}

\noindent{ \textbf{\omdefii{Convolutional} Neural Networks (CNNs).} \geraldo{CNNs~\omdefii{\cite{rastegari2016xnor,he2020sparse,krizhevsky2012imagenet}} are used in many classification tasks \omdefi{such as} image and handwriting classification. \nastaran{CNNs are often} \omdefii{computationally} \nastaran{intensive} as they use many general-matrix-multiplication (GEMM) operations \omdefii{using} \omdefii{floating}-point operations for each convolution. 
Prior works~\cite{rastegari2016xnor, lin2017towards, he2020sparse} demonstrate that \omdefi{instead of the costly \omdefii{floating-point} multiplication operations, convolutions can be performed} 
\nastaran{using a series of bitcount, addition, shift, and XNOR operations}. In this work, we use the XNOR-NET~\cite{rastegari2016xnor} implementations of VGG-13, VGG-16, and LeNET provided by \cite{he2020sparse}, \omdefi{to evaluate the functionality of SIMDRAM}.  We modify \omdefi{these} implementations to make use of SIMDRAM's \omdefiii{bitcount}, addition, shift, and XNOR operations. We evaluate \omdefi{all} three networks \revBMicro{for inference} using two different datasets: VGG-13 and VGG-16 (using CIFAR-10~\cite{krizhevsky2010convolutional}), and LeNet-5 (using MNIST~\cite{deng2012mnist}).}}

\geraldo{\noindent{\textbf{k-Nearest Neighbor Classifier (kNN).} 
\omdefii{We use a kNN classifier to solve the handwritten digits recognition problem~\cite{Lecun95learningalgorithms}. The kNN classifier finds a group of k objects in the input set using a simple distance \omdefiii{algorithm} such as Euclidean distance~\cite{friedman2001elements}.} 
\omdefi{In our evaluations, we} use SIMDRAM to implement the Euclidean distance algorithm \omdefi{entirely} in DRAM. We evaluate a kNN algorithm using the MNIST dataset~\omdefii{\cite{deng2012mnist}} with 3000 training images and 1000 testing images.  We quantize the inputs using an 8-bit representation. }}

\geraldo{\noindent{\textbf{Database.} We evaluate \omdefi{SIMDRAM using} two different database workloads. First, 
we evaluate a simple table scan query \texttt{‘select count(*) from T where c1 <= val <= c2'} using \omdefii{the} BitWeaving algorithm~\cite{li2013bitweaving}. 
Second, we evaluate the performance of the TPC-H~\cite{tpch} scheme using query 01, \omdefi{\omdefii{which} executes many arithmetic operations, including addition and multiplication}. \omdefi{For our evaluation,} we follow the column-based data layout employed in \cite{santos2017operand} and use a scale factor of 100. }}

\geraldo{\noindent{\textbf{Brightness.} We use a simple image brightness algorithm~\omdefii{\cite{foley1996computer}} to demonstrate the benefits of \omdefii{the} SIMDRAM predication operation. The algorithm evaluates if a given brightness value is larger than 0. If so, it increases the pixel value of the image by the brightness value. Before assigning the new brightness value to the pixel, the algorithm verifies if the new pixel value is between 0 and 255. In our SIMDRAM implementation, we use both addition and predication operations. }}

\end{document}